\documentclass[a4paper,11pt]{article}
\pdfoutput=1 

\usepackage{jheppub} 

\usepackage[T1]{fontenc} 

\usepackage{float}
\usepackage{graphicx}
\usepackage{subcaption}
\usepackage{bbm}
\usepackage{siunitx}
\usepackage{listings}

\usepackage{bbold}

\usepackage{graphicx,epsf}
\usepackage{amsmath,amsfonts,amssymb,amsbsy}
\usepackage{mathtools}
\usepackage{physics}

\usepackage[dvipsnames]{xcolor}

\usepackage{hyperref}
\usepackage{xcolor}

\usepackage{booktabs} 
\usepackage{arydshln} 
\usepackage{adjustbox}

\usepackage{array}
\newcolumntype{C}[1]{>{\centering\arraybackslash}m{#1}}

\usepackage{multirow}
\usepackage{here}
\usepackage{stackrel}
\usepackage{tikz}
\usetikzlibrary{calc}

\usepackage{ifdraft}
\usepackage[obeyFinal]{easy-todo}

\usepackage{placeins}
\usepackage{slashed} 

\usepackage[utf8]{inputenc}

\usepackage{graphicx}
\usepackage{pdfpages}

\usepackage{cleveref}

\renewcommand{\imath}{\mathrm{i}}

\newcommand{\higgs}{\mathrm{H}}
\newcommand{\gluon}{\mathrm{g}}

\newcommand{\eps}{\epsilon}

\newcommand*{\gs}{g_\mathrm{s}}
\newcommand*{\alphas}{\alpha_\mathrm{s}}

\def\tr{{\operatorname{tr}}}
\newcommand*{\tracep}[1]{\operatorname{tr}\left( #1 \right)}

\def\NC{{N_{\!\!\:c}}}
\def\NF{{N_{\!\!\:f}}}

\def\SUN{{\mathrm{SU}(\NC)}}

\def\fig#1{fig.~{\ref{#1}}}

\def\tab#1{table~{\ref{#1}}}

\def\nn{\nonumber}

\def\spa#1.#2{\left\langle#1\,#2\right\rangle}
\def\spb#1.#2{\left[#1\,#2\right]}
\def\spash#1.#2{\spa{\smash{#1}}.{\smash{#2}}}
\def\spbsh#1.#2{\spb{\smash{#1}}.{\smash{#2}}}
\def\sand#1.#2.#3{%
\left\langle\smash{#1}{\vphantom1}^{-}\right|{#2}%
\left|\smash{#3}{\vphantom1}^{-}\right\rangle}
\def\sandpp#1.#2.#3{%
\left\langle\smash{#1}{\vphantom1}^{+}\right|{#2}%
\left|\smash{#3}{\vphantom1}^{+}\right\rangle}
\def\sandpm#1.#2.#3{%
\left\langle\smash{#1}{\vphantom1}^{+}\right|{#2}%
\left|\smash{#3}{\vphantom1}^{-}\right\rangle}
\def\sandmp#1.#2.#3{%
\left\langle\smash{#1}{\vphantom1}^{-}\right|{#2}%
\left|\smash{#3}{\vphantom1}^{+}\right\rangle}

\def\trFive{{\rm tr}_5}

\def\mH{m_H}
\def\mtop{m_t}

\def\Caravel{{\sc Caravel}}

\newcommand*{\Kira}{\texttt{Kira}}
\newcommand*{\FIRE}{\texttt{FIRE}}

\DeclareUnicodeCharacter{27E8}{\langle}
\DeclareUnicodeCharacter{27E9}{\rangle}
\DeclareUnicodeCharacter{0394}{\Delta}
\DeclareUnicodeCharacter{27EA}{\langle\langle}
\DeclareUnicodeCharacter{27EB}{\rangle\rangle}

\newcolumntype{C}[1]{>{\centering\arraybackslash}m{#1}}
\newcolumntype{M}{>{\centering\arraybackslash$}l<{$}}

\title{Two-loop leading-color QCD corrections for Higgs plus two-jet production in the heavy-top limit}

\preprint{\begin{tabular}{l} ZU-TH 18/26 \\ SLAC-PUB-260429\end{tabular}}

\author[1]{Giuseppe~De~Laurentis,}
\author[2,3]{Harald~Ita,}
\author[2,3]{Viktor~Kuschke,}
\author[4]{Michael~Ruf,}
\author[5]{Vasily~Sotnikov}

\affiliation[1]{Higgs Centre for Theoretical Physics, University of Edinburgh, Edinburgh, EH9 3FD, United Kingdom}
\affiliation[2]{Paul Scherrer Institut, CH-5232 Villigen PSI, Switzerland}
\affiliation[3]{Department of Astrophysics, University of Zurich, Winterthurerstrasse 190, 8057 Zurich, Switzerland}
\affiliation[4]{SLAC National Accelerator Laboratory, Stanford University, Stanford, CA 94309, USA}
\affiliation[5]{Physik-Institut, University of Zurich, Winterthurerstrasse 190, 8057 Zurich, Switzerland}

\emailAdd{
  giuseppe.delaurentis@ed.ac.uk,
  harald.ita@psi.ch,
  viktor.kuschke@psi.ch,
  mruf@slac.stanford.edu,
  vasily.sotnikov@physik.uzh.ch
}

\abstract{
  We present the leading-color two-loop QCD corrections for Higgs-boson
  production in association with two jets through gluon fusion in the
  heavy-top effective theory. We provide analytic expressions for the
  finite remainders of the 
  helicity amplitudes, written in terms of one-mass pentagon functions
  with spinor-helicity coefficients. These expressions are obtained by
  reconstructing the amplitudes from numerical finite-field samples
  computed within the numerical unitarity framework. The
  reconstruction is made possible by several advances in exploiting
  the analytic structure of the amplitudes, which both reduce the
  number of required samples and lead to compact representations. In
  particular, we introduce a new algorithm for multivariate partial
  fraction decomposition, based on a generic bivariate slice and a
  simplified treatment of ideal intersections. Using the resulting
  analytic expressions, we provide an efficient and stable
  implementation of their numerical evaluation, ready for
  phenomenological applications. Finally, we study the singularity
  structure of the remainders and confirm the existence of a threshold
  at non-degenerate physical momentum configurations, usually
  associated with massive virtual particle exchanges.
}

\begin{document}
\maketitle


\newpage
\section{Introduction}

The precise determination of the properties of the Higgs boson is one of the
central objectives of the LHC physics program.  A particularly important
production mode is weak vector-boson fusion (VBF), which gives direct access to
the couplings of the Higgs boson to the $W$ and $Z$ bosons.  The VBF process is
characterized by the presence of two well separated hard jets in the final state.
Here we are concerned with the main irreducible QCD background to this process, 
namely Higgs production in association with two jets through gluon fusion 
mediated by the top-quark Yukawa coupling (ggF $Hjj$).
Selection cuts, such as a large dijet invariant mass, can be used to enhance
the VBF contribution relative to ggF.  Nevertheless, a precise description of
ggF $Hjj$ production is essential for reducing the uncertainties associated
with disentangling the two production mechanisms~\cite{Chen:2025whf}.
The ggF $Hjj$ production is an important signal process in its own right,
as it provides sensitivity to the $\mathcal{CP}$ structure of the 
Higgs couplings~\cite{Klamke:2007cu,Demartin:2014fia,Bahl:2023qwk}.
Although current measurements are still statistically 
limited~\cite{CMS:2022wpo,ATLAS:2022fnp,ATLAS:2023pwa,CMS:2025fdx,ATLAS:2026ovz}, 
higher-order perturbative QCD
corrections will become increasingly important during the high-luminosity LHC
phase and at potential future hadron colliders~\cite{Chen:2025whf,Huss:2025nlt}.

The VBF contribution can be efficiently described in an approach where QCD 
corrections factorize into independent corrections to the two quark lines~\cite{Han:1992hr},
and is known through next-to-next-to leading order 
(NNLO)~\cite{Bolzoni:2010xr,Cacciari:2015jma,Cruz-Martinez:2018rod} and inclusively 
through N\textsuperscript{3}LO~\cite{Dreyer:2016oyx}.
The ggF $Hjj$ production, on the other hand, is loop induced, which makes 
the computation of perturbative corrections substantially more challenging.
At leading order (LO) \cite{DelDuca:2001fn,Neumann:2016dny,Ellis:2018hst,Budge:2020oyl} 
it is mediated by diagrams with closed top-quark loops.
And already at next-to-leading order (NLO), the
virtual corrections require two-loop five-point amplitudes with internal
masses, which are currently beyond the reach of available computational
methods. These are currently included in NLO computations via reweighting 
of LO with full mass dependence~\cite{Maltoni:2014eza,Chen:2021azt}.
However, for many observables, relevant energy scales remain sufficiently 
below the top-quark mass.
In such cases, a good approximate description 
\cite{Harlander:2012hf, Greiner:2016awe, Lindert:2018iug, Andersen:2018kjg, Jones:2018hbb, Chen:2021azt, Bonciani:2022jmb,Graudenz:1992pv,Czakon:2021yub}
can be obtained 
in the heavy-top effective field theory (EFT) \cite{Wilczek:1977zn,Shifman:1979eb,Inami:1982xt}, 
where the top quark is integrated out and its coupling to the Higgs boson and 
gluons is replaced by a local effective interaction.
Equivalently, amplitudes in this EFT can be viewed as the heavy-top limit (HTL) 
of the corresponding SM amplitudes.
In the heavy-top EFT, ggF $Hjj$ production is currently known at 
NLO~\cite{Campbell:2006xx,Campbell:2012am,vanDeurzen:2013rv,Greiner:2015jha,Andersen:2017kfc,Andersen:2018tnm}.

In this work, we compute the double-virtual NNLO QCD corrections to ggF $Hjj$
production in the heavy-top EFT.  In addition to the HTL approximation, we employ
the leading-color approximation (LCA), treating the number of colors and massless quark
flavors as $\NF \sim \NC\gg 1$.  We present, for the first time, analytic results
for the two-loop helicity amplitudes in all partonic channels contributing to
this process: the four-gluon, two-quark two-gluon, and four-quark channels.
Our results also provide the double-virtual corrections for the top-Yukawa
contribution to $pp \to H b\bar{b}$ production with massless bottom quarks,
which were very recently presented in ref.~\cite{Hartanto:2026xjz}.  In
addition, they can be analytically continued to the decay region~\cite{Chen:2026jxf} to obtain the
double-virtual QCD corrections for Higgs decay into four jets.

Compared to pure QCD, the LCA in the heavy-top EFT receives contributions from
non-planar Feynman diagrams.  Moreover, since the EFT is
not power-counting renormalizable, loop-momentum tensors of higher rank appear than in the
corresponding QCD amplitudes.  These features place the computation at the
current frontier of high-multiplicity two-loop amplitude calculations.  Indeed,
the complete set of required Feynman integrals and special functions has only
recently become available~\cite{Abreu:2023rco}.
And the two-loop amplitudes with
five-particle one-mass kinematics involving non-planar diagrams have so far
been derived only for a limited number of 
processes~\cite{Guo:2024bsd,Dixon:2024yvq,Badger:2025uym,Badger:2024sqv,Badger:2024mir},
including some of the partonic channels relevant for ggF $Hjj$
production~\cite{Hartanto:2026xjz}.

Our main result consists of the analytic expressions for the UV-renormalized
and IR-subtracted $pp\to Hjj$ two-loop amplitudes, the finite remainders.
To tackle the complexity
of the computation we employ functional reconstruction in finite fields 
\cite{vonManteuffel:2014ixa,Peraro:2016wsq},
which leverages numerical evaluations to recover the full analytic form of the
remainders.  
For the numerical amplitude computation we use the \Caravel{} program
\cite{Abreu:2020xvt} which implements the multi-loop numerical unitarity method
\cite{Ita:2015tya,Abreu:2017xsl,Abreu:2017idw,Abreu:2017hqn,Abreu:2018jgq,Abreu:2019odu}.
In this method, generalized unitarity cuts 
\cite{Bern:1994zx,Bern:1994cg,Britto:2004nc,Bern:2007ct,Buchbinder:2005wp} 
are represented by tree amplitudes and matched to an integrand parametrization
of master and surface terms. To calculate the trees, we extended the
Berends--Giele recursion \cite{Berends:1987me} of \Caravel{} to
the heavy top EFT. 
To accomplish the highly non-trivial construction of surface terms we employ 
unitarity-compatible integration-by-parts relations \cite{Gluza:2010ws}
combined with a linear-algebra algorithm~\cite{Abreu:2023bdp,Klinkert:2023cyd}.

We reconstruct the analytic form of the two-loop finite remainders in 
spinor-helicity variables, building on the method of 
refs.~\cite{DeLaurentis:2023nss,DeLaurentis:2023izi,DeLaurentis:2025dxw}.
To handle the substantial increase of complexity of the present computation, 
we refine the method through several key modifications.
The initial choice of transcendental functions \cite{Chicherin:2021dyp,
Abreu:2023rco} dictates the rational coefficient functions to be reconstructed.
Linear transformations to a simplified rational basis effectively decouple this
basis choice from the reconstruction
\cite{Abreu:2018zmy,DeLaurentis:2023nss,DeLaurentis:2025dxw}. Specifically,
this approach identifies relations among coefficient functions via numerical
evaluation on univariate phase-space slices to find representatives with fewer
poles. To reduce these evaluations, we exploit symmetries, empirically exclude
physical poles, and optimize the search strategy for functions with the most
poles.
We then reconstruct the rational functions in the improved basis by solving 
linear systems for coefficients of tailored partial-fraction 
ansätze \cite{Laurentis:2019bjh,DeLaurentis:2022otd,DeLaurentis:2025dxw}.
Constructing these partial-fraction decompositions (PFDs) previously relied on empirical
data of correlated poles \cite{Abreu:2021oya,DeLaurentis:2025dxw} and
numerical evaluations in singular limits 
\cite{Laurentis:2019bjh,DeLaurentis:2022otd}. To systematize this, we introduce
a new multivariate PFD algorithm based on a generic bivariate slice
evaluation in phase space. As in previous studies
\cite{DeLaurentis:2022otd,Abreu:2023bdp,DeLaurentis:2025dxw}, determining this
PFD prior to full analytic reconstruction allows us to exploit its structure
and significantly reduce the required random sample size to determine 
the ansatz coefficients.
The key difference is that in this work we obtain similar PFDs directly from finite-field samples.
Finally, we optimize the step of lifting these expressions to the rational numbers.
Instead of requiring recomputation of the linear basis change and the PFD fit
in additional finite fields~\cite{DeLaurentis:2025dxw},
we are able to perform the lift using only a small number of additional 
phase-space evaluations in a second finite field. 

The two-loop finite remainders computed in this work have a rich analytic
structure, and we showcase two interesting aspects.
First of all, we observe cancellations leading to surprising simplifications.
Specifically, the rational coefficients contain only denominator
factors that are already present in planar amplitudes with identical kinematics.
This is in contrast to the general expectation that all rational factors
entering the differential equations for the special functions may also appear
as denominator factors in the rational coefficients~\cite{Abreu:2018zmy}.
Secondly, we observe a new non-analyticity near a threshold configuration.
An unusual analytic behavior of some non-planar two-loop integrals
was observed in ref.~\cite{Abreu:2023rco}: they become non-analytic when crossing
a certain surface inside the physical scattering region, even though this
surface is not associated with any vanishing Gram determinant of external momenta.  
We examine whether this non-analyticity survives at the level of the
amplitudes, and find that it does not cancel completely, thereby constituting an
``anomalous'' threshold.  While such thresholds are well known to occur when
virtual exchanges of massive particles are involved, to the best of our
knowledge, an explicit example in \emph{massless} scattering has not been
previously reported.

We make our analytic results available in ancillary files and provide a
\texttt{C++} library for their fast and stable numerical evaluation.  This
implementation will enable phenomenological applications of ggF $Hjj$
production at NNLO accuracy.  In addition, combined with the three-loop results
of ref.~\cite{Chen:2025utl}, our amplitudes provide the ingredients needed for
N\textsuperscript{3}LO QCD predictions for $Hj$ production in gluon fusion.

This paper is organized as follows.  The notation and conventions for the
amplitudes we compute are collected in section~\ref{sec:notation}.  In
section~\ref{sec:numComputation} we describe how we obtain numeric finite-field
samples of the two loop amplitudes.  Our method for the analytical
reconstruction from the numerical samples is explained in
section~\ref{sec:analyticReconstruction}.  The main results are presented in
section~\ref{sec:results} that include discussion of analytic structure of the
obtained expressions, and presentation of the numerical evaluation of our
results in a \texttt{C++} code.  Conclusions are given in
\cref{sec:conclusions}.

\section{Setup and conventions}
\label{sec:notation}
In this section we describe the basic setup used in this article.

\subsection{Heavy top-quark effective field theory}

\begin{figure}[ht] \centering \begin{subfigure}[c]{0.44\textwidth} \centering
\includegraphics[width=0.6\linewidth]{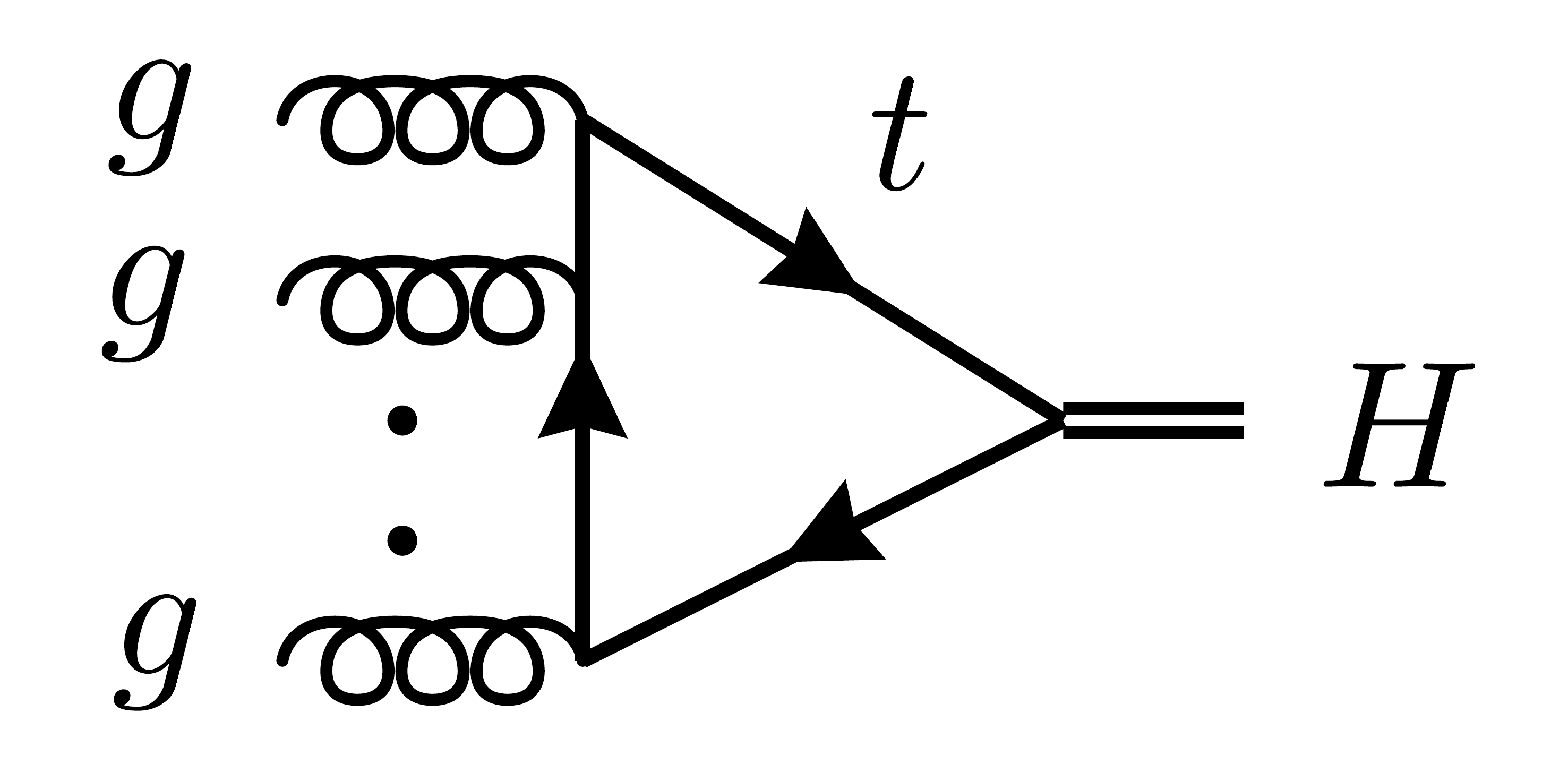} \end{subfigure}
\begin{subfigure}[c]{0.09\textwidth} \centering
\includegraphics[width=0.45\linewidth]{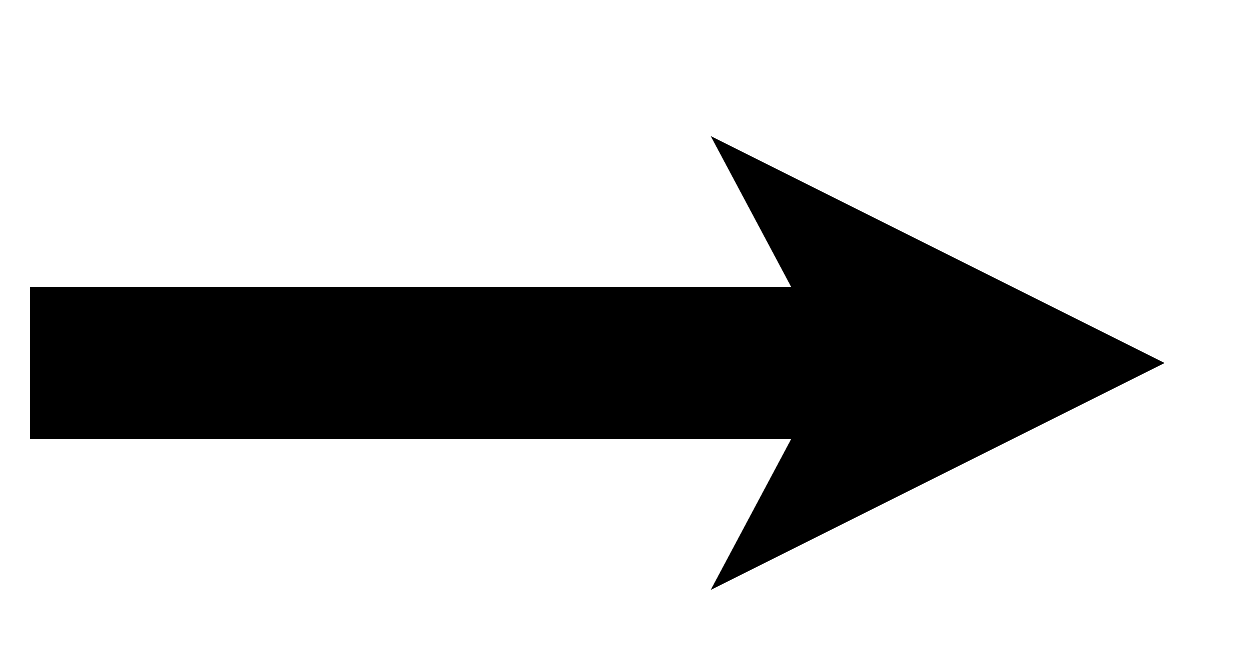} \end{subfigure}
\begin{subfigure}[c]{0.44\textwidth} \centering
\includegraphics[width=0.45\linewidth]{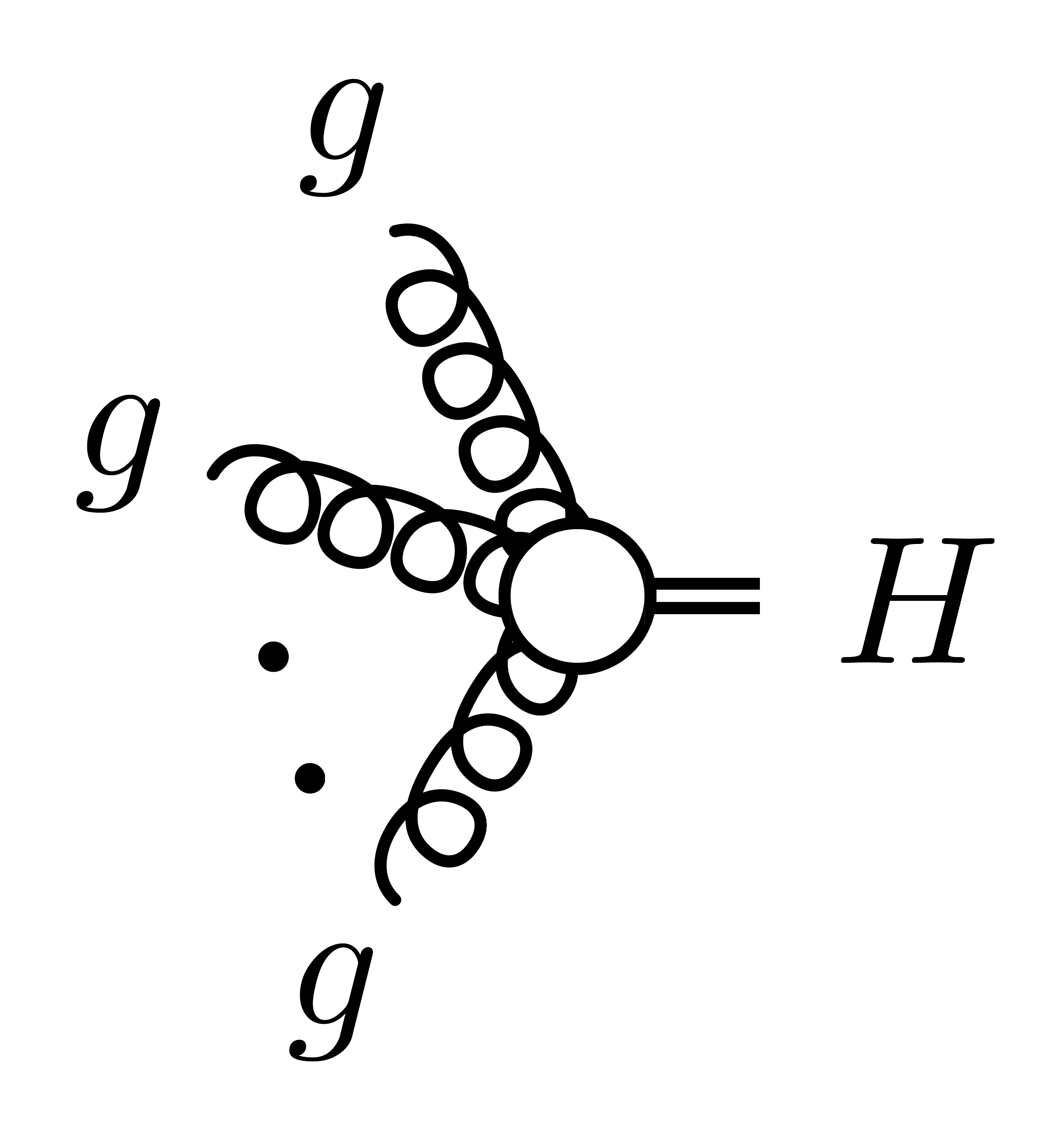} \end{subfigure} 
\caption{In the heavy-top limit the top-quark loops (left) are replaced by an effective
interaction vertex (right).
} \label{fig:eft} 
\end{figure}

To obtain scattering amplitudes for the production of a Higgs boson in
association with two jets we apply a number of well known approximations within
the perturbative regime of QCD.  First of all, we exploit a large separation of
scales between the scattering energies and the masses of the light quarks.
This justifies working with five massless quark flavors ${u,d,s,c,b}$ and
solely the top quark $t$ retains its mass $\mtop$.
In this approximation, the Higgs boson ($H$) couples at tree level via Yukawa
coupling only to the top quark.  And consequently, at higher orders,
considering no top-quark final states, the $H$ boson interacts via top-quark
loops only.
Furthermore, we focus on the regime with energy transfers below the top
threshold, which justifies to work in the heavy-top-quark
limit (HTL). In this limit we consider the expansion in inverse powers 
of $\mtop$ and keep the leading contribution only.
This amounts to integrating out the top-quark field, leading to effective
operators describing interactions of the Higgs-boson with
gluons~\cite{Wilczek:1977zn,Shifman:1979eb,Inami:1982xt}.
A diagrammatic representation of the gluon-Higgs-boson interaction in the HTL
and its origin in the Standard Model (SM) from top-quark loops is shown in
\cref{fig:eft}.
%
%
\newline At leading order in the HTL the effective couplings $\mathcal{L}_\mathrm{eff}$
between the Higgs-field and the gluon field-strength tensor
($G^a_{\mu\nu}=\partial_\mu G^a_\nu - \partial_\nu G^a_\mu  + g_s f_{abc} G^b_\mu G^c_\nu $)
as part of the Lagrangian read~\cite{Wilczek:1977zn,Shifman:1979eb,Inami:1982xt},
\begin{equation} \label{eq:eft-lagrangian}
	\mathcal{L} = \mathcal{L}_{\mathrm{QCD}} + \mathcal{L}_{\mathrm{eff}},\qquad  \mathcal{L}_{\mathrm{eff}} = - \frac{C}{4v}\, H\, G^{a,\mu\nu}G_{a,\mu\nu} \,,
\end{equation}
with the standard QCD Lagrangian $\mathcal{L}_{\mathrm{QCD}}$. $v$ is the vacuum 
expectation value of the Higgs field and $C$ is the Wilson coefficient of the 
effective operator $\mathcal{L}_{\mathrm{eff}}$.
Matched to the SM, the Wilson coefficient $C$ is given as a series in 
$\alphas$~\cite{Djouadi:1991tka, Dawson:1990zj,  Kniehl:1995tn, Chetyrkin:1997un,Chetyrkin:1997iv,Kramer:1996iq}.\footnote{This disregards electroweak effects as we do here.}
Currently, it is known up to four-loop order~\cite{Gerlach:2018hen, Chetyrkin:2025qpc}.
Similar effective interaction terms with more than two gluon field-strength
tensors are suppressed by inverse powers of the top mass.  In this paper we
neglect such terms and consider only the leading effective interaction terms in
the HTL as spelled out in \cref{eq:eft-lagrangian}.
This heavy top-quark approximation is valid if the energy of the produced
gluons lies below top-quark mass \cite{Ellis:1987xu,Baur:1989cm}, which was
demonstrated in refs.~\cite{Jones:2018hbb,Campbell:2006xx}.

\begin{figure}[ht]
\centering
\begin{subfigure}[c]{0.41\textwidth}
        \centering
        \includegraphics[width=0.9\linewidth]{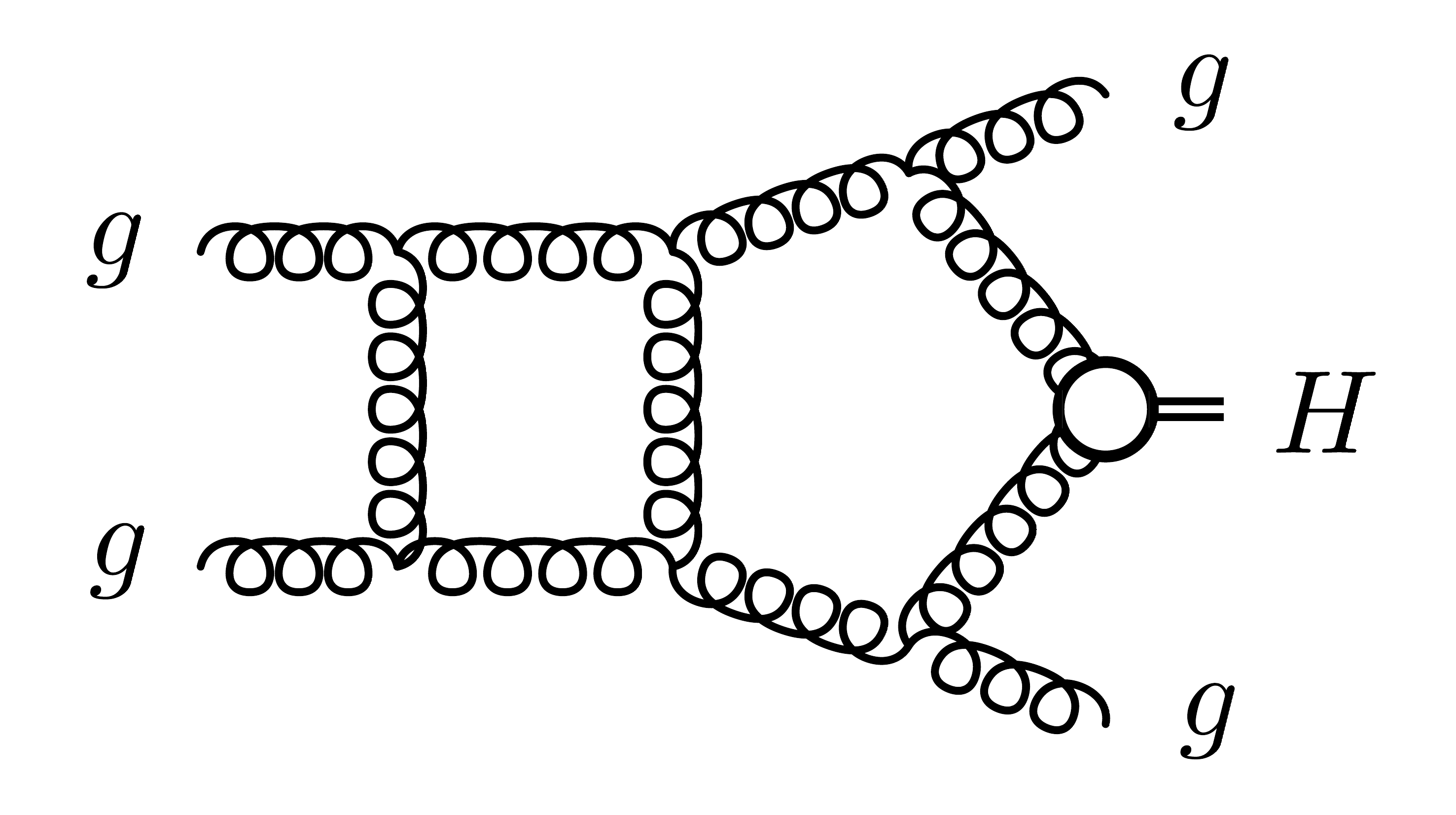}
  \end{subfigure}
  \begin{subfigure}[c]{0.41\textwidth}
        \centering
        \includegraphics[width=0.9\linewidth]{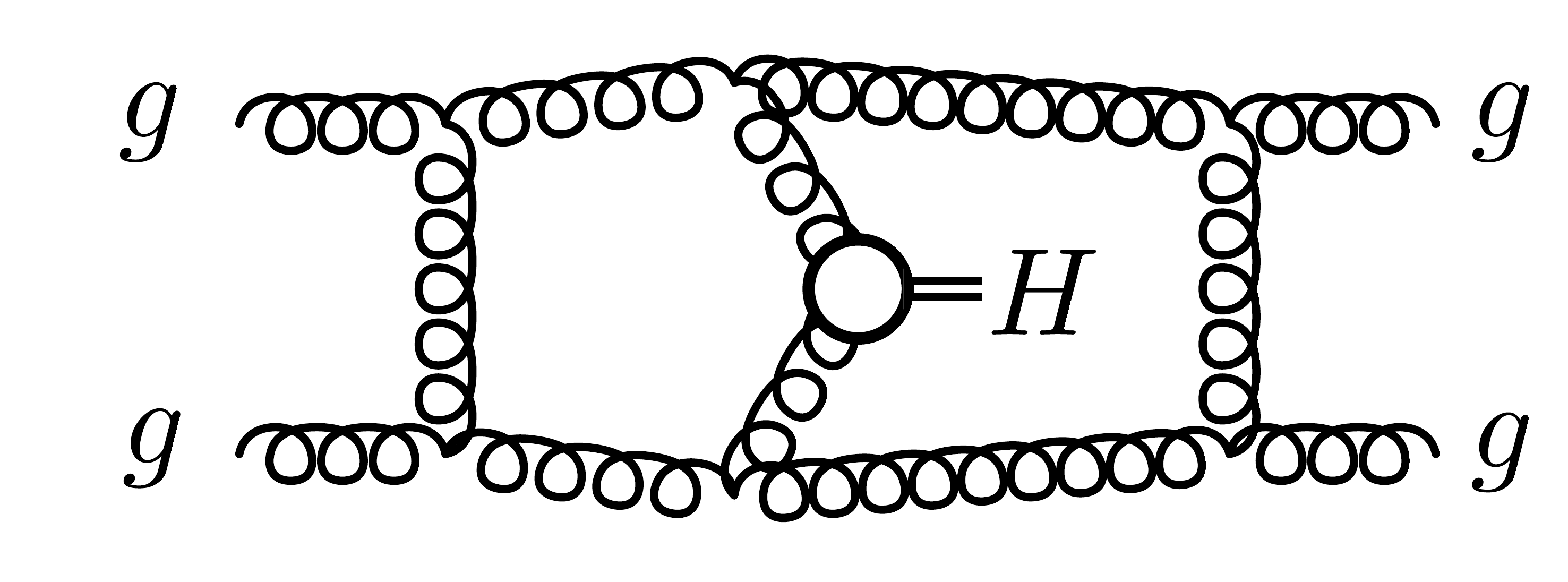}
  \end{subfigure}
\caption{The leading-color four-gluon scattering with a Higgs boson includes planar (left) and non-planar (right) two-loop diagrams.}
\label{fig:nonplanar}
\end{figure}

Secondly, we adopt the leading-color (LC) approximation (LCA), which keeps the
leading terms in the large number of colors $\NC$ limit, with the ratio of
$\NC$ to the number of light flavors $\NF$ fixed.  Unlike the LCA in QCD where
only planar diagrams contribute, Higgs amplitudes in the HTL include
contributions from non-planar diagrams.  An example of such a leading-color
non-planar contribution is shown in \cref{fig:nonplanar}.
Nevertheless, the computation in the LCA is simplified because only
diagrams with a strict ordering of the colored external states contribute.

\subsection{Partonic channels}

\begin{figure}[ht]
\centering
\begin{subfigure}[c]{0.31\textwidth}
        \centering
        \includegraphics[height=0.55\linewidth]{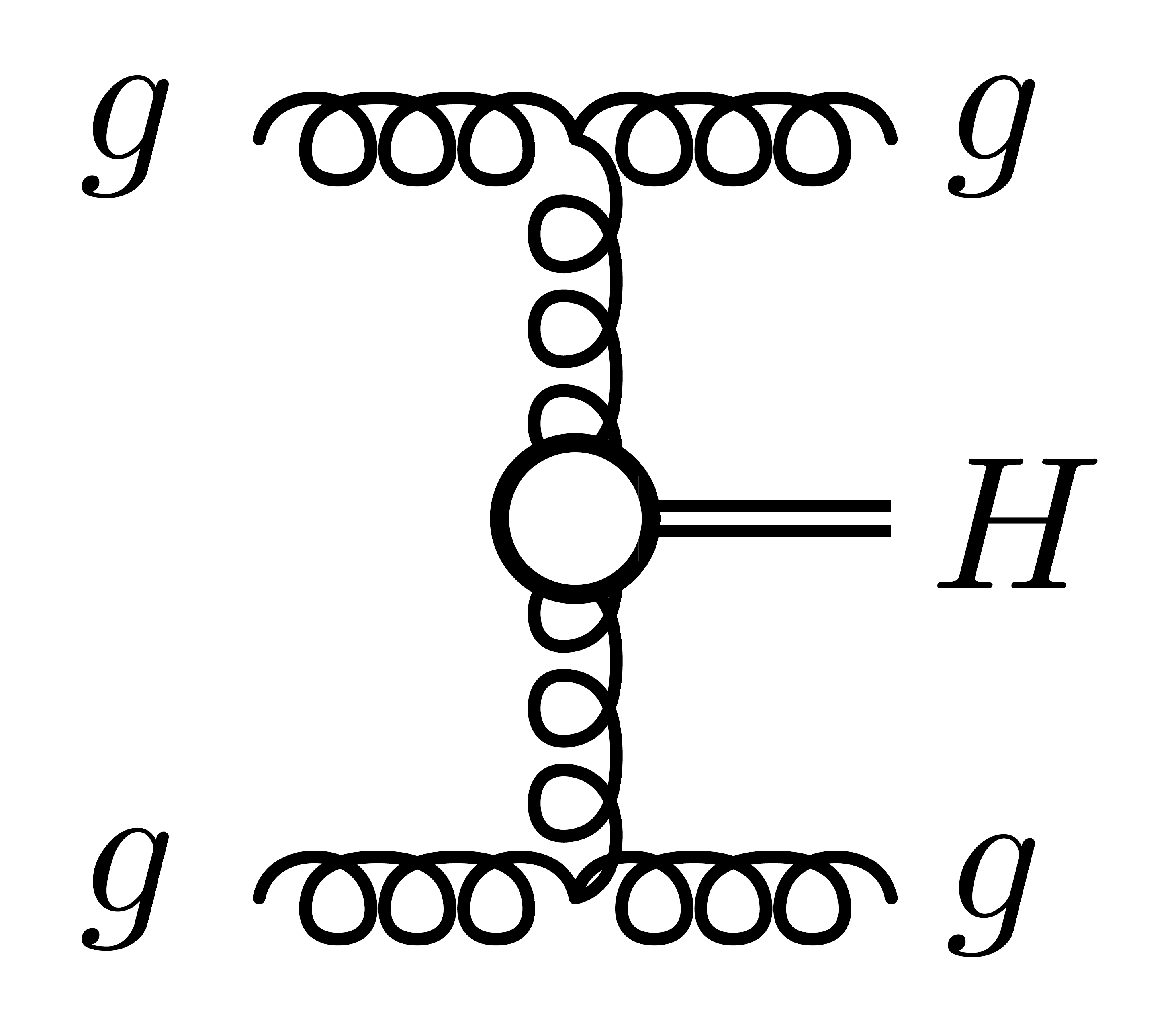}
  \end{subfigure}
  \begin{subfigure}[c]{0.31\textwidth}
        \centering
        \includegraphics[height=0.55\linewidth]{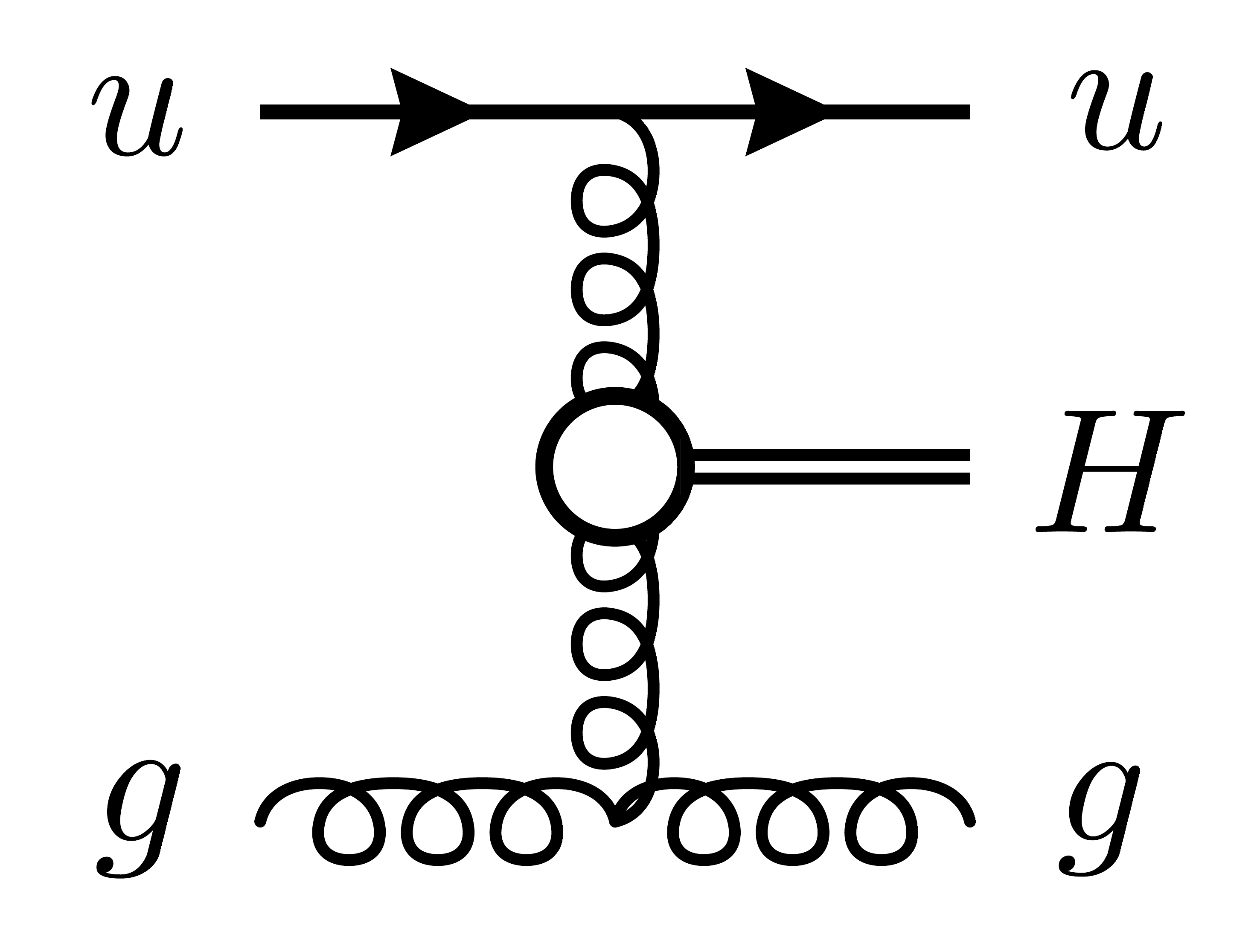}
  \end{subfigure}
  \begin{subfigure}[c]{0.31\textwidth}
        \centering
        \includegraphics[height=0.55\linewidth]{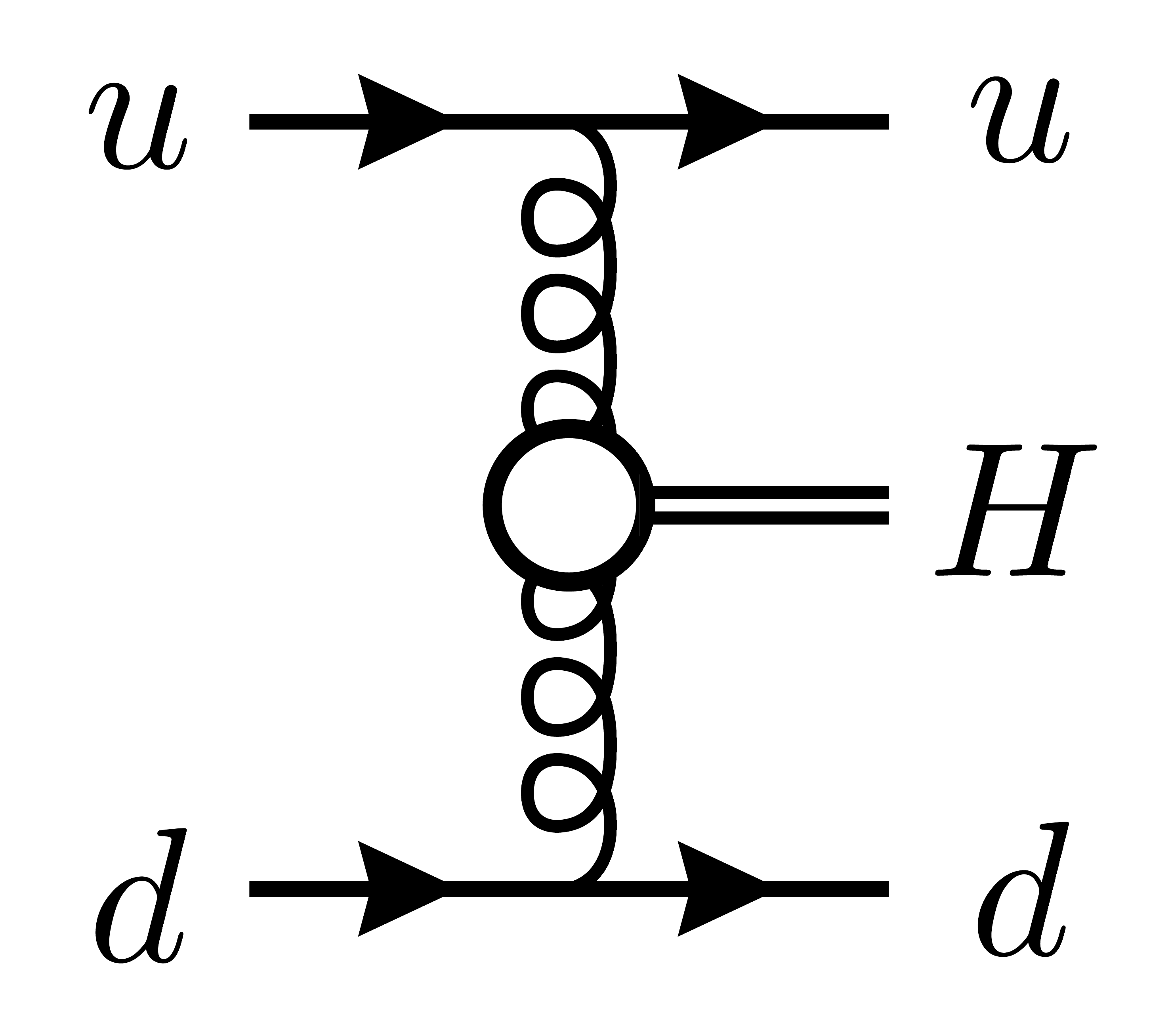}
  \end{subfigure}
\caption{Representative leading-order QCD contributions to four-parton Higgs-boson interaction. 
}
\label{fig:LO}
\end{figure}

The individual partonic channels for the scattering of a Higgs boson with four
partons, up to crossings between initial and final states, are
\begin{subequations} \label{eq:channels}
\begin{align}
    \gluon_1\,\gluon_2  &\rightarrow   \gluon_3\,\gluon_4 \, \higgs_5\,, \\
    u_1\,\gluon_2  &\rightarrow   u_3\,\gluon_4 \, \higgs_5\,, \\
    u_1\,d_2  &\rightarrow   u_3\,d_4 \, \higgs_5\,, \\
    u_1\,u_2  &\rightarrow   u_3\,u_4 \, \higgs_5 \,.
\end{align}
\end{subequations}
Here we focus on production channels of the Higgs boson.\footnote{The
Higgs-boson-decay channels can be obtained from our results by employing the
analytic continuation of the related special functions into the decay
kinematics derived in ref.~\cite{Chen:2026jxf}.} In total there are 12 partonic
channels when considering only the interactions of the effective theory and QCD in
the scattering.  The full list can be found in \cref{sec:referenceEvaluations}
in \tab{tab:Hjj-benchmark}.  As the processes here do not depend on the
electric charges of the quarks and all quarks are treated massless, it is
sufficient to work with at most two different external quark flavors, which,
without loss of generality, are picked to be $u$ and $d$.

\begin{figure}[t]
  \centering

  \begin{subfigure}{\textwidth}
    \begin{subfigure}[c]{0.31\textwidth}
      \centering
      \includegraphics[width=0.85\linewidth]{figures/4gH}
    \end{subfigure}
    \begin{subfigure}[c]{0.31\textwidth}
      \centering
      \includegraphics[width=0.85\linewidth]{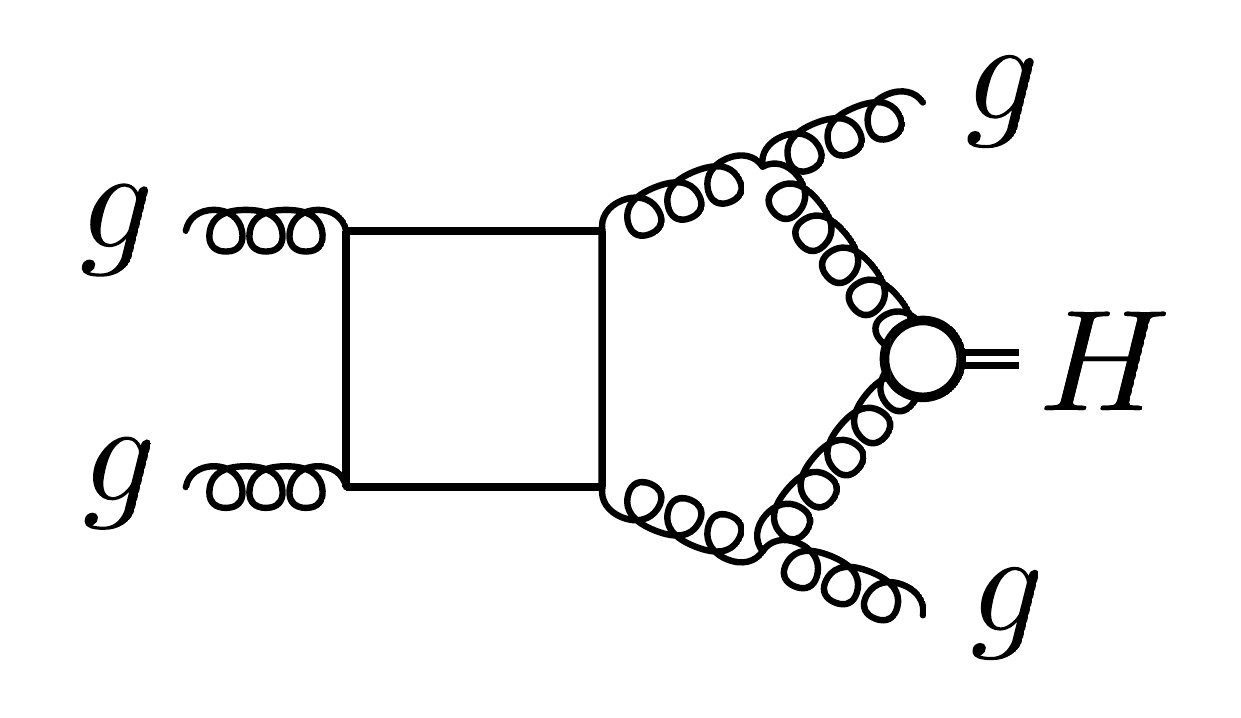}
    \end{subfigure}
    \begin{subfigure}[c]{0.31\textwidth}
      \centering
      \includegraphics[width=0.85\linewidth]{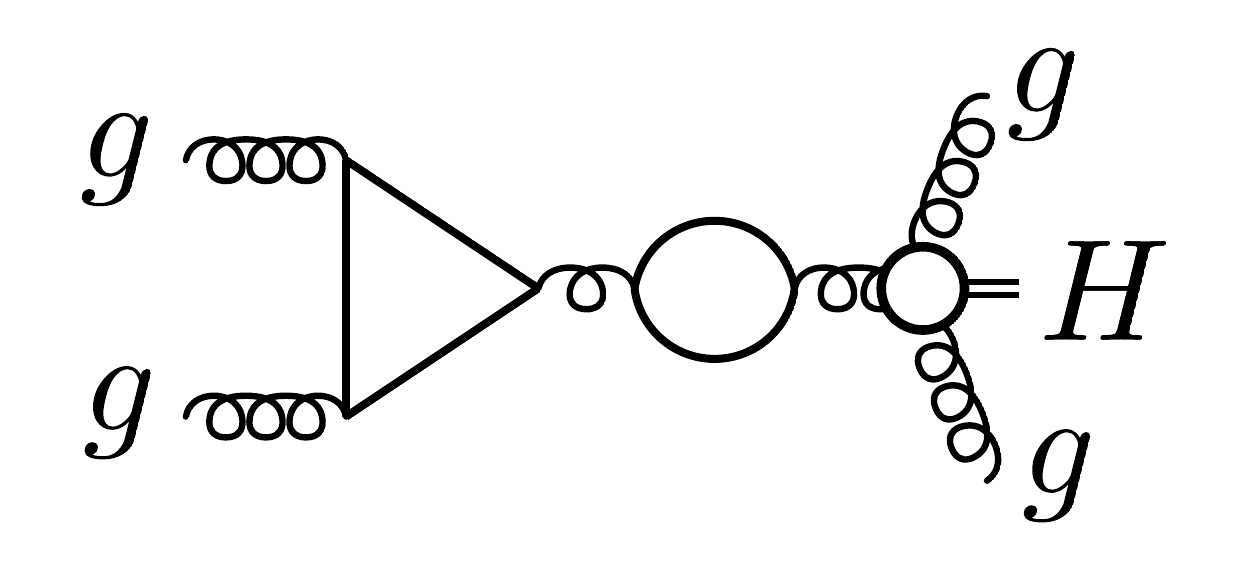}
    \end{subfigure}
    \caption{Four-gluon channel}
    \label{fig_parents4g}
  \end{subfigure}

  \begin{subfigure}{\textwidth}
    \begin{subfigure}[c]{0.31\textwidth}
      \centering
      \includegraphics[width=0.85\linewidth]{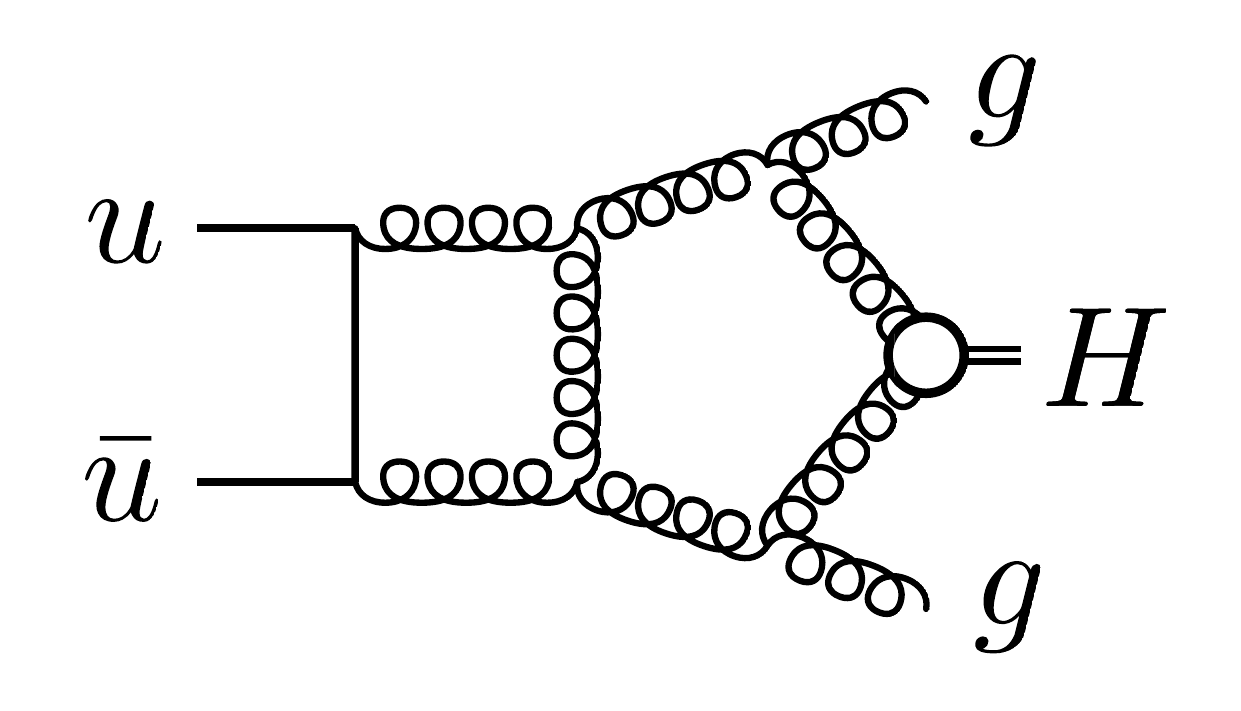}
    \end{subfigure}
    \begin{subfigure}[c]{0.31\textwidth}
      \centering
      \includegraphics[width=0.85\linewidth]{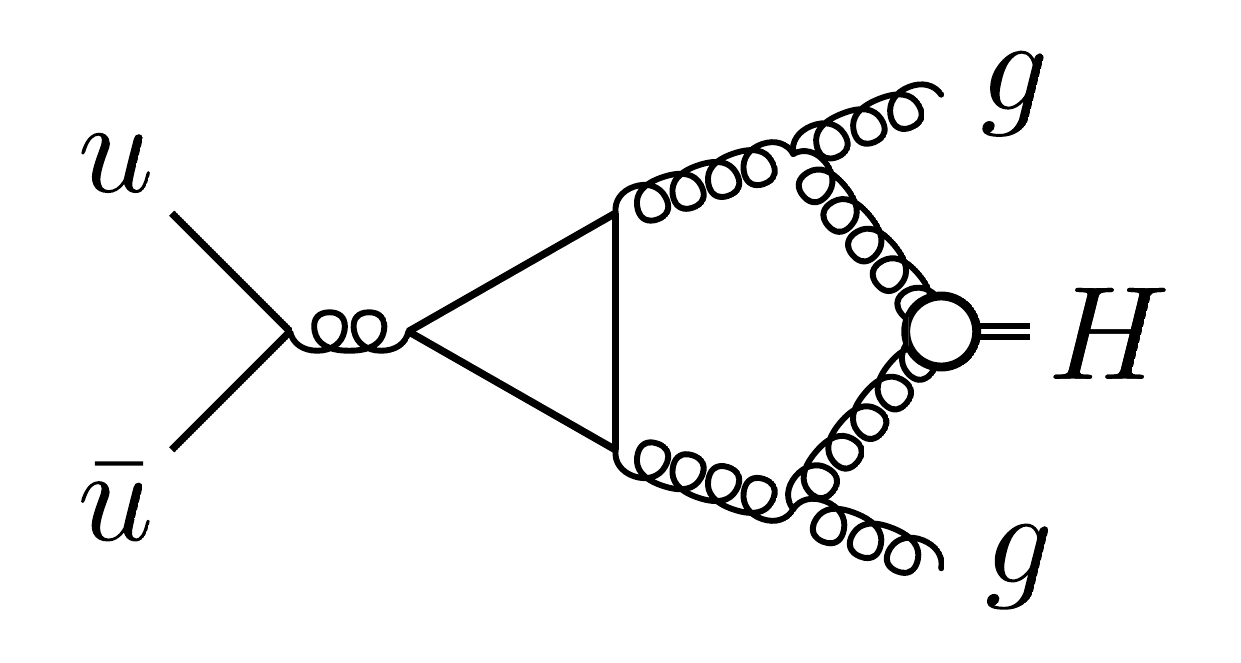}
    \end{subfigure}
    \begin{subfigure}[c]{0.30\textwidth}
      \centering
      \includegraphics[width=0.85\linewidth]{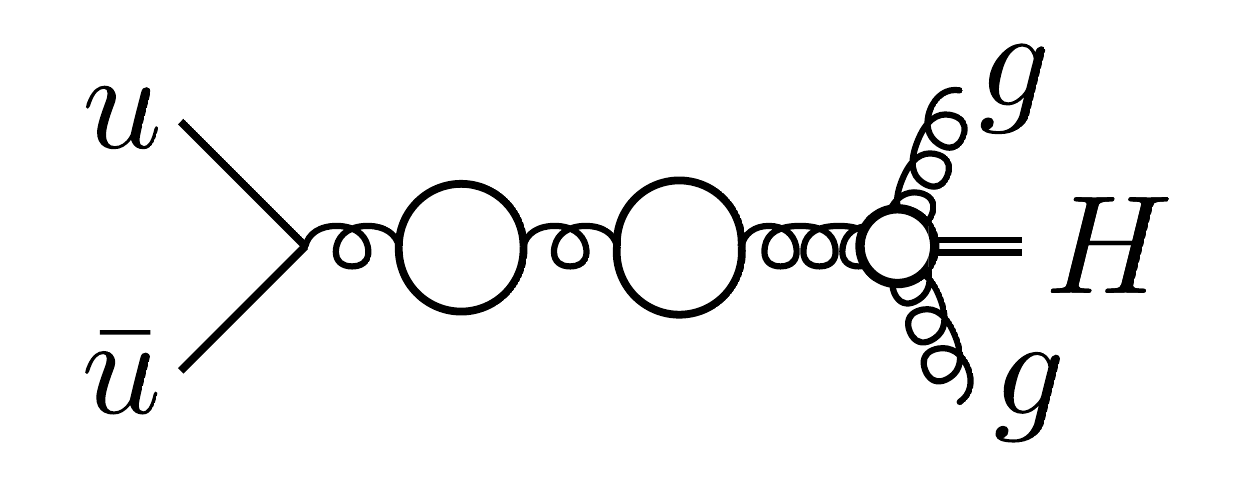}
    \end{subfigure}
    \caption{Two-gluon two-quark channel}
    \label{fig_parents2q}
  \end{subfigure}

  \begin{subfigure}{\textwidth}
    \begin{subfigure}[c]{0.31\textwidth}
      \centering
      \includegraphics[width=0.85\linewidth]{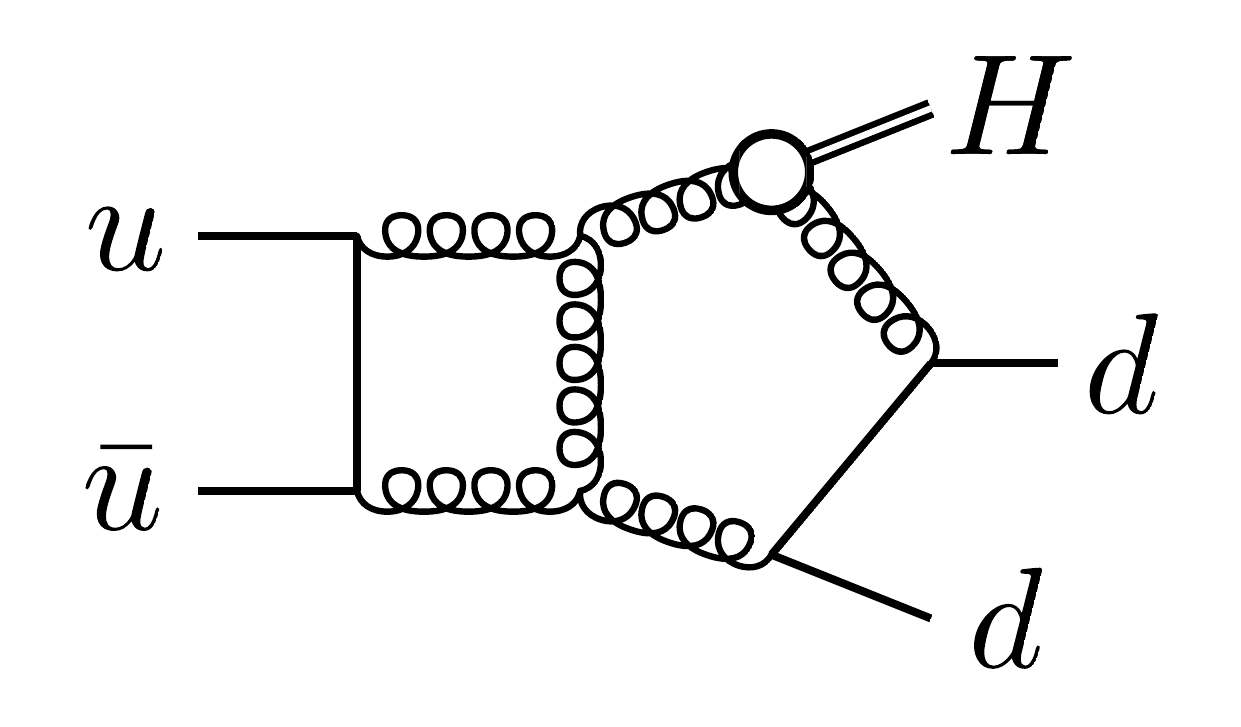}
    \end{subfigure}
    \begin{subfigure}[c]{0.31\textwidth}
      \centering
      \includegraphics[width=0.85\linewidth]{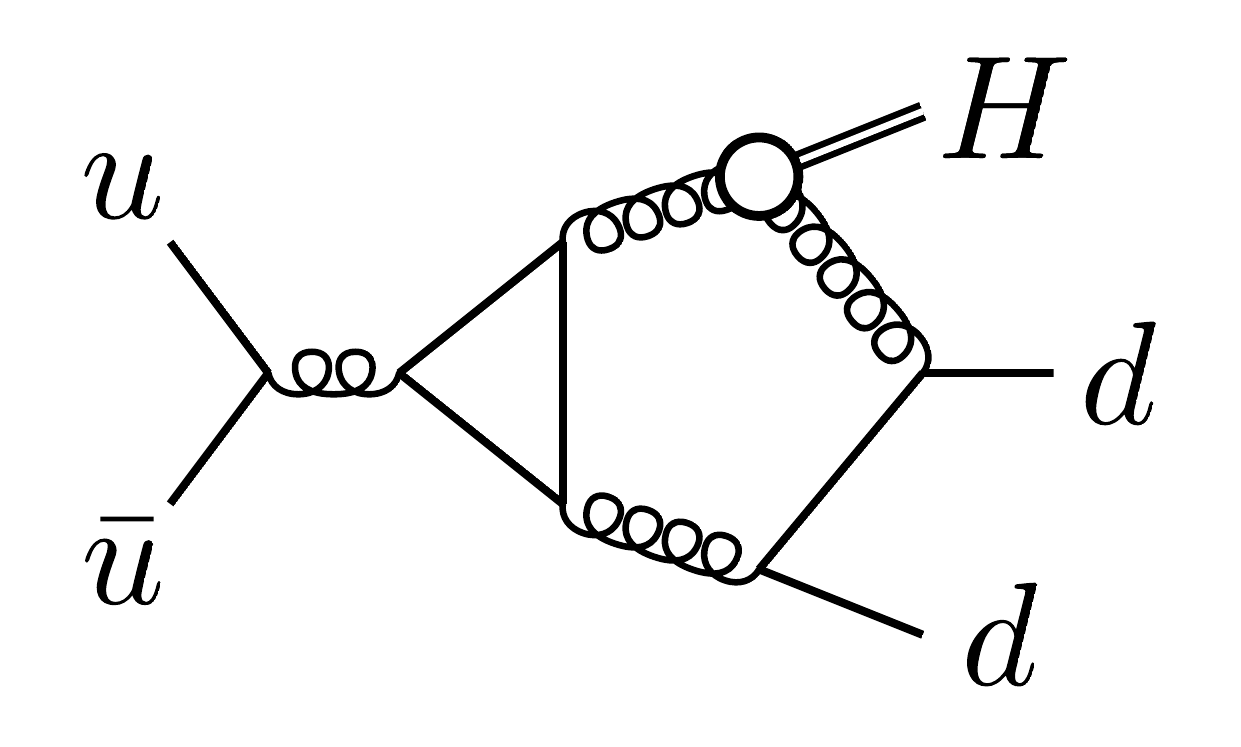}
    \end{subfigure}
    \begin{subfigure}[c]{0.31\textwidth}
      \centering
      \includegraphics[width=0.85\linewidth]{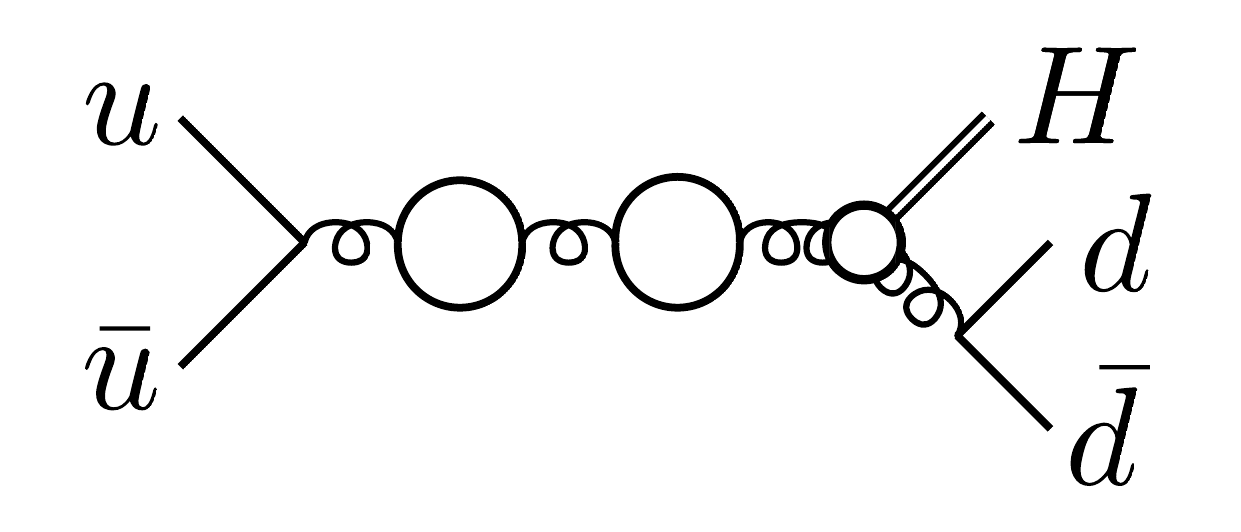}
    \end{subfigure}
    \caption{Four-quark channel}
    \label{fig_parents4q}
  \end{subfigure}
\caption{Representative diagrams for the Higgs-boson four-parton scattering.
The columns show contributions with zero, one and two closed fermion loops, respectively.}
\end{figure}

Representative leading-order diagrams are shown in~\fig{fig:LO}. The new
results of this work include computations of two-loop QCD diagrams like the
ones shown in~\cref{fig_parents4g,fig_parents2q,fig_parents4q}.

\subsection{Kinematics}

All processes considered here involve four massless and one massive
external particle.  We will follow closely the conventions used in
ref.~\cite{DeLaurentis:2025dxw}.  The first four momenta $p_i,\, i=1,2,3,4$ are
assigned to the four massless partons, and $p_5$ is reserved for the Higgs
boson. The momenta are all treated as outgoing, s.t.~momentum conservation
and on-shell conditions read
\begin{equation}
	\sum_{i=1}^5 p_i^\mu = 0 \qquad \text{and} \qquad p_{i\leq4} ^2 = 0, \qquad p_{5}^2 = \mH^2 \,,
\end{equation}
with the Higgs-boson mass $\mH$.  Consequently, the momenta $p_1$ and
$p_2$ which we will use for the two incoming particles have negative energy
components.

From the Mandelstam variables $s_{ij}=(p_i+p_j)^2$, the six Lorentz invariants
\begin{equation}
	\big\{s_{12}, s_{23}, s_{34}, s_{45}, s_{51}, p_5^2\big\}
\end{equation}
together with the parity-odd contraction of four momenta
\begin{equation}
\trFive=4\imath\,\varepsilon_{\alpha\beta\gamma\delta}\, p^\alpha_1 p^\beta_2 p^\gamma_3 p^\delta_4\quad\mbox{with}\quad \varepsilon_{0123}=1
\end{equation}
are sufficient to fully specify the kinematic configuration.
To describe the helicity states of the partons we use the spinor helicity formalism following the conventions of ref.~\cite{Maitre:2007jq}.
From the two-component spinors $\lambda_{i}^{a}$ and $\tilde\lambda_{i}^{\dot a}$, defined through
\begin{equation}
	p_{i, \mu} \sigma^{\mu, \dot a a} = \tilde\lambda_i^{\dot a} \lambda_i^a \quad\mbox{for}\quad  i\in \{1,2,3,4\} \,,
\end{equation}
invariant spinor brackets are formed as
\begin{equation}
\langle i j \rangle = \epsilon_{a b} \lambda_{i}^{a} \lambda_{j}^{b} \quad \text{and}
	\quad [ij] =  \epsilon^{\dot a \dot b}  \tilde\lambda_{i, \dot a} \tilde\lambda_{j, \dot b}\,
	\quad \text{for}  \quad i,j\in \{1,2,3,4\} \,.
\end{equation}
Here, the conventions for the antisymmetric tensors $\epsilon_{a b}$ and
$\epsilon^{\dot a \dot b}$ are $\epsilon^{12}=-\epsilon_{12}=1$.  The
Mandelstam invariants of the massless momenta are related to the spinor
brackets by $s_{ij} = \langle ij\rangle [ji]$.  Moreover, it is convenient to
use an expression of $\trFive$ in terms of spinor brackets
\begin{equation}\label{eq:trFive}
  \trFive = [12]\langle23\rangle[34]\langle41\rangle-\langle12\rangle[23]\langle34\rangle[41] \, .
\end{equation}

Several spinor chains will occur, which we write as
\begin{equation}\label{eq:spinor3chain}
  \langle i | j \pm k | l ] = \langle i j \rangle [j l] \pm \langle ik \rangle [kl] \, .
\end{equation}
If the massive momentum $p_5$ appears in a chain, one has to use
\begin{equation}\label{eq:spinor3chainmassive}
	\langle i | 5 | j ] = \lambda_{i}^a \bigl( p_{5, \mu}\, \bar\sigma^{\mu}_{a \dot a}\bigr) \tilde\lambda_j^{\dot a}\,,
\end{equation}
in the evaluation.
Longer spinor chains are defined recursively, for instance we encounter
\begin{equation}
	\langle i | j | 5 | k | l ] =
	\langle i j \rangle [j|5|k\rangle [k l] \,,
\end{equation}
for $i,j,k$ and $l$ being labels of massless momenta, and the spinor chains \eqref{eq:spinor3chain} and 
\eqref{eq:spinor3chainmassive} are equal to their
reversed form, \emph{e.g.}\  $[ i | j\pm k | l \rangle = \langle l | j\pm k | i ]$.
It is also convenient to introduce the notation
\begin{equation}
  \tr_{ij|kl} = (i+j)^{a\dot a}(k+l)_{\dot a a}
  = s_{ik} + s_{il} + s_{jk} + s_{jl} \, ,
\end{equation}
and, in terms of this, the three-mass triangle Gram determinant can be
written as
\begin{equation}\label{eq:trigram}
  \Delta_{ij|kl|5} = \frac{1}{4}(\tr_{ij|kl})^2 - s_{ij}s_{kl} \, .
\end{equation}

The Poincaré-invariance of the S-matrix implies a covariant
transformation of the scattering amplitudes under little-group
transformations of helicity states. Under a little-group
transformation of leg $i$,
\begin{equation}\label{eq:little-group-weight-def}
  \mathcal{A}(\ldots,z_i\lambda_i,\tilde\lambda_i/z_i,\ldots) =
  z_i^{-2h_i}\,\mathcal{A}(\ldots,\lambda_i,\tilde\lambda_i,\ldots) \,
  .
\end{equation}
The exponent of $z_i$ will be referred to as the little-group weight.
All quantities we consider also have uniform mass dimension,
\emph{i.e.}\ under the uniform rescaling of all external spinors,
\begin{equation}\label{eq:mass-dim-def}
  \mathcal{A}(z\lambda,z\tilde\lambda) =
  z^{2\,\deg(\mathcal{A})}\,\mathcal{A}(\lambda,\tilde\lambda) \, ,
\end{equation}
where for the present amplitudes we have $\deg(\mathcal{A})=0$. This matches
the actual mass dimension of these amplitudes after removing the couplings.

\subsection{Helicity amplitudes}

The scattering amplitudes for the processes in \cref{eq:channels} depend on the
helicity and color charges of the external-state partons.  
Making these quantum numbers manifest, we compute
the complete set of helicity amplitudes which we denote by
\begin{subequations} \label{eq:ScatteringAmplitudes}
\begin{align}
	&\mathcal{A}^{a_1,\dots,a_4}(\gluon_1^{h_1}, \gluon_2^{h_2}, \gluon_3^{h_3}, \gluon_4^{h_4},\higgs_5)\,,\\
	&\mathcal{A}^{\bar{i}_2; a_3,a_4}_{i_1}(u_1^{h}, \bar u_2^{-h}, \gluon_3^{h_3}, \gluon_4^{h_4},\higgs_5)\,, \\
	&\mathcal{A}^{\bar{i}_2\bar{i}_4}_{i_1 i_3}(u_1^{h}, \bar u_2^{-h}, d_3^{h'}, \bar d_4^{-h'},\higgs_5) \,.
\end{align}
\end{subequations}
The momentum assignments correspond to the subscripts of the particle labels.
The superscripts denote the particles helicities.
The labels $a_i$ denote the color indices of the adjoint representation of $\mathrm{SU}(\NC)$ of the gluon states, while $i_n$ denote the fundamental indices of the quark states and $\bar{i}_n$ the anti-fundamental indices of the anti-quark states.

In the channels involving external quarks, helicity is conserved along quark
lines, s.t.~helicity amplitudes for which a quark has the same helicity as the
anti-quark (of the same quark flavor) vanish exactly and do not have to be
considered. Also, it is not necessary to separately compute amplitudes for the equal
quark flavor case, as these can be obtained by anti-symmetrization of the
distinct flavor case, for example
\begin{equation}
\begin{aligned}
    \mathcal{A}^{\bar{i}_2\bar{i}_4}_{i_1 i_3}(u_1^{h_1}, \bar u_2^{h_2}, u_3^{h_3}, \bar u_4^{h_4},\higgs_5) &=  \mathcal{A}^{\bar{i}_2\bar{i}_4}_{i_1 i_3}( u_1^{h_1}, \bar u_2^{h_2}, d_3^{h_3}, \bar d_4^{h_4},\higgs_5)\, - \\ &\qquad\qquad \mathcal{A}^{\bar{i}_4\bar{i}_2}_{i_1 i_3}( u_1^{h_1}, \bar u_4^{h_4}, d_3^{h_3}, \bar d_2^{h_2},\higgs_5)\,,
\label{eq:AmplitudeEqualQuarkFlavors}
\end{aligned}
\end{equation}
Since we are working in the approximation of massless light quarks, the amplitudes of other light-quark flavors are identical to those presented for $u,d$ quarks.
Finally, due to crossing relations, the amplitudes in~\eqref{eq:ScatteringAmplitudes} and~\eqref{eq:AmplitudeEqualQuarkFlavors} are sufficient to cover all different subchannels of the processes listed in~\eqref{eq:channels}.

\subsection{Color decomposition}
\label{sec:colorDecomposition}

By exploiting properties of the scattering amplitudes w.r.t.~the gauge group
$\SUN$ their intricate analytic properties can be disentangled, which often 
simplifies the computations. We will now give our approach and conventions.

We work with generators $(T^{a})_j^{\,\,\,\bar i}$ of the fundamental
representation of $\SUN$ normalized to
\begin{equation}
	\tr\bigl(T^a T^b\bigr) = \delta^{ab}.
\end{equation}
Consequently, their Lie-algebra relations read
\begin{equation}
	\big[ T^a,T^b \big] = \imath \sqrt{2} f_{abc} T^c \, ,
\end{equation}
with $f_{abc}$ the structure constants of $\SUN$.
For the color decomposition of the helicity amplitudes we use traces and products of fundamental generators, which are referred to as color structures.
The color structures $\mathrm{C}_{k}$ appearing in the processes here are
\begin{equation} \label{eq:colorStructures}
\begin{split}
\mathrm{C}_k \in \Big\{ &
	\tr\bigl( T^{a_1} T^{a_2} T^{a_3} T^{a_4} \bigr),
	\tr\bigl(T^{a_1} T^{a_2}\bigr) \tr\bigl( T^{a_3} T^{a_4}\bigr)\,, \\
		&  \bigl(T^{a_3} T^{a_4}\bigr)_{i_1}^{\,\,\,\bar{i}_2} \,,
		\tr\bigl(T^{a_3} T^{a_4}\bigr)\, \delta_{i_1}^{\bar{i}_2}\,,
		\delta_{i_1}^{\bar{i}_4}\, \delta_{i_3}^{\bar{i}_2} \,,
		\delta_{i_1}^{\bar{i}_2}\, \delta_{i_3}^{\bar{i}_4}
\Big\} \,.
\end{split}
\end{equation}
An amplitude ${\cal A}$ is decomposed to partial amplitudes $A_k$ as
\begin{equation}\label{eq:colorDec}
    \mathcal{A} = \sum_{k,\, \sigma\,\in\,\mathcal{S}_k} \sigma\bigl( \mathrm{C}_{k}\, A_{k}(1,\dots,5) \bigr)\,, 
\end{equation}
where we suppressed momentum and helicity labels.  For a given particle
assignment only the subset of the color structures contributes, which matches
the color charges of the particles.  The permutations $\sigma$ act on the
particle labels in a passive way $\sigma(i) = \sigma_i$.  The set
$\mathcal{S}_k$ for a given color structure $\mathrm{C}_k$ includes exactly those
permutations such that the sum runs over a complete set of independent color
structures.  The explicit decompositions will be given below.

We now collect the color decomposition of the amplitudes for each channel in
turn.  The four-gluon Higgs scattering amplitudes are constructed from two
partial amplitudes,
\begin{multline}\label{eq:4g-color-dec}
	\mathcal{A}^{a_1,\dots,a_4}(\gluon_1^{h_1}, \gluon_2^{h_2}, \gluon_3^{h_3}, \gluon_4^{h_4},\higgs_5) = \sum_{\sigma\, \in\, S_3(2,3,4) }\sigma\bigl(\tracep{T^{a_1} T^{a_2} T^{a_3} T^{a_4}}\, A_{1}(1^{h_1},2^{h_2},3^{h_3},4^{h_4}, 5) \bigr) \\
    + \sum_{\sigma\,\in\, C_3(2,3,4) }\sigma\bigl(\tracep{T^{a_1} T^{a_2}}\,\tracep{T^{a_3} T^{a_4}}\, A_2(1^{h_1},2^{h_2},3^{h_3},4^{h_4}, 5) \bigr)\, ,
\end{multline}
where $S_n$ stands for the symmetric group and $C_n$ for the cyclic permutations of
$n$ objects, which are trivially extended such that $\sigma_j = j$ for labels $j$
not appearing in their arguments. 
For the scattering of two quarks, two gluons and one Higgs boson, the color
decomposition reads
\begin{equation} \label{eq:2q2g-color-dec}
\begin{split}
	\mathcal{A}^{\bar{i}_2; a_3,a_4}_{i_1}(u_1^{h}, \bar u_2^{-h}, \gluon_3^{h_3}, \gluon_4^{h_4},\higgs_5) = & \sum_{\sigma\, \in\, S_2(3,4)} \sigma\bigl(\bigl(T^{a_3} T^{a_4}\bigr)^{\,\,\,\bar{i}_2}_{i_1}\, A_3(1^h,2^{-h}, 3^{h_3},4^{h_4}, 5) \bigr) \\
    & + \tracep{T^{a_3} T^{a_4}}\, \delta^{\bar{i}_2}_{i_1}\, A_4(1^h,2^{-h}, 3^{h_3},4^{h_4},5) \,.
\end{split}
\end{equation}
Finally, for the scattering of two quark pairs and a Higgs boson, the following
color decomposition applies,
\begin{equation}
\begin{aligned}
	\mathcal{A}^{\bar{i}_2\bar{i}_4}_{i_1 i_3}(u_1^{h}, \bar u_2^{-h}, d_3^{h'}, \bar d_4^{-h'},\higgs_5) &= 
	 \delta^{\bar{i}_4}_{i_1}\delta^{\bar{i}_2}_{i_3}\, A_5(1^{h},2^{-h},3^{h'},4^{-h'},5)\,
	 + \\ &\qquad\qquad \delta^{\bar{i}_2}_{i_1} \delta^{\bar{i}_4}_{i_3}\, A_6(1^{h},2^{-h},3^{h'},4^{-h'},5)\,.
\label{eq:4q-color-dec}
\end{aligned}
\end{equation}

The partial amplitudes $A_k$ are expanded as series in the bare strong coupling
constant $\alphas^0 = (\gs^0)^2/(4\pi)$
\begin{equation}
	A_{k} = (\gs^0)^2\, \frac{C^0}{2v} \sum_{\ell\, =\, 0} \biggl(\frac{\alphas^0}{2\pi} \biggr)^{\!\ell}\,A^{(\ell)}_{k}
\label{eq:perturbative expansion Amp}
\end{equation}
where $\ell$ corresponds to the number of loops.  The overall factor $C^0$ is
the bare Wilson coefficient for the effective Higgs-gluon coupling
(\ref{eq:eft-lagrangian}).  Furthermore, the $A^{(\ell)}_k$ are decomposed in
terms of color-stripped amplitudes $A^{(\ell), (n_c, n_f)}_k$, which makes factors of $\NC$
and $\NF$ manifest
\begin{equation}
    A^{(\ell)}_k = \sum_{n_c, n_f} \NC^{n_c} \NF^{n_f} A^{(\ell), (n_c, n_f)}_k \, .
\end{equation}
In the LCA, only the leading terms in $\NC$ and $\NF$ are kept in the sum.
At the $\ell$-th loop level these have powers $n_c + n_f = \ell$.
Only the partial amplitudes $A_1, A_3, A_5$ contribute with such leading terms.
Explicitly, in the LCA, 
\begin{align}
	A_k^{(0)} &= A_k^{(0),(0,0)}\,,\\
	A_k^{(1)} &= \NC A_k^{(1),(1,0)} + \NF A_k^{(1),(0,1)}\,,\\
	A_k^{(2)} &= \NC^2 A_k^{(2),(2,0)} + \NC \NF A_k^{(2),(1,1)} + \NF^2 A_k^{(2),(0,2)}\,
\end{align}
for $k=1,3,5$.
Representative diagrams for these different contributions are shown in figs.~\ref{fig_parents4g}--\ref{fig_parents4q}.

Due to symmetries of the scattering amplitudes under exchange of gluons,
charge- and parity-conjugation, it is not necessary to independently compute
the helicity amplitudes for all
momentum and helicity assignments.  One can
map these to a generating set by appropriate permutations of momenta,
helicities and parity conjugation.  The generating set we consider is
\begin{equation} \label{eq:listHelConfs}
  \begin{split}
&A_1(1^+,2^+,3^+,4^+,5)\,, \quad A_3(1^+,2^-,3^+,4^+,5)\,,\quad A_5(1^+,2^-,3^+,4^-,5)\,,\\
&A_1(1^+,2^+,3^+,4^-,5)\,, \quad A_3(1^+,2^-,3^+,4^-,5)\,,\quad A_5(1^+,2^-,3^-,4^+,5)\,, \\
&A_1(1^+,2^+,3^-,4^-,5)\,, \quad A_3(1^+,2^-,3^-,4^+,5) \,,\\
&A_1(1^+,2^-,3^+,4^-,5)\,,
  \end{split}
\end{equation}
For convenience, we provide the corresponding tree amplitudes in
Appendix~\ref{app:trees}.

\subsection{Finite remainders}
\label{sec:renormalization}

We employ the 't~Hooft--Veltman scheme of dimensional regularization with
dimension $D=4-2\epsilon$ to regularize ultraviolet (UV) and infrared (IR)
divergences of loop amplitudes.  In this scheme external states are kept in a
four-dimensional subspace.  We define helicity amplitudes with
external fermions as described in ref.~\cite{Abreu:2018jgq}.  The divergences manifest
as poles in the dimensional regulator $\eps$.  Their universal structure allows
one to define the amplitudes' finite remainders, which have poles subtracted in
the following way.

\paragraph{UV renormalization}
The divergences of UV origin are cancelled by renormalizing the bare strong
coupling $\alphas^0$ and the bare Wilson coefficients $C^0$.  We work in the
$\overline{\text{MS}}$ scheme and collect the expressions for the bare
couplings in terms of renormalized ones, $\alphas(\mu)$ and $C$, for
completeness below.  For the strong coupling one has
\begin{equation}\label{eq:uvRenAlpha}
	\alphas^0 \mu_0^{2\eps} S_{\epsilon} = \alphas(\mu) \,\mu^{2\eps} \left(1-\frac{\beta_0}{\epsilon}\bigg(\frac{\alphas(\mu)}{2 \pi}\bigg)+\bigg(\frac{\beta_0^2}{\epsilon^2}-\frac{\beta_1}{2 \epsilon}\bigg)\bigg(\frac{\alphas(\mu)}{2 \pi}\bigg)^2+\mathcal{O}\bigl(\alphas^3\bigr)\right)\,,
\end{equation}
where $\mu$ is the renormalization scale and $S_{\epsilon} = (4 \pi)^{\epsilon}
e^{-\epsilon \gamma_{\mathrm{E}}}$.  The regularization scale $\mu_0$ was introduced to
keep the bare coupling dimensionless in the regularization procedure.  The
normalization of the bare Wilson coefficient $C^0$ reads~\cite{Harlander:2001is}
\begin{equation}\label{eq:uvRenC}
    C^0 \mu_0^\epsilon = C(\mu) \mu^\epsilon\, \left(1 - \frac{\beta_0}{\epsilon}\frac{\alphas(\mu)}{2 \pi} + \bigg(\frac{\beta_0^2}{\epsilon^2}-\frac{\beta_1}{\epsilon}\bigg)\bigg(\frac{\alphas(\mu)}{2 \pi}\bigg)^2+\mathcal{O}\bigl(\alphas^3\bigr)\right)\,.
\end{equation}
The coefficients $\beta_0$, $\beta_1$ of the QCD $\beta$-function are
\begin{equation}
	\beta_0 = \frac{11 \NC}{6} - \frac{\NF}{3}, \qquad \beta_1 = \frac{17 \NC^2}{6} - \frac{13 \NC \NF}{12} +  \frac{\NF}{4 \NC}.
\end{equation}

\paragraph{IR subtraction}
Finally, remaining $\eps$-poles associated to IR divergences are cancelled by
application of an operator
$\mathbf{I}$~\cite{Catani:1998bh,Sterman:2002qn,Becher:2009cu,Gardi:2009qi} on
the amplitude, resulting in the finite remainder $\mathcal{R}$
\begin{equation}\label{eq:finite-remainder}
	\mathcal{R}(\mu) = \lim_{\epsilon\to 0} ~ {\bf I}(\mu, \epsilon) \mathcal{A}(\epsilon)\,.
\end{equation}
The operator ${\bf I}(\mu, \epsilon)$ acts formally on the color indices of the amplitudes.
It is perturbatively expanded as
\begin{equation}
	\mathbf{I}(\mu, \epsilon) = \mathbb{1} + \frac{\alphas}{2\pi} \mathbf{I}^{(1)}(\mu, \epsilon) + \left(\frac{\alphas}{2\pi}\right)^2 \mathbf{I}^{(2)}(\mu, \epsilon) +\mathcal{O}(\alphas^3)
\end{equation}
and we adopt the scheme of Catani for the $\mathbf{I}^{(\ell)}$ defined
in ref.~\cite{Catani:1998bh}.  In the LCA the IR operator is diagonal in color space and acts
like as normalization factors on the individual partial amplitudes.  The explicit
expressions of $\mathbf{I}^{(\ell)}$ for the three different channels are
given in the appendix~\ref{sec:CataniOperators}.

The finite remainders for the different channels admit a color decomposition in
terms of the color structures in~\cref{eq:colorStructures} analogously to the
amplitudes \eqref{eq:colorDec},
\begin{align}\label{eq:colorDecR}
    \mathcal{R} = \sum_{k,\sigma} \sigma\bigl( \mathrm{C}_{k}\, R_{k} \bigr) \,,
\end{align}
where the $R_{k}$ are the partial remainders.  For simplicity, we
suppressed momentum and helicity labels and we follow the notation
of~\cref{eq:colorDecR}.  Just like for the amplitude, the $R_{k}$ are
perturbatively expanded using the $\overline{\text{MS}}$ renormalized $\alphas$
and further decomposed to color-stripped remainders with factors of $\NC$ and $\NF$ made explicit
\begin{align} 
	R_{k} &= \gs^2 \frac{C}{2v} \left( {R}_k^{(0)} + \frac{\alphas}{2\pi} {R}_k^{(1)} + \left(\frac{\alphas}{2\pi}\right)^2 {R}_k^{(2)} + \mathcal{O}\bigl(\alphas^3\bigr) \right)\,, \\
	{R}_k^{(\ell)} &= \sum_{j=0}^\ell \NC^{j}\NF^{\ell - j} {R}_k^{(\ell), (j,\ell - j)}\,,
\label{eq:colorstrfinRem}
\end{align}
only showing the LC contributions in the second line.
The symmetries of the amplitude, which permit any amplitude being mapped to the
generating set shown in~\eqref{eq:listHelConfs}, are preserved under
renormalization.  We therefore choose a generating set of finite
remainders analogously to the one of the amplitudes.  These are the objects
that we finally compute, as will be described in the next section.

\paragraph{Scale dependence}
The dependence of the finite remainder on the renormalization scale $\mu$ stems
from the renormalized couplings and the scale-dependent IR operator.  For the
scattering processes involving $4$ partons and one Higgs boson the finite remainder obeys
the scaling laws
\begin{subequations}\label{eq:scaling-remainders}
\begin{align}
	\frac{\partial \mathcal{R}^{(1)}}{\partial\log(\mu)} &= 4 \beta_0  \mathcal{R}^{(0)} \,, \\
	\frac{\partial \mathcal{R}^{(2)}}{\partial\log(\mu)} &= 6 \beta_0  \mathcal{R}^{(1)} + \left(6\beta_1 + 4 \mathbf{H}^{(2)}_{\mathrm{R.S.}}\right) \mathcal{R}^{(0)}  \,,
\end{align}
\end{subequations}
where $\mathbf{H}^{(2)}_{\mathrm{R.S.}}$ is a process-dependent
operator related to the collinear singularities as part of the IR operator
$\mathbf{I}^{(2)}$.  Its explicit expressions are given in the
appendix~\ref{sec:CataniOperators}.

\section{Amplitude computation over finite fields in \textsc{Caravel}}
\label{sec:numComputation}

We obtain analytic one- and two-loop scattering amplitudes of a Higgs boson and four partons
from numerical fitting in a finite field.
The numerical evaluations of the amplitudes using the numerical unitarity method~\cite{Ita:2015tya,Abreu:2017xsl,Abreu:2017idw}, which is
implemented in the \texttt{C++} framework
\Caravel~\cite{Abreu:2020xvt}, is the subject of this section.

For a given particle, color-structure and helicity assignment, \Caravel{} returns
bare amplitudes represented as coefficients of a predefined master integral basis, which for this application
consists of the canonical five-point one-mass integrals of refs.~\cite{Abreu:2020jxa,Abreu:2021smk,Abreu:2023rco}.
The coefficients are output as rational functions in the dimension parameter $D$ for a given numerical phase-space point in a finite field.
It is then straight forward to expand the coefficients in the dimensional 
regulator $\eps=(4-D)/2$ and combine them with the corresponding expansion of the master integrals in terms of integral functions \cite{Chicherin:2021dyp, Abreu:2023rco}.
Applying the above UV-renormalization (\ref{eq:uvRenAlpha}, \ref{eq:uvRenC}) and IR-subtraction \eqref{eq:finite-remainder} to these semi-numerical bare amplitudes, we obtain the generating set of finite 
remainders $R_k$ in eq.~\eqref{eq:colorstrfinRem}. 

The finite remainders admit a 
representation in terms of pentagon functions 
\cite{Chicherin:2021dyp,Abreu:2023rco}
and rational coefficient functions $r_{k, i}^{\ell,(n_c,n_f)}$,
\begin{equation}\label{eq:remainder}
  R^{(\ell),(n_c, n_f)}_k(1^{h_1},2^{h_2},3^{h_3},4^{h_4},5) = \sum_i r_{k, i}^{(\ell),(n_c,
    n_f),\vec h} \, G_i \,,
\end{equation}
where $\vec h=(h_1,h_2,h_3,h_4)$ denotes the helicity labels.
For completeness, we added various labels in the expression, such as the loop order $\ell$, color structure
labelled by $n_c$ and $n_f$, and color structure label $k$, but will suppress them in the following.
For a suitable choice of $G_i$ the coefficient functions $r_i$ depend in a rational way on holomorphic and
anti-holomorphic spinors ($\lambda$ and $\tilde\lambda$) separately, while the
functions $G_i$ depend only on the four momenta.
We choose most of the $G_i$ to be pure transcendental functions $f_i$
(or products thereof), but some include square roots in their
definition,
\begin{align}
  G_i = \frac{f_i}{\sqrt{q}} \quad &\text{with} \quad \sqrt{q} \in \Delta_3^{(i)} \, , \label{eq:tripent} \\
  G_i  = \sqrt{q} \, f_i \quad &\text{with} \quad \sqrt{q} \in \Sigma_5^{(i)} \, , \label{eq:sig5pent}
\end{align}
where the corresponding polynomials $\Delta_3^{(i)}$ and
$\Sigma_5^{(i)}$ are defined in
refs.~\cite{Abreu:2021smk,Abreu:2023rco}. The former differs by a
factor of 4 from the definition given in eq.~\eqref{eq:trigram}. For
the remaining square root $\sqrt{\Delta_5}$, we exploit the fact that
it is rationalized in spinor-helicity variables and use the unimodular normalization
\begin{equation}
  G_i=\left(\frac{\trFive}{\sqrt{\Delta_5}}\right) f_i \, ,
  \label{eq:del5pent}
\end{equation}
where the branch choice  of $\sqrt{\Delta_5}$ is defined in ref.~\cite{Abreu:2023rco}.
These normalizations imply that the coefficients are in fact rational in
spinor variables and have mass dimension $2$, $-4$, and $0$ for the
functions in eqs.~\eqref{eq:tripent}, \eqref{eq:sig5pent}, and
\eqref{eq:del5pent}, respectively. One might expect that the
coefficients of the functions associated with $\Sigma_5^{(i)}$ and
$\Delta_5$ would carry denominator factors of $\Sigma_5^{(i)}$ and
$\trFive$, respectively, but we observe that such factors are
absent. In fact, our finite remainders are manifestly free of any
spurious poles in $\Sigma_5^{(i)}$ and $\trFive$. We come back to this point in \cref{sec:analyticStructure}.

We now discuss the challenges encountered in the current amplitude computation
and highlight the recent extensions to \Caravel{} that enabled these numerical
results.
In the applied numerical unitarity method, the generalized unitarity cuts   
of loop integrands are constructed from tree amplitudes.
Numerical values for such tree amplitudes are obtained in \Caravel{} by an
efficient Berends--Giele recursion~\cite{Berends:1987me}.
We thus extended the QCD Feynman rules in \Caravel{}
with the $ggH$, $gggH$, and $ggggH$ vertices of the effective Higgs-gluon
interaction \eqref{eq:eft-lagrangian}. In this implementation the
embedding-space dimension $D_s$ can be set to even integer values.

To implement the 't~Hooft--Veltman scheme of dimensional regularization, we
require the analytic dependence on the state-counting parameter $D_s$, which is
subsequently set to $D_s = 4 - 2\epsilon$.
Our numerical approach allows the use of dimensional reconstruction \cite{Giele:2008ve} to obtain the
analytic $D_s$ dependence by sampling over $D_s\in\{6,8,10\}$.
However, to reduce computational costs, we extended the dimensional reduction method
\cite{Anger:2018ove,Abreu:2018jgq,Abreu:2019odu,Sotnikov:2019onv}
for Higgs-gluon Feynman rules, allowing the $D_s$ dependence to be
tracked analytically via auxiliary scalar and fermionic states.
Compared to sampling over $D_s$, we observed with this approach
shorter runtimes by $\{10\%, 45\%, 80\%\}$
and memory-use reduction by $\{25\%, 55\%, 75\%\}$
in the four-gluon, two-quark and four-quark amplitudes, respectively.

Assembling unitarity cuts in \Caravel{} requires input data that specifies a
partial amplitude's cut-diagram hierarchy and its expansion in $\NC$ and $\NF$.
The cut diagrams are generated with \texttt{qgraf}~\cite{Nogueira:1991ex} and
further organized into their cut hierarchy using a private \texttt{Mathematica} library, which
we have extended to incorporate the effective Higgs theory.

The reduction of the amplitude is then performed by matching its cuts to a
parametrization of the loop integrands in terms of master and surface
terms~\cite{Ita:2015tya}, which we extended in \Caravel.
Required planar integrands \cite{Abreu:2020jxa} were taken from earlier work~\cite{Abreu:2021asb}.
In the leading-color approximation, the only non-planar integral families
required were one Hexa-Box ({HBzzz}) and one Double-Pentagon ({DPzz}) family as introduced in refs.~\cite{Abreu:2021smk,Abreu:2023rco}.
These families include exactly the non-planar topologies
that become planar by removing the massive Higgs leg.
The definitions for some master integrals involve doubled propagators,
which we eliminate by using \Kira{}~\cite{Klappert:2020nbg} to obtain
a unitarity-compatible representation.
Also, we implemented a new set of surface terms for the non-planar
topologies, and extended all planar surface terms~\cite{Abreu:2021asb} to
cover higher loop-momentum powers originating in the non-renormalizability of the Higgs-gluon interaction.
Surface terms correspond to integration-by-parts identities from
suitably chosen unitarity-compatible generating vectors
\cite{Gluza:2010ws,Schabinger:2011dz}. Each generating vector
serves as a template for surface terms of different power-counting degree,
which is adjusted by multiplication with suitable monomials.
Most generating vectors were derived by solving a syzygy problem in
algebraic geometry \cite{Gluza:2010ws} following the approach of
ref.~\cite{Abreu:2017hqn}.
To obtain the more complicated vectors, linear systems were generated
by sampling syzygies numerically on phase-space points~\cite{Abreu:2023bdp,Klinkert:2023cyd}.
The new set of surface terms covers the integrand space up to seventh power in propagator variables
for the non-planar eight-propagator topologies.

\section{Analytic reconstruction}
\label{sec:analyticReconstruction}

In this section, we explain the analytic reconstruction method
we use to obtain analytic expressions for the rational coefficients of the finite remainders.
We build on ideas that were explored in refs.~\cite{vonManteuffel:2014ixa, Peraro:2016wsq},
starting with an ansatz for the coefficient functions, which is then fit to
numerical evaluations in a finite-field.
Reconstructions in different finite-fields are then combined to obtain the coefficient functions over the rational numbers.
In principle, the reconstruction could be performed with sufficiently
many numerical samples using multivariate functional reconstruction
techniques~\cite{Peraro:2016wsq} (see also
refs.~\cite{Klappert:2020nbg,Magerya:2022hvj,Belitsky:2023qho,Liu:2023cgs})
based on Newton and Thiele's interpolation algorithms, or through
solutions to Vandermonde systems
\cite{Press:2007ipz,Ellis:2007br,Klappert:2019emp,Abreu:2021asb}.
However, given the complexity of the present functions, a dedicated
algorithm is required to successfully complete the reconstruction and
to obtain a sufficiently compact final result.

In this computation we will follow our earlier work: we analyze the
analytic form of the finite remainder coefficients in spinor-helicity variables
\cite{Laurentis:2019bjh,DeLaurentis:2022otd}
in order to constrain the ansatz.
Thereby we reduce the number of samples required to fit all free parameters.
The reconstruction follows the recent approach \cite{DeLaurentis:2023nss,DeLaurentis:2025dxw}
which is further optimized for the present computation.

\subsection{Spaces of rational coefficients}
We combine all rational functions $r_{k,i}$ appearing in the finite
remainders $R_k$ up to two loops into the vector spaces
\begin{equation}
  B_k(1^{h_1},2^{h_2},3^{h_3},4^{h_4},5)
  =
  \text{span} \left\{
    r_{k,i}^{\ell,(n_c,\ell-n_c),\vec h}
    \;\text{ for }\;
    0 \le \ell \le 2,\;
    0 \le n_c \le \ell
  \right\} \, .
\end{equation}
We then identify and merge those spaces whose helicities are related
by permutations. Specifically, we consider
\begin{align}
  B_{4g}^{++} &= B_1(1^+,2^+,3^+,4^+,5) \,,\\
  B_{4g}^{+-} &= B_1(1^+,2^+,3^+,4^-,5) \,,\\
  B_{4g}^{--} &= B_1(1^+,2^+,3^-,4^-,5) + B_1(1^+,3^-,2^+,4^-,5) \,, \\[2.5mm]
  B_{2q2g}^{+} &= B_3(1^+,2^-,3^+,4^+,5) \,, \\
  B_{2q2g}^{-} &= B_3(1^+,2^-,3^+,4^-,5) + B_3(1^+,2^-,4^-,3^+,5) \,, \\[2.5mm]
  B_{4q} &= B_5(1^+,2^-,3^+,4^-,5) + B_5(1^+,2^-,4^-,3^+,5) \,.
\end{align}
We have labelled the $B_{4g}$ spaces by the helicities of $g_3$ and
$g_4$, and the $B_{2q2g}$ spaces by the helicity of $g_4$.
It is convenient to combine the vector spaces in this manner, as the
dimension of their sum is smaller than the sum of the individual
dimensions, implying that fewer functions need to be determined.

\begin{table}[t]
\begin{tabular}{C{1.2cm}@{\hspace{8mm}}C{2.0cm}C{0.9cm}@{\hspace{8mm}}
                C{1.4cm}C{1.4cm}@{\hspace{8mm}}C{1.5cm}C{1.7cm}}
\toprule 
Vector &
Subsp. &
Gen. & 
\multicolumn{2}{c}{\hspace{-8mm}Common denom.} &
\multicolumn{2}{c}{Partial fractions} \\
space &
dims. &
set &
\(\deg(\mathcal{B}^{\rm deg})\) &
\(\deg(\mathcal{B}^{\text{imp}})\) &
\(\deg(\mathcal{B}^{\text{imp}})\) &
Nbr.~terms  \\
\midrule
\(B_{4g}^{--}\) & \{330, 7, 0\} & 93 & 168 & 82 & 14 & 48 \\
\(B_{4g}^{+-}\) & \{179, 10, 1\} & 104 & 135 & 59 & 12 & 59 \\
\(B_{4g}^{++}\) & \{76, 4, 2\} & 24 & 63 & 50 & 14 & 16 \\
\midrule
\(B_{2q2g}^{-}\) & \{277, 7, 0\} & 159 & 106 & 66 & 13 & 26 \\
\(B_{2q2g}^{+}\) & \{150, 8, 1\} & 148  & 76 & 37 & 13 & 28 \\
\midrule
\(B_{4q}\) & \{151, 5, 0\} & 47 & 76 & 46 & 10 & 14 \\
\bottomrule
\end{tabular}
\centering
\caption{\label{tab:VSdata} Number of $\mathbb{Q}$-linearly
  independent rational functions appearing in the $pp \to Hjj$
  amplitudes, given as dimensions of the three disjoint subspaces of
  mass dimension $0$, $2$, and $-4$, together with the size of the
  generating set under symmetries. The next two columns give the
  maximum mass dimension, or equivalently polynomial degree, of the
  numerator polynomials before and after the basis change. The final
  two columns show the maximum mass dimension of the numerators after
  partial fractioning, and the maximum number of fractions in a
  multivariate partial fraction decomposition.
  }
\end{table}

We now wish to determine the linear dependencies of the elements of the spaces
$B_\kappa^h$. At this point we do not have the analytic expressions of the
coefficient functions available and we used evaluations on a random set of
phase-space points to determine their linear dependence \cite{Abreu:2018rcw,Abreu:2019odu}.
To this end, the values of the coefficient functions are
represented as vectors, whose linear dependence is determined using linear
algebra.
The dimensions of the spaces constructed in this manner are shown in
the second column of \cref{tab:VSdata}. Due to the presence of square roots,
$\sqrt{q}$, the resulting vector spaces split into up to three
disjoint subspaces, graded by mass dimension $0$, $2$, and $-4$ according to
the degree of the polynomial $q$.
Interestingly, the helicity configurations one might expect to be the
most complex, namely those with an equal or nearly equal number of
positive- and negative-helicity, while indeed containing the largest
number of functions overall, have fewer functions in the mass-dimension-$2$
and $-4$ subspaces, the latter being empty in half of the vector spaces
considered.

The dimensions of the vector spaces also provide further justification
for the choice made in eq.~\eqref{eq:del5pent} for handling
$\sqrt{\Delta_5}$. With that choice, the corresponding coefficients
are assigned to the mass-dimension-$0$ subspace rather than to a
distinct mass-dimension-$4$ subspace. As a result, the dimensions of
the pure-gluon spaces are reduced by $7$ for $B_{4g}^{++}$ and by $16$
for both $B_{4g}^{+-}$ and $B_{4g}^{--}$.

\subsection{Improved rational-coefficient bases}\label{sec:impbases}

We exploit the freedom to select a basis of coefficient functions to
simplify their subsequent functional reconstruction. To identify the
least complex functions, we analyze their least common denominator
(LCD) form,
\begin{equation} \label{eq:ratFunc}
  r_i(\lambda,\tilde \lambda) = \frac{\mathcal{N}_i(\lambda,\tilde \lambda)}{\prod_j \mathcal{D}_j^{q_{ij}} (\lambda,\tilde \lambda) } \, ,
\end{equation}
where $\mathcal{N}_i$ is a polynomial numerator, the $\mathcal{D}_j$
are irreducible denominator factors, and the $q_{ij}$ are integers.
As observed in ref.~\cite{Abreu:2018zmy}, the denominator factors are
distinguished polynomials that can be inferred from the symbol
alphabet of the relevant integral functions, here the pentagon
functions \cite{Abreu:2023rco}. Explicitly, the irreducible
denominator factors are
\begin{equation}\label{eq:denomfactors}
  \mathcal{D}_m \in \bigl\{ \langle ij\rangle, [ij], s_{i5}, \langle
  i|5|i], \langle i|j{+}k|i], \langle i|j{+}k|l], \Delta_{ij|kl|5},
        \langle i|j|5|k|i]-\langle j|l|5|k|j] \bigr\} \, ,
\end{equation}
where the labels $i,j,k,l \in \{1,2,3,4\}$ refer to massless momenta,
while momentum $5$ is understood to be massive. The last denominator factor is the
only new one at two loops and, perhaps surprisingly, only those that appear in (permutations of) planar amplitudes appear. That is, the set appearing
in \cref{eq:denomfactors} is identical to that of
ref.~\cite{DeLaurentis:2025dxw}.

Next, a small set of dedicated numerical evaluations of the finite
remainders is used to constrain the denominator structure of the
coefficient functions. In particular, the exponents $q_{ij}$ in
\cref{eq:ratFunc} are obtained from a univariate reconstruction of the
coefficient functions on a slice
$\{\lambda_i(t),\tilde\lambda_i(t)\}_{i=1,4}$ in the parameter $t$
\cite{Abreu:2018zmy}, which we review in \cref{sec:slices}. Once the
exponents $q_{ij}$ are known, the LCDs of the rational coefficients
are determined.

To characterize the complexity of a rational coefficient, we consider
the polynomial degree of its numerator in spinor brackets, denoted by
$\deg_{\mathcal{N}}(r_i)$. This coincides with the mass dimension
defined by the uniform rescaling in eq.~\eqref{eq:mass-dim-def}. Since
the overall mass dimension of the coefficients is fixed, the numerator
degree is related to the denominator degree by at most a constant shift.  The
LCDs therefore determine the numerator degrees
$\deg_{\mathcal{N}}(r_i)$ and, with them, the number of evaluations
required in the reconstruction.

Based on this analysis, we select the subset of linearly independent
coefficients $r_i$ of lowest numerator degree as basis ${\mathcal
  B}_\kappa^{h,{\rm deg}}$ of the vector space $B_\kappa^h$
\begin{align}
	r_i = \sum_{ r_j\,\in\,{\mathcal B}_\kappa^{h,{\rm deg}}} \widehat{M}_{ij} r_j \,, \quad r_i \in B_\kappa^h\,,
\end{align}
which is done in turn for each of the vector spaces $B_{4g}^{++}$,
$B_{4g}^{+-}$, $B_{4g}^{--}$, $B_{2q2g}^{+}$, $B_{2q2g}^{-}$ and
$B_{4q}$. Once a basis of functions is identified, we will not require
the matrix $\widehat{M}_{ij}$ explicitly. As we discuss in
\cref{sec:functionalReconstruction}, the $r_i$ will instead be
expressed directly in terms of a different, improved set of functions,
with a corresponding new matrix.

To assess the complexity of functional reconstruction, we list in
\cref{tab:VSdata} the maximal degree of the basis functions for the
different vector spaces, which we define by
\begin{equation}
  \deg(\mathcal{B}) = {\rm max}_{r_i \in \mathcal{B} }  \big( \deg_\mathcal{N}(r_i) \big) \,,
\label{eq:degreeVS}
\end{equation}
for a basis $\mathcal{B}$ of a vector space $B$. We observe that the
numerator of the highest degree of all the $\mathcal{B}^{h, {\rm deg}}_\kappa$ has mass dimension $168$, giving an
estimate of approximately~$8\cdot 10^7$ free parameters. A naive
functional reconstruction would be intractable in practice, both
because of the average evaluation times and because it would lead to
prohibitively complex expressions. We therefore apply an improved
reconstruction algorithm, which we summarize below.

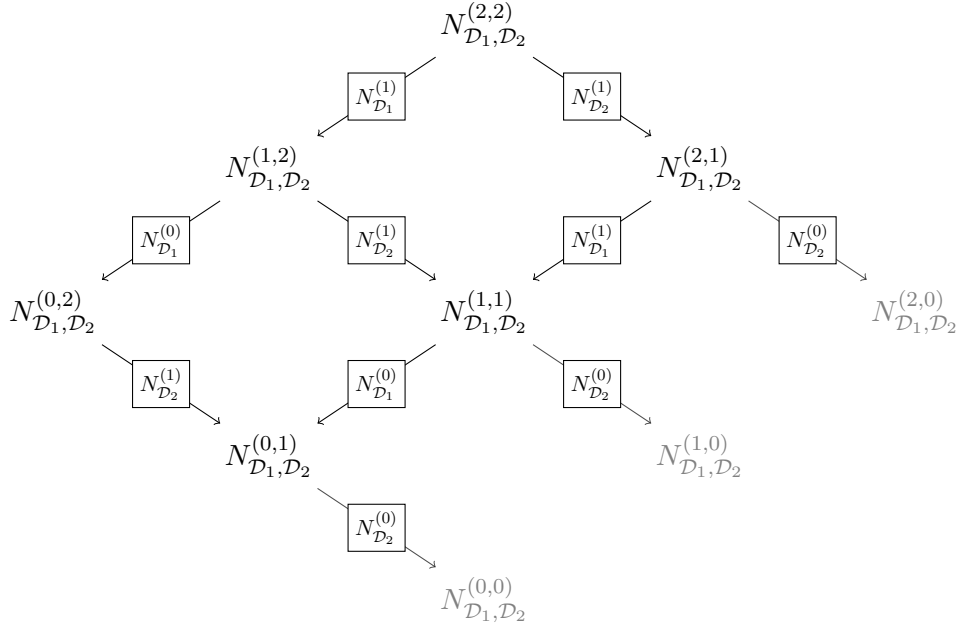
\begin{figure}
\centering
\begin{tikzpicture}
[
  scale=0.95, level 1/.style = {draw, ->, black, sibling distance = 6cm},
  level distance = 20mm
]

\hspace{-18mm}
\node {$N^{(2,2)}_{\mathcal{D}_1, \mathcal{D}_2}$}
child {
  node {$N^{(1,2)}_{\mathcal{D}_1, \mathcal{D}_2}$}
  child {
    node {$N^{(0,2)}_{\mathcal{D}_1, \mathcal{D}_2}$}
    child{
      node [opacity=0] {$N^{(-1,1)}_{\mathcal{D}_1, \mathcal{D}_2}$}
      edge from parent [draw=none]
    }
    child{
      node [opacity=1]{$N^{(0,1)}_{\mathcal{D}_1, \mathcal{D}_2}$}
      edge from parent node [midway, fill=white, draw, scale=0.75] {$ N_{\mathcal{D}_2}^{(1)}$}
    }
    edge from parent node [midway, fill=white, draw, scale=0.75] {$ N_{\mathcal{D}_1}^{(0)}$}
  }
  child {
    node [opacity=1] {$N^{(1,1)}_{\mathcal{D}_1, \mathcal{D}_2}$}
    child{
      node [opacity=0]{$N^{(0,1)}_{\mathcal{D}_1, \mathcal{D}_2}$}
      child{
        node [opacity=0] {$N^{(-1,1)}_{\mathcal{D}_1, \mathcal{D}_2}$}
        edge from parent [draw=none]
      }
      child{
        node [opacity=0.5]{$N^{(0,0)}_{\mathcal{D}_1, \mathcal{D}_2}$}
        edge from parent [opacity=0.7] node [midway, fill=white, draw, opacity=1, scale=0.75] {$N_{\mathcal{D}_2}^{(0)}$}
      }
      edge from parent node [midway, fill=white, draw, scale=0.75] {$ N_{\mathcal{D}_1}^{(0)}$}
    }
    child{
      node [opacity=0.5]{$N^{(1,0)}_{\mathcal{D}_1, \mathcal{D}_2}$}
      edge from parent [opacity=0.7] node [midway, fill=white, draw, opacity=1, scale=0.75] {$N_{\mathcal{D}_2}^{(0)}$}
    }
    edge from parent node [midway, fill=white, draw, scale=0.75] {$N_{\mathcal{D}_2}^{(1)}$}
  }
  edge from parent node [midway, fill=white, draw, scale=0.75] {$N_{\mathcal{D}_1}^{(1)}$}
}
child {
  node {$N^{(2,1)}_{\mathcal{D}_1, \mathcal{D}_2}$}
  child {
    node [opacity=0] {$N^{(1,1)}_{\mathcal{D}_1, \mathcal{D}_2}$}
    edge from parent node [midway, fill=white, draw, scale=0.75] { $N_{\mathcal{D}_1}^{(1)}$}
  }
  child {
    node[opacity=0.5] {$N^{(2,0)}_{\mathcal{D}_1, \mathcal{D}_2}$}
    child {
      node [opacity=0] {$N^{(1,0)}_{\mathcal{D}_1, \mathcal{D}_2}$}
      child {
        node [opacity=0] {$N^{(0,0)}_{\mathcal{D}_1, \mathcal{D}_2}$}
        edge from parent [draw=none] 
      }
      child{
        node [opacity=0] {$N^{(1,-1)}_{\mathcal{D}_1, \mathcal{D}_2}$}
        edge from parent [draw=none]
      }
      edge from parent [draw=none] 
    }
    child{
      node [opacity=0] {$N^{(2,-1)}_{\mathcal{D}_1, \mathcal{D}_2}$}
      edge from parent [draw=none]
    }
    edge from parent [opacity=0.7] node [midway, fill=white, draw, opacity=1, scale=0.75] {$N_{\mathcal{D}_2}^{(0)}$}
  }
  edge from parent node [midway, fill=white, draw, scale=0.75] {$N_{\mathcal{D}_2}^{(1)}$}
};
\end{tikzpicture}

\caption{\label{FigSearchTree}Example of a simple search graph for
finding a replacement $\tilde r_i$ for the function $r_i$ which has denominator factors $\mathcal{D}^2_1\mathcal{D}_2^2$.  Any node represents
  the intersection of the space entering the edge with the space along
  the edge. Faded vector spaces are empty or do not cover linear combination involving $r_i$.
  This example has a global minimum $N_{\mathcal{D}_1,\mathcal{D}_2}^{(0,1)}$ w.r.t.\ the denominator factors $\tilde r_i$ could have.
  It is quicker to reach it from the bottom up rather than from top down.}
\vspace{-2mm}
\end{figure}

To start with we follow the strategy of
refs.~\cite{DeLaurentis:2023nss,DeLaurentis:2025dxw} to find a  linear
transformation to an improved basis $\mathcal{B}_\kappa^{h,{\rm imp}}$
starting from the basis $\mathcal{B}^{h,{\rm deg}}_\kappa$
\begin{equation}
  \tilde r_i = \sum_j O_{ij} r_{j} \, ,
\quad \mbox{with}\quad 
\tilde r_i \in \mathcal{B}^{h,{\rm imp}}_\kappa
\quad \mbox{and}\quad  
r_j \in \mathcal{B}^{h,{\rm deg}}_\kappa \,,
\end{equation}
with reduced numerator degrees. 
The linear transformation matrix has constant rational entries $O_{ij} \in \mathbb{Q}$. Here we will
determine it in a single finite field, which suffices for the reconstruction.
We will return to this topic in \cref{sec:functionalReconstruction}.

The construction of the linear
transformation exploits the pole structure of the rational functions $r_i$
in eq.~\eqref{eq:ratFunc}. At the vanishing locus of the denominator factors $\mathcal{D}_j$
the coefficients develop poles, some of which are absent in the combined
remainder function $R_k$ (\ref{eq:remainder}). We refer to these poles as spurious
poles in the following. The spurious poles cancel because their residues become
linearly dependent and the integral function basis degenerates simultaneously at the poles.
Such correlations between residues can be
used \cite{Abreu:2019odu,DeLaurentis:2023nss,DeLaurentis:2025dxw}
to find linear combinations of coefficient functions with
reduced pole degrees, which is the central idea for constructing the basis change.
Consequently, the new functions $\tilde r_i$ are simpler since they have fewer denominator factors
and thereby lower numerator degree.

We now discuss technical aspects of the computation of the basis
change matrix $O_{ij}$.  To highlight improvements w.r.t.~our older
work, we introduce only the essential objects.  A more detailed
explanation of the base algorithm can be found in
ref.~\cite{DeLaurentis:2023nss, DeLaurentis:2025dxw}.  For a given
denominator factor $\mathcal{D}_k$ and pole degree $\nu$, we denote by
$N_{\mathcal{D}_k}^{(\nu)}$ the vector space of coefficient vectors
$\vec\alpha = (\alpha_i)_{i=1}^{|\mathcal{B}^{h,{\rm deg}}_\kappa|}$
that satisfy
\begin{align}\label{eq:nullSpace}
	\sum_{r_i\,\in\,\mathcal{B}^{h,{\rm deg}}_\kappa} \alpha_i r_i  = \mathcal{O}{\left(\frac{1}{\mathcal{D}^\nu_k}\right)} \,,
\end{align}
\emph{i.e.}~$N_{\mathcal{D}_k}^{(\nu)}$ covers all linear combinations
of functions $r_i$ for which the pole order of $\mathcal{D}_k$ is less
than or equal to $\nu$.  Thereby, a vector $\vec\alpha \in
N_{\mathcal{D}_k}^{(\nu)}$ describes a linear dependency of residues
of a pole $\mathcal{D}_k^\rho$, for some order $\rho > \nu$. Here, by
residue we mean the coefficient of a pole in $\mathcal{D}_k$ of
arbitrary order, and we refer to $N_{\mathcal{D}_k}^{(\nu)}$ as the
corresponding residue space.

The linear combination in eq.~\eqref{eq:nullSpace} is by itself a
candidate for a simplified coefficient function $\tilde r_i$ or,
equivalently, $\vec \alpha$ is a candidate row of $O_{ij}$.
Furthermore, one can find linear combinations, which remove multiple
poles; this amounts to finding vectors $\vec \alpha$ which lie in the
intersection of distinct vector spaces,
\emph{e.g.}~$N_{\mathcal{D}_k}^{(\nu)}$ and $N_{\mathcal{D}_{k'}}^{(\nu')}$.
For such intersections of two or more $N^{(\nu)}_{\mathcal{D}}$'s, we
introduce the notation
\begin{equation}
N^{\vec \nu}_{\vec{\mathcal{D}}} = N^{(\nu_1)}_{\mathcal{D}_{k_1}} \cap \dots \cap N^{(\nu_m)}_{\mathcal{D}_{k_m}}\, .
\end{equation}
In this notation, one naturally has $\vec\alpha_{r_i} \in N^{(q_{i1},q_{i2},\dots)}_{(\mathcal{D}_{1}, \mathcal{D}_{2},\dots)}$, where $\vec\alpha_{r_i}$ is the coefficient vector ($\alpha_{r_i,j} = \delta_{ji}$) representing the coefficient $r_i$ with denominator factors $\prod_j \mathcal{D}_j^{q_{ij}}$.
By constructing intersections like
\begin{equation}
	N^{(\nu_j-1)}_{\mathcal{D}_j} \cap N^{\vec\nu}_{\vec{\mathcal{D}}} = N^{(\nu_1,\dots,\nu_j - 1,\dots)}_{\vec{\mathcal{D}}}\, ,
\end{equation}
one can inspect if there is a linear combination of coefficients involving $r_i$ with reduced pole degree of $\mathcal{D}_j$.
Consequently, simplifying the basis $\mathcal{B}^{\rm deg}$ amounts to
finding optimal intersections of vector spaces, as discussed in refs.~\cite{DeLaurentis:2023nss,
DeLaurentis:2025dxw}.

The numerical construction of the basis change matrix $O_{ij}$ involves two key steps:
\begin{enumerate}
\item First, we need to construct the vector spaces
$N_{\mathcal{D}_k}^{(\nu)}$ which is tantamount to finding vectors $\vec
\alpha$ in \cref{eq:nullSpace}. This analysis relies on the
analytic reconstruction of the coefficient functions on sets of univariate slices
\cite{DeLaurentis:2023nss, DeLaurentis:2025dxw}. The univariate rational form
of the coefficient functions is used to extract residues on denominator
poles $\mathcal{D}_k$. By repeating the residue computation for a set of slices,
the linear dependence of the residues (\ref{eq:nullSpace})
and thus the vectors $\vec \alpha$ can be determined. It is important to note that the number
of slices that are required depends on the number of independent residue
functions associated to a particular pole.
\item The second computational task is to search for
a vector space $N_{\vec{\mathcal{D}}}^{\vec \nu}$, which gives the simplest
non-trivial denominator poles for a new basis function $\tilde r_i$, which replaces $r_i$,
\begin{equation}
\tilde r_i = \sum_j \alpha_j r_j \,, \quad \mbox{with} \quad \alpha_j \in N_{\vec{\mathcal{D}}}^{\vec \nu}\quad \mbox{and} \quad \alpha_i\neq 0 \,.
\end{equation}
For this search it is convenient to construct a search graph with the
vector spaces $N_{\vec {\mathcal{D}}}^{\vec \nu}$ as nodes
and associate the edges to the $N_{\mathcal{D}_k}^{(\nu)}$
\cite{DeLaurentis:2025dxw}.
An example search graph is shown in \cref{FigSearchTree}.
The node at the top is the vector space corresponding to the denominator exponents of the coefficient $r_i$,
which is to be replaced by a simpler function $\tilde r_i$.
$N_{\vec {\mathcal{D}}}^{\vec \nu}$ with smaller exponents are ordered below.
For each node we find a path from the top to reach it. The edges in turn determine the
vector spaces $N_{\mathcal{D}_k}^{(\nu)}$, which need to be intersected to
reach a given node.
The task is then to find a node with a non-empty space $N_{\vec
{\mathcal{D}}}^{(\vec \nu)}$ with lowest exponents $\vec \nu$ in an efficient way.
The search for such admissible denominators is often time-consuming and
requires a dedicated search algorithm.
\end{enumerate}
We applied this method to obtain the basis change $O_{ij}$ for all processes,
but with two computational modifications compared to previous calculations.
\begin{enumerate}
\item The first modification concerns the type of poles
  $\mathcal{D}_k$ that are included in the construction of the basis
  change. Empirically, it is more efficient to omit residues
  associated with some of the $\langle ij\rangle$ or $[ij]$ invariants
  from the analysis, since the high dimension of their residue spaces
  $N^{(\nu)}_{\langle ij\rangle}$ and $N^{(\nu)}_{[ij]}$ would require
  too many samples while providing only marginal additional
  simplification of the denominators.

  To validate this approach, for $B_{4g}^{++}$ we explicitly tested
  the effect of omitting poles, namely in this case the physical
  singularities $\langle 1\,2\rangle$, $\langle 2\,3\rangle$, $\langle
  3\,4\rangle$ and $\langle 1\,4\rangle$, confirming that the
  resulting gain in degree is only marginal.  Furthermore, we
  quantified the improvement for $B_{2q2g}^{+}$. In this case the
  largest retained residue space is the double pole in
  $\langle3\,4\rangle$, with dimension $26$, whereas the dimension of
  the corresponding simple pole exceeds $50$ and we omitted it.
  Including it would have required at least doubling the set of
  sampled slices, with a simplification by at most one unit in the
  numerator degree.

  Finally, we collect the excluded poles:
  \begin{align*}
    &B_{4g}^{++}: && \bigl\{\langle1\,2\rangle, \langle2\,3\rangle,
    \langle3\,4\rangle, \langle1\,4\rangle \bigr\}, \\
    &B_{4g}^{+-}: && \bigl\{\langle1\,2\rangle, \langle2\,3\rangle,
                      [2\,4], \langle1\,3\rangle^2\bigr\},\\
    &B_{4g}^{--}: && \bigl\{\langle1\,3\rangle, [1\,3],
                      \langle1\,4\rangle, [1\,4], \langle2\,3\rangle,
                                        [2\,3], \langle2\,4\rangle,
                                        [2\,4], \langle1\,2\rangle^2,
                                        [3\,4]^2\bigr\}, \\
    &B_{2q2g}^{+}: && \bigl\{\langle3\,4\rangle\bigr\},\\
    &B_{2q2g}^{-}: && \bigl\{\langle1\,3\rangle, [2\,4],
                                        \langle3\,4\rangle,
                                                          [3\,4]\bigr\},
  \end{align*}
  where an exponent denotes the degree of the pole. For the remaining
  vector space $B_{4q}$, we included all denominator factors in the
  analysis.

\item Secondly, we adjust the search algorithm for finding an
  admissible denominator structure using vector space intersection
  \cite{DeLaurentis:2025dxw}. In the original work, we used a search
  strategy starting from the top node and consecutively constructing
  spaces with lower pole orders $\vec\nu$, while prioritizing
  denominator factors of the same overall pole order $\sum_i \nu_i$
  (top-down breadth-first search), to find admissible denominators.
  Despite several optimizations, this search algorithm becomes
  prohibitively time and memory consuming for the most complex
  functions. We therefore make two adjustments:
\begin{enumerate}
\item A first mitigation of this combinatorial growth exploits
  symmetries of the remainders. For example, if a remainder such as
  $R_1(1^+,2^+,3^-,4^-,5)$ is symmetric under the permutation $1234
  \to 2143$, so that
  \begin{equation}
    R_1(1^+,2^+,3^-,4^-,5)=R_1(2^+,1^+,4^-,3^-,5)\,,
  \end{equation}
  then multiple functions can be related by such permutations. In
  favorable cases, an already reconstructed $\tilde r_i$ can, under
  permutation, provide functions that would otherwise require the
  basis change of a more complicated $r_i$. In such situations, the
  construction of a basis change for some functions can be avoided
  entirely. In practice, we run the analytic reconstruction
  (see~\cref{sec:PFD}) in parallel with the basis change, and once we
  have a set of functions $\tilde r_i$ which spans $B_{\kappa}^{h}$,
  we stop computing additional rows of $O_{ij}$, providing an
  effective early termination criterion. The effect of exploiting
  symmetries is reflected in table~\ref{tab:VSdata} by the difference
  between the vector-space dimensions and the sizes of the generating
  sets.
 
\item A second mitigation addresses the search strategy. The top-down
  breadth-first search becomes particularly costly when, at
  intermediate depths, the number of candidate intersections grows
  rapidly. This typically occurs for functions for which many poles
  can be dropped. We observe maximum breadths of the graph in excess
  of $100\,$k nodes, whereas the toy example in
  figure~\ref{FigSearchTree} has maximum breadth~3. In such cases, we
  perform a bottom-up breadth-first search instead. We begin by
  identifying the poles that can be reduced by one order. We then
  assume that all such poles can be simultaneously reduced,
  corresponding to the optimal scenario, even though this usually
  yields an empty intersection of null spaces. This amounts to testing
  whether the bottom node of the graph is admissible. From there, we
  progressively relax this assumption by allowing an increasing number
  of poles to remain, \emph{i.e.}\ we consider intersections in which
  all but one pole are reduced, then all but two, and so on, until a
  non-trivial intersection is found. In this way, the search moves
  upwards through the graph of denominators, avoiding the bottleneck
  associated with the large intermediate breadth.
\end{enumerate}
\item Finally, we computed the basis change matrices in a single finite-field only.
The final result was lifted to the rational numbers at the very end as discussed
in \cref{sec:functionalReconstruction}.
\end{enumerate}
The impact of the basis change is demonstrated in columns three and four of \cref{tab:VSdata}.
The numerator dimension is significantly reduced in the improved $\mathcal{B}^{\rm imp}$
compared to the initial basis $\mathcal{B}^{\rm deg}$.
The modifications 1 and 2 are of heuristic nature and we leave the development
of a more systematic algorithm to future work.

\subsection{Partial-fraction ansatz from a bivariate slice}
\label{sec:PFD}

Having obtained the optimized basis of coefficient functions $\tilde
r_i$, the next step is the reconstruction of their analytic form.
Despite the significant improvements by lowering mass-dimensions, the
numerators of the LCD form still have too many free parameters, so
that a direct determination from numerical evaluations is
unfeasible. Instead, we first improve the ansatz to obtain a partial
fractioned form for every basis element, which is then reconstructed.

Multivariate PFDs have been widely studied in commutative algebra and
symbolic computation, and have also found applications in high-energy
physics
\cite{Abreu:2019odu,Heller:2021qkz,deKorte:2026zrl,raichev2012leinartas,
  leinartas1978factorization,Bendle:2021ueg,Meyer:2016slj,
  beck2005partialfractionsmethodcountingsolutions}. The present
problem differs from these settings in two important ways: first, the
full multivariate numerator is not known analytically, and only
numerical sampling is available; second, the construction must work
effectively with spinor variables.\footnote{More precisely, the
construction needs to work in polynomial quotient rings, where the
chosen variables satisfy algebraic relations.}  Our goal is therefore
to determine, from numerical data, a compact multivariate
partial-fraction ansatz suitable for subsequent analytic
reconstruction. In this way, the complexity is controlled more by the
compact final representation, rather than by potentially large
starting expressions.

Explicitly, we first find a partial-fractioned form of the basis
elements
\begin{equation} \label{eq:pfd}
  \tilde r_i(\lambda,\tilde \lambda) = \sum_{k=1}^{k_{\rm max}} 
	\frac{\mathcal{N}_{ik}(\lambda,\tilde \lambda)}{\prod_j \mathcal{D}_j^{q_{ijk}} (\lambda,\tilde \lambda) } 
	\quad \mbox{with} \quad q_{ijk} \le q_{ij} \,,
\end{equation}
For better readability we suppress the index $i$ labeling the
coefficient functions from here on. In a second step, we apply an
analytic reconstruction procedure, which determines the free ansatz
parameters in the numerators $\mathcal{N}_{ik}$ using evaluations on
random phase-space points. This approach was used already in amplitude
computations \cite{Abreu:2023bdp,DeLaurentis:2025dxw} and we now
present a new method to determine the PFD from numerical evaluations.
The central novelty is the use of a bivariate analytic form of the
coefficient functions, which we obtain from functional reconstruction
on a bivariate slice.  Given this form we exploit the analytic
dependence on two variables to obtain a PFD \eqref{eq:pfd} with
denominator factors $\mathcal{D}_k$.

The central property of the PFD is that the originally common
denominator factors are distributed between distinct terms.  Here we
use a more stringent property which we assume to hold for coefficient
functions; namely, we assume that certain pairings of denominator
factors are excluded in the PFD. The observation that certain
denominator factors can be ``cleanly separated'' has already been used
to construct PFDs
\cite{Abreu:2023bdp,DeLaurentis:2022otd,DeLaurentis:2025dxw}. Here we
present a method to systematically determine excluded pairings and
construct a corresponding numerator ansatz.

We now describe the general structure underlying the partial-fraction
ansatz, before turning to its numerical implementation on the
bivariate slice. The key observation is that the exclusion of
denominator pairings can be formulated as an ideal-membership
criterion \cite{DeLaurentis:2022otd}: two denominator factors
$\mathcal{D}_i$ and $\mathcal{D}_j$ are split between terms in a PFD
if the numerator is an element of the ideal generated by the same
factors,
\begin{equation}\label{eq:idealMembership}
  \mathcal{N} \in \big\langle \mathcal{D}_i, \mathcal{D}_j \big\rangle \, ,
\end{equation}
where angle brackets $\langle \ldots \rangle$ denote a
polynomial ideal.
To give an example for this criterion, suppose that $\mathcal{N}\in \big\langle \mathcal{D}_1, \mathcal{D}_2 \big\rangle$, meaning that
\begin{equation}
	\mathcal{N} = \mathcal{N}_1 \mathcal{D}_1 + \mathcal{N}_2 \mathcal{D}_2\,,
\end{equation}
for some polynomials $\mathcal{N}_i$. This implies that the following fraction has the PFD
\begin{equation}
	\frac{\mathcal{N}}{\mathcal{D}_1 \mathcal{D}_2} = \frac{\mathcal{N}_1 \mathcal{D}_1 + \mathcal{N}_2 \mathcal{D}_2}{\mathcal{D}_1 \mathcal{D}_2} = \frac{\mathcal{N}_2}{\mathcal{D}_1} + \frac{\mathcal{N}_1}{\mathcal{D}_2}.
\end{equation}
More generally, if multiple pairs of factors (with general exponents) split in the partial fraction
decomposition, the numerator $\mathcal{N}$ is simultaneously element of
multiple ideals,
\begin{align}\label{eq:idealIntersection}
\mathcal{N} \in \mathcal{I}=
\big\langle \mathcal{D}^{a_1}_{i_1}, \mathcal{D}^{a_2}_{i_2} \big\rangle \cap 
\big\langle \mathcal{D}^{a_3}_{i_3}, \mathcal{D}^{a_4}_{i_4} \big\rangle \cap \ldots \cap\
\big\langle \mathcal{D}^{a_{2n-1}}_{i_{2n-1}}, \mathcal{D}^{a_{2n}}_{i_{2n}} \big\rangle \,,
\end{align}
where we allowed generic exponents $a_i$ of the denominator factors
and intersected multiple two-generator ideals.  Computing such an
ideal intersection is often computationally challenging.  Instead of
computing the ideal intersection, we construct (see below) elements of
the ideal which are monomials in the $\mathcal{D}_i$,
\eqref{eq:idealIntersection}
\begin{align}
m_k(\mathcal{D}) = \prod_j \mathcal{D}_j^{\hat q_{jk}} \in \mathcal{I}  \quad\mbox{with}\quad k=1,\ldots, k_{\rm max}
\end{align}
to construct an ansatz for the numerator
\begin{align}
\mathcal{N} = \sum_{k=1}^{k_{\rm max}} \mathcal{N}_k \,  m_k(\mathcal{D} ) \,.
\end{align}
Provided that this ansatz holds, the polynomials $\mathcal{N}_k$ are
the desired numerators of the PFD,
\begin{equation}\label{eq:PFD_from_Iroll}
  \tilde r = \frac{\mathcal{N}}{\prod_j \mathcal{D}_j^{q_j}} =
  \sum_{k=1}^{k_{\rm max}} \frac{\mathcal{N}_k }{\prod_j \mathcal{D}_j^{q_{jk}}}
	\quad\mbox{with} \quad \prod_j \mathcal{D}_j^{q_{jk}} = \frac{\prod_j \mathcal{D}_j^{q_{j}}}{m_k(\mathcal{D})} \, .
\end{equation}
The new numerators $\mathcal{N}_k$ are of lower degree than the
initial polynomial $\mathcal{N}$ and may now be determined from evaluations
on random phase-space points. In the denominators appearing in
eq.~\eqref{eq:PFD_from_Iroll}, those excluded pairs whose constraints
can be imposed simultaneously are manifestly absent. We show the
improved numerator degrees and the number of terms $k_{\rm max}$ in
the last two columns of \cref{tab:VSdata}, respectively. Together,
these quantities provide a measure of the number of unknown parameters
in the PFD.

\paragraph{Numerical implementation}
So far we discussed an analytic ideal membership criterion
\eqref{eq:idealMembership}, which implies a PFD. We will now implement
this criterion numerically.  In earlier work \cite{Laurentis:2019bjh,
  DeLaurentis:2022otd} the membership was detected through singular
evaluations near the vanishing surface of a pair of factors
$\mathcal{D}_i$ and $\mathcal{D}_j$.  In contrast, here the key
observation is that we can validate the ideal membership analytically
on a generic two-dimensional subspaces of parameter space.  To start
with, we determine the coefficient functions $\tilde r$ on a generic
two-dimensional surface in parameter space,
\begin{equation}
	\{\lambda_i(u,v),\tilde\lambda_i(u,v)\}_{i=1,\dots,4}\,,
\end{equation}
which we refer to as a bivariate slice. For an explicit phase-space
parametrization we refer to \cref{sec:slices}.
As the denominators of the $\tilde r_i$ were already found, the analytic
reconstruction on a bivariate slice amounts to determining the numerator
polynomials,
\begin{equation}
  \mathcal{N}(u, v) = \tilde r(u,v)\, {\prod_j \mathcal{D}_j^{q_{j}} (u,v) }
\end{equation}
via bivariate Newton interpolation.\footnote{See
\emph{e.g.}~\texttt{bivariate\_Thiele\_on\_slice\_given\_LCD} available at
\texttt{antares.scaling.slicing} \cite{giuseppe_de_laurentis_2026_18894183}.} 
For numerator polynomials of
degree $\deg(\mathcal{N}_i)$, this step requires
$(\deg(\mathcal{N}_i)+1)^2$
sampling points for the used slice parametrization (see \cref{sec:slices}).
With the analytic form of the coefficients on the bivariate slice available
\begin{equation}
  \tilde r(u, v) = \frac{\mathcal{N}(u,
    v)}{ \prod_j \mathcal{D}_j^{q_{j}} (u,v)  } \, ,
\end{equation}
a partial fraction ansatz can now be constructed using the ideal membership approach above.
The ideal membership criterion on the bivariate slice reads
\begin{equation} \label{eq:idealMembershipSlice}
  \mathcal{N}(u,v) \stackrel{?}{\in} \mathcal{J}_{ij}^{a_ia_j}:=\big\langle
    \mathcal{D}_i(u,v)^{a_i},
    \mathcal{D}_j(u,v)^{a_j}
  \big\rangle
  \quad\, \text{for} \quad\,
  1 \le a_i \le q_i,\;\; 1 \le a_j \le q_j \, .
\end{equation}
This provides a systematic way to identify which pairs of denominator
factors, and with which exponents, may be separated in a PFD.  We
collect all ideals for which the ideal membership criterion
\eqref{eq:idealMembershipSlice} is fulfilled,
\begin{equation}\label{eq:Jtot}
\mathcal{J}^{\rm tot} = \bigl\{
\big\langle \mathcal{D}^{a_1}_{i_1}, \mathcal{D}^{a_2}_{i_2} \big\rangle \,, 
\big\langle \mathcal{D}^{a_3}_{i_3}, \mathcal{D}^{a_4}_{i_4} \big\rangle \,, \ldots \,,
\big\langle \mathcal{D}^{a_{2n-1}}_{i_{2n-1}}, \mathcal{D}^{a_{2n}}_{i_{2n}} \big\rangle \bigr\} \,.
\end{equation}
For later purposes it is useful to sort the ideals, so that the most
complicated ones appear first. This is achieved by reverse lexicographic
ordering with respect to the list in \cref{eq:denomfactors}.

\paragraph{Simplified ideal intersection}
A non-trivial issue arises when attempting to combine the pairwise
constraints \eqref{eq:idealMembershipSlice} by ideal intersection
\eqref{eq:idealIntersection}.  Ideal intersection can be
computationally intensive and might produce complicated results.  Here
we instead construct a simple subset of the ideal intersection, which
suffices for our purposes: given two ideals
$\mathcal{J}_{ij}^{a_ia_j}$ and $\mathcal{I}_{\rm s}=\langle
m_1,...,m_r\rangle$ we introduce a simplified intersection operation
$\cap_{s}$,
\begin{equation}
\begin{aligned}\label{eq:RS_merge_generalised}
 \mathcal{I}_{\rm s}'&=
	\mathcal{I}_{\rm s} \cap_{s} \mathcal{J}_{ij}^{a_ia_j}:=
  \big\langle
    \mathrm{lcm}(m_1,n_1), \mathrm{lcm}(m_1,n_2), \ldots,
    \mathrm{lcm}(m_r,n_1), \mathrm{lcm}(m_r,n_2)
  \big\rangle \\ 
& \quad \mbox{with}\quad 
	n_1=\mathcal{D}_i(u,v)^{a_i}
	  \quad\mbox{and}\quad
	n_2=\mathcal{D}_j(u,v)^{a_j}\,,
\end{aligned}
\end{equation}
where $\mathrm{lcm}(m,n)$ denotes the least common multiple of the
polynomials $m$ and $n$.  The resulting ideal is in general only a
sub-ideal of the actual ideal intersection,
\begin{equation}\label{eq:RS_merge_subset}
  \mathcal{I}_{\rm s}' 
\subseteq \mathcal{I}_{\rm s} \cap \mathcal{J}_{ij}^{a_ia_j}\,,
\end{equation}
so this construction does not retain the full information of the
intersection. In practice, however, it is sufficient for our purposes.

Our procedure is then as follows.  We start with an ideal
$\mathcal{I}_{\rm s}$, initially set to the first ideal in
$\mathcal{J}^{\rm tot}$
\begin{equation}
\mathcal{I}_{\rm s} = \mathcal{J}^{\rm tot}_1 \,.
\end{equation}
Then, we intersect $\mathcal{I}_{\rm s}$ with the next candidate
two-generator ideal, $\mathcal{J}^{\rm tot}_i$ with $i=2$
\begin{equation}\label{eq:merge}
  \mathcal{I}_{\rm s}'= \mathcal{I}_{\rm s} \cap_{s} \mathcal{J}^{\rm tot}_i \, .
\end{equation}
After this merge, redundant generators are removed in
$\mathcal{I}_{\rm s}'$ by passing to a minimal basis, \emph{i.e.}\ by
discarding any generator divisible by another. Since $\cap_s$ retains
only a sub-ideal of the full intersection, the condition
$\mathcal{N}(u,v)\in\mathcal{I}_{\rm s}'$ is no longer automatic and
must be checked explicitly:
\begin{equation}
  \mathcal{N}(u,v) \stackrel{?}{\in} \mathcal{I}_{\rm s}' \, .
\end{equation}
This is done by reducing $\mathcal{N}(u,v)$ modulo a Gr\"obner basis
of $\mathcal{I}_{\rm s}'$. If the test succeeds, the candidate pair
is accepted and $\mathcal{I}_{\rm s}$ is updated
\begin{align}
\mathcal{I}_{\rm s}=\mathcal{I}_{\rm s}'\,.
\end{align}
Otherwise, it is discarded and we attempt to merge the subsequent
two-generator ideal $\mathcal{J}^{\rm tot}_{i+1}$, going back to
\cref{eq:merge}. The final result is the ideal $\mathcal{I}_{\rm s}$,
which consists of monomials $m_k$ in the factors $\mathcal{D}_i$,
\begin{equation} \label{eq:simplIdeal}
\mathcal{I}_{\rm s}=\langle m_1(\mathcal{D}),\ldots,m_r(\mathcal{D})\rangle \quad 
\mbox{with} \quad \mathcal{N} \in \mathcal{I}_{\rm s} \,.
\end{equation}
Compared to the full ideal intersection, this simplified procedure may
miss constraints on the numerator that are not captured by monomials
in the denominator factors, \emph{i.e.}\ by additional generators of
the true intersection \eqref{eq:idealIntersection}. The resulting ansatz may therefore not be
optimal, but in practice it is more than sufficient to achieve a
substantial simplification. 

\paragraph{Partial fraction ansatz}
At this stage, given $\mathcal{N}(u,v) \in \mathcal{I}_{\rm s}$, there
exist polynomials $\mathcal{N}_k(u,v)$ such that
\begin{equation} \label{eq:decompositionRelation}
  \mathcal{N}(u,v) = \sum_{k=1}^{k_{\rm max}} \mathcal{N}_k(u,v)\, m_k(\mathcal{D}) \, .
\end{equation}
This relation is established on the bivariate slice, and we then test
whether it lifts to the full multivariate setting in
\cref{sec:functionalReconstruction}. In that case it is equivalent to
the multivariate PFD in \cref{eq:PFD_from_Iroll}, where the $m_k$ are
multivariate monomials in the denominator factors.
Given the construction of the simplified intersection procedure $\cap_s$,
each quotient in the denominator of \cref{eq:PFD_from_Iroll} is
guaranteed to be a polynomial, which implies a PFD with identical or reduced
denominator degrees compared to the starting LCD form of $\tilde r$ in \cref{eq:PFD_from_Iroll}.

Besides the already mentioned subtlety in lifting the bivariate decomposition
to the full multivariate ring (see \cref{sec:functionalReconstruction}),
another limitation is that two-variables slices do not capture all
ways to partial fraction the coefficient functions. 
We leave the investigation of a more fine-grained denominator
decomposition, such as utilizing higher-dimensional slices, to future work.

In summary, we never need to compute the true intersection
\eqref{eq:idealIntersection} explicitly. Instead, we only test whether the
considered numerator can be written in terms of the monomials in the simplified
ideal $\mathcal{I}_{\rm s}$ \eqref{eq:simplIdeal}, and hence whether it admits
the decomposition of \cref{eq:PFD_from_Iroll}. This yields a practical
iterative construction of a PFD basis from two-generator ideals.

\subsection{Functional reconstruction}
\label{sec:functionalReconstruction}

Having determined a simplified basis $\mathcal{B}^{\text{imp}}$, with
basis functions $\tilde r_i$, and their PFD form, as in
\cref{eq:PFD_from_Iroll}, we have achieved sufficient simplification
to perform their analytic reconstruction with numerical coefficients
kept in a finite field.  To this end the numerators $\mathcal{N}_{ik}$
are parametrized through a similar spinor ansatz in six-point
kinematic as employed for the $pp\rightarrow Vjj$ process
\cite{DeLaurentis:2025dxw}. The six-point kinematic has the vector
boson momentum parametrized by its massless decay-product momenta
$p_V=p_5'+p_6'$. Kinematically this momentum parametrization applies
as well to the Higgs boson with $p_H=p_5=p_5'+ p_6'$.  However, when
considering spinor polynomials one has to impose zero little-group
weights in the auxiliary spinors $\lambda_{5}'$, $\lambda_{6}'$,
$\tilde\lambda_{5}'$ and $\tilde\lambda_{6}'$.  Once available, a
spinor parametrization can then be recast trivially from the massless
six-point to the actual five-point one-mass notation.

Given the parametrization for the numerators $\mathcal{N}_{ik}$, we
determine their coefficients from evaluations at generic phase-space
points. As anticipated in the discussion of the PFD ansatz, not every
decomposition that is valid on the bivariate slice lifts to a valid
multivariate PFD. Empirically, this issue is associated with including
the invariants $\langle a\,b\rangle$ and $[a\,b]$ in the two-factor
ideals of $\mathcal{J}^{\rm tot}$ in eq.~\eqref{eq:Jtot}. Excluding
such two-factor ideals consistently yields valid PFDs in all cases
considered here. Even with this restriction, the resulting PFDs remain
sufficiently compact for practical use, with the largest ansatz,
corresponding to the function with numerator mass dimension $82$,
requiring $12\,$k phase-space evaluations to fit.

\paragraph{Lifting to rational numbers}
The next step is to lift the finite-field coefficient functions
$\tilde r_i$ to the rational numbers. In principle, the whole
computation can be repeated in further finite fields to facilitate
this lift to rational numbers. In practice, we find a more efficient
approach.

To avoid sampling on univariate slices for distinct finite fields, we
determine the basis-change matrix $O_{ij}$ in a single finite field,
without lifting it to the rationals. This reduces by more than a
factor of two the number of sampled kinematic points required to
compute the basis change. However, it also implies that the $\tilde
r_i$ cannot be evaluated directly in different finite fields, so some
additional care is required when lifting their coefficients to
rational numbers.  For most functions this does not pose a problem, as
a single finite field already suffices to reconstruct $\tilde r_i$,
since their parameters are often simple rational numbers. In some
cases, this requires reconstructing $\tilde r_i$ normalized by one of
its non-zero numerical coefficients rather than reconstructing $\tilde
r_i$ itself. This is acceptable, since the goal is to determine an
arbitrary basis of the space. We can verify validity over the
rationals by testing whether the lifted $\tilde r_i$ belongs to the
function vector space in a different finite field. This requires only
as many phase-space points as the dimension of the space
$|\mathcal{B}^\text{deg}|$.

However, for a few $\tilde r_i$ in each remainder, we
find it necessary to sample in a second finite field. In such cases,
we compare against the original basis
$r_{i}\in\mathcal{B}^{\rm deg}$, while modding out the remaining
functions, analogously to the procedure 
in ref.~\cite{Abreu:2023bdp}, \emph{i.e.}~we consider
\begin{equation}\label{eq:basis-changed-ansatz-against-original-basis}
  r_i = \sum_{r_j\,\in\,\mathcal{B}^{\rm deg},\, j\neq i }  c_j r_j + 
  \sum_{k=1}^{k_{\rm max}} \frac{\mathcal{N}_{ik}}{\prod_j \mathcal{D}_{j}^{q_{ijk}}} \, ,
\end{equation}
and solve for $c_j$ and the parameters in $\mathcal{N}_{ik}$.
Here, some of the functions $r_j\in \mathcal{B}^{\rm deg}$
can also be replaced by functions $\tilde r_i$ that have already been
reconstructed. In other words, the
relevant row of $O_{ij}$ is recomputed at fitting time using random
phase-space points rather than univariate slices, by determining the
coefficients $c_j$. Even when a second finite field is required, we
find that most coefficients in $\mathcal{N}_{ik}$, being simple rational numbers, 
are already correct from the first field,
so that only $\mathcal{O}(10^2\!-\!10^3)$ additional points are needed
in the second field. The correctness of the result is validated using an
evaluation in an additional finite field in the end.

Attempting to fit rational functions directly from the outset using
\cref{eq:basis-changed-ansatz-against-original-basis} can lead to
degenerate systems, in which all $\mathcal{N}_{ik}$ vanish
identically on the sampled points. This is likely related to
degeneracies in the basis change: if the intersection of null spaces
has dimension greater than one, multiple distinct linear combinations
of $r_i\in\mathcal{B}^{\rm deg}$ can produce the same denominator for
$\tilde r_i$, leading to underconstrained systems. Deferring this step
until the basis $\tilde r_i$ has been partially reconstructed and some 
simple parameters in $\mathcal{N}_{ik}$ and some $c_j$ 
have been determined avoids this issue.

After fitting, we subject each function to an automated brute-force
search that attempts to lower the degree of each pole in each
fraction. 
In principle, this search could be computationally
expensive, but in practice it is made efficient by caching both target
evaluations and phase-space points. As a result, each trial only
requires rebuilding the corresponding linear system and performing the
associated row reduction, both of which are GPU accelerated. The
reconstruction logic, including caching, is implemented in a private
development version of \textsc{Antares}
\cite{giuseppe_de_laurentis_2026_18894183}, while the linear-system
solving is performed with \textsc{Linac}, whose implementation will be
described in a forthcoming paper \cite{DeLaurentis:Linac}. 

Lastly, for some of the functions, we further apply iterated fitting
in singular limits using $p\kern0,2mm$-adic evaluations
\cite{Laurentis:2019bjh,DeLaurentis:2022otd,Chawdhry:2023yyx}. We
observe that this procedure typically yields expressions that are up
to a factor of two more compact than those obtained with the present
approach. The advantage of the present method, however, is that it
does not require evaluations in singular limits and is more easily
automated. 

\section{Results}
\label{sec:results}

We provide analytic expressions for the finite remainders
\eqref{eq:remainder} corresponding to the color-stripped amplitudes listed in
\cref{eq:listHelConfs} through two loops. They are presented in
the form
\begin{equation}\label{eq:finRemainderAnalytic}
	R^{(\ell),(n_c, n_f)}_k = \sum_{i,j} \tilde{r}_{k,i} M^{(\ell),(n_c, n_f)}_{k,ij} G_j \, ,
\end{equation}
where $\tilde{r}_i$ are rational functions in spinor-helicity
variables, $M_{ij}$ are sparse matrices of rational numbers, and $G_j$
are (monomials of) five-point one-mass pentagon functions. The latter are implemented
in \textsc{PentagonFunctions-cpp} \cite{PentagonFunctions}. Both
\texttt{Mathematica} and \texttt{Python} versions of the results are
provided: the former as ancillary files to the arXiv submission, and
the latter on \texttt{GitHub} in the \textsc{antares-results}
repository \cite{antaresresults}, whose releases are archived on
\texttt{Zenodo} \cite{giuseppe_de_laurentis_2026_18896068}. Further
details on the format of the analytic results are given in
appendix~\ref{app:ancillaries}. Overall, the rational functions
$\tilde r_i$ occupy about $1.5\,$MB in plain-text \LaTeX{} format,
whereas the matrices $M_{ij}$ take about $22\,$MB in plain-text
coordinate format.

\begin{table}[t]
\begin{tabular}{cC{2.6cm}C{2.6cm}C{2.2cm}C{2.8cm}}
\toprule
Vector space & Univ.~slices & Biv.~slices & Random & Total \\
\midrule
\(B_{4g}^{--}\) & \(60 \times 171\) & \(85 \times 85\) & 20k & \(\approx 55\)k \\
\(B_{4g}^{+-}\) & \(55 \times 138\) & \(62 \times 62\) & 25k & \(\approx 45\)k \\
\(B_{4g}^{++}\) & \(79\times 66\) & \(53 \times 53\) & 2k & \(\approx 12.5\)k \\
\midrule
\(B_{2q2g}^{-}\) & \(55 \times 109\) & \(69 \times 69\) & 10k & \(\approx 22.5\)k \\
\(B_{2q2g}^{+}\) &  \(50\times 79\) & \(40 \times 40\) & 5k & \(\approx 12.5\)k \\
\midrule
\(B_{4q}\) & \(50 \times 79\) & \(49 \times 49\) & 3k & \(\approx 10\)k \\
\bottomrule
\end{tabular}
\caption{\centering\label{tab:NbrPoints} Number of points collected for
  reconstruction across each of the vector spaces, including a breakdown
  of the main sources.}
\end{table}

The analytic expressions are obtained through the analytic
reconstruction procedure described in section
\ref{sec:analyticReconstruction}, leveraging numerical evaluations of
the amplitude over a finite field in \textsc{Caravel}, see
section~\ref{sec:numComputation}. A summary of the number of points
collected for each vector space is summarized in \cref{tab:NbrPoints},
together with a breakdown into the main sources.

The first of these are the slices required to determine the residues
for the basis change, as presented in \cref{sec:impbases}. The exact
number of points required for each slice is $(\deg(\mathcal{B}^{\rm
  deg}) + 3)$,\footnote{To remind the reader, the degree was
introduced in eq.~\eqref{eq:degreeVS}} as the termination criterion in
our implementation is two consecutive zeros.\footnote{See
\texttt{(multivariate\_)Newton\_polynomial\_interpolation} available
at \texttt{pyadic.interpolation}
\cite{giuseppe_de_laurentis_2026_18881428}.}  We need several slices
for the residues, depending on the dimension of their spaces, leading
to a total number of sampled points equal to $\big(\text{max dim.~of
  residue spaces } N^{(\nu)}_{\mathcal{D}_j}\big) \times
\big(\deg(\mathcal{B}^{\rm deg}) + 3\big)$. The dimension of the
residue spaces is at first estimated, and then discovered
numerically. As explained in section \ref{sec:impbases}, some poles
are also omitted if the associated dimension is too high.  After the
basis change, we require a single bivariate slice to determine the
multivariate partial fraction decomposition, as presented in
\cref{sec:PFD}. This bivariate reconstruction requires
$\big(\deg(\mathcal{B}^{\text{imp}})+3\big) \times
\big(\deg(\mathcal{B}^{\text{imp}})+3\big)$ points to be sampled. The
final source consists of a few thousand random samples.

The total number of collected points exceeds the sum of the main
contributions listed above for two reasons. First, it includes
additional points required at intermediate stages of the
reconstruction, for example to determine the vector-space dimensions
and least common denominators, verify symmetries, and lift rational
functions and matrices to $\mathbb{Q}$. While lifting the rational
functions typically required only a couple of finite-field values,
lifting the matrices for the most complex remainders required
up to eight 32-bit primes. Second, for the gluon vector spaces, some
points were collected in early stages of the computation that were
ultimately not needed in the final reconstruction strategy.

Let us comment on the number of finite-field samples required for the analytic reconstruction of the $B_{2q2g}^{\pm}$ vector spaces in this work compared to the corresponding requirements in ref.~\cite{Hartanto:2026xjz},
which is based on a substantially different methodology.
While a direct comparison is not meaningful due to these methodological differences, we estimate that our approach requires approximately two orders of magnitude fewer samples.
In making this estimate, we take into account that each sample quoted in ref.~\cite{Hartanto:2026xjz} includes a univariate reconstruction,
while in our work a sample refers to fully numerical external kinematics.

\subsection{Analytic structure}
\label{sec:analyticStructure}

\newcommand{\sif}[1]{\Sigma_5^{(#1)}}
\newcommand{\sqrtsi}[1]{\sqrt{\sif{#1}}}

We now discuss the analytic structure of the two-loop finite remainders that we have calculated in this work.
We remind the reader that in the heavy-top EFT non-planar diagrams contribute in LCA.
The properties of non-planar Feynman integrals \cite{Abreu:2021smk,Abreu:2023rco}, which have notably richer analytic structure than their planar counterparts \cite{Abreu:2020jxa,Chicherin:2021dyp}, are therefore imprinted on the analytic structure of the finite remainders.
New higher-degree polynomials appear as denominators in the differential equations satisfied by the former, and consequently in the $\dd\log$ integration kernels (letters).
In particular, a new (permutation orbit of) quartic polynomials $\Sigma_5^{(i)}$ that define a variety on which the related integrals showcase puzzling non-analytic behavior within the physical scattering region \cite{Abreu:2023rco}, including a logarithmic divergence through $\Sigma_5^{(3)} = 0$, with%
\footnote{
  Note that in ref.~\cite{Abreu:2023rco} momenta $\{4,5\}$ are incoming, while in our convention particles with momenta $\{1,2\}$ are in the initial state.
  The two conventions are related by reversing the order of momenta.
}
\begin{equation}
  \Sigma_5^{(3)} \, = \, \big(s_{12}s_{24}+s_{13}(s_{14}+s_{24})+s_{123}(s_{12}+s_{14})\big)^2-4s_{13}s_{14}s_{123}s_{124} \, .
\end{equation}
Some integrals must be normalized by $\sqrt{\Sigma_5^{(i)}}$ to satisfy the canonical differential equations, which gives rise to functions that are odd under the sign flip $\sqrt{\Sigma_5^{(i)}} \to -\sqrt{\Sigma_5^{(i)}}$, which we will refer to as $\mathcal{F}_{\Sigma_5}^-$.

As conjectured in ref.~\cite{Abreu:2023rco} and already observed in previous two-loop results with the same kinematics \cite{Badger:2024sqv,Badger:2024mir}, 
the pentagon functions that are divergent at $\Sigma_5^{(3)} = 0$, while present in the intermediate stages of the computation, cancel at the level of bare amplitudes.
Furthermore, we observe that all functions involving letters $\dd\log\Sigma_5^{(i)}$, as well as the odd functions $\mathcal{F}_{\Sigma_5}^-$ with weight higher than two cancel.
We also observe that pentagon functions involving the letter $\dd\log\Delta_5$ cancel out in the finite remainders (see the related discussions in ref.~\cite{Chicherin:2020umh,Bossinger:2022eiy}).
This highlights the importance of deriving a transcendental function basis that manifests such cancellations analytically \cite{Chicherin:2021dyp,Abreu:2023rco}.

One may wonder, however, if non-analytic behavior survives in the form of square-root like divergences or discontinuities of derivatives.
Indeed, some of the pentagon functions that involve letters algebraic in $\Sigma_5^{(i)}$ are not smooth across this threshold \cite{Abreu:2023rco}.
Consider, for example, the weight-two function $f^{(2)}_{32}$ that behaves near the threshold like
\begin{equation} \label{eq:f232-local}
  f^{(2)}_{32} \xrightarrow{\sif{3} \to 0} - 4 \pi^2 + \order{\sqrtsi{3}} \,.
\end{equation}
An associated master integral to which the amplitude is reduced over rationals is proportional at order $\epsilon^2$ to $f^{(2)}_{32}/\sqrtsi{3}$.
This implies that the amplitude should have a $1/\sqrtsi{3}$ divergence, unless the rational coefficients in \cref{eq:remainder} conspire to factorize $\sif{3}$.
We observe that the latter is indeed happening, giving the motivation for the definition of respective transcendental functions with $\sqrtsi{i}$ in the numerator in \cref{eq:sig5pent}.
These cancellations ensure that the amplitudes (and finite remainders) are continuous through $\Sigma_5^{(3)} = 0$.
Let us emphasize that \cref{eq:sig5pent}, together with the local behavior of the pentagon functions, ensure that 
the finite remainders are smooth everywhere else within the physical region.

\begin{figure}[ht]
  \centering
  \includegraphics[width=0.8\textwidth]{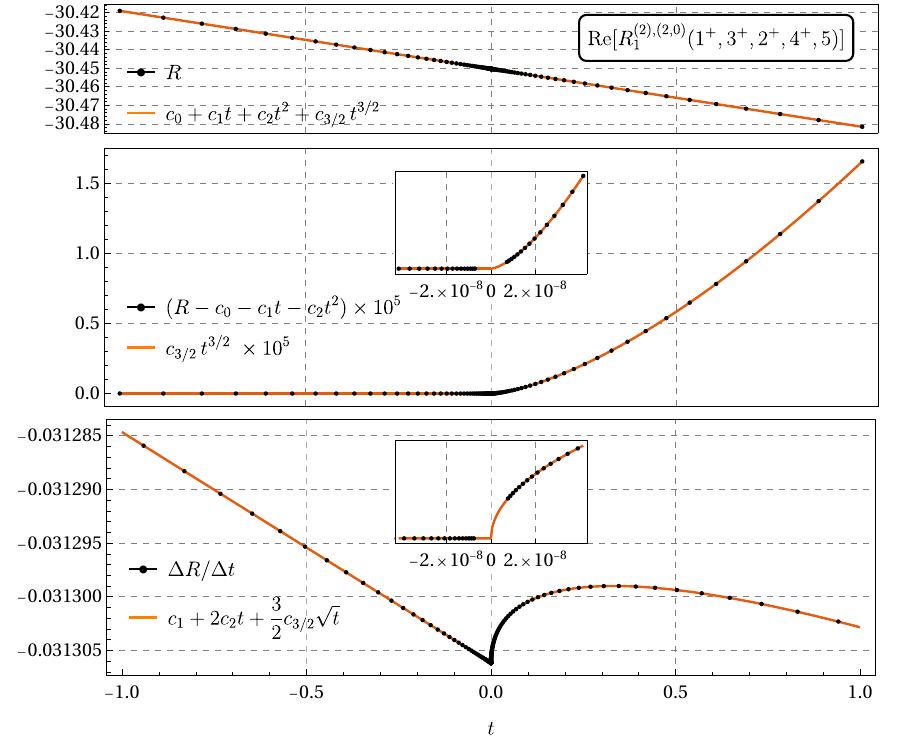}
  \caption{
    The real part of $R = R_1^{(2),(2,0)}(1^+,3^+,2^+,4^+,5) / A^{(0)}_1(1^+,3^+,2^+,4^+,5)$ evaluated along the path $\gamma(t)$ that crosses $\sif{3}(t)=0$ at $t=0$.
    The numerical evaluations are shown as black points, and the least-squares fit \eqref{eq:ls-fit} is shown in orange.
    In the middle panel, the analytic part of the fit is subtracted and the result is rescaled for the purpose of presentation.
    In the bottom panel, the derivative of the fit is compared to the numerical derivative from finite differences.
    The inset plots show the evaluations closest to $t=0$, and the extrapolated fits.
  }
  \label{fig:sigma5-threshold}
\end{figure}

This leaves the possibility of a discontinuity in derivatives. Indeed, from \cref{eq:f232-local} one expects that the finite remainder should behave locally as $c_0 + c_{1/2}\,\sqrtsi{3} + \cdots{}$.
To confirm the existence of such discontinuities, we perform a numerical analysis of the local behavior along a randomly chosen line in Mandelstam invariants $\gamma: t \in [-1,1] \mapsto (s_{12},s_{23},s_{34},s_{45},s_{15}, p_5^2)$ which crosses the $\sif{3}=0$ surface and lies fully within the physical scattering phase-space of the $s_{12}$ channel.
As an example, we show in \cref{fig:sigma5-threshold} the real part of $R_1^{(2),(2,0)}(1^+,3^+,2^+,4^+,5)$ evaluated on a line, which is normalized to the respective tree amplitude.
The ends of the line
\begin{equation*}
\begin{aligned}
  \nonumber
  \gamma(-1) &\simeq (5.4063947,-1.6848293,0.59976334,2.9808814,-2.8128731,0.57894543)\,,\\
  \gamma(1) &\simeq (5.4057359,-1.6847056,0.59918791,2.9808731,-2.8132074,0.57876760)\,,
\end{aligned}
\end{equation*}
are chosen such that the line crosses the vanishing surface of $\sif{3}$ at $t=0$.
We evaluate the finite remainder using our implementation described in \cref{sec:numeval}, and perform a least squares fit to the ansatz
\begin{equation} \label{eq:ls-fit}
  R_\text{fit}(t) =  c_0 + c_1 t + c_2 t^2 + c_{1/2}\,t^{1/2} + c_{3/2}\,t^{3/2}.
\end{equation}
To verify robustness of the fit, we perform separate fits on both sides of the threshold with differently spaced numerical samples, and observe excellent convergence.
We also verify that a fit to a quadratic polynomial in $t$ alone leads to an unstable coefficient $c_2$.
We observe that $c_{1/2}$ is consistent with zero, and that $c_{3/2}$ is non-vanishing. 
To further confirm this behavior we also compare the derivative of the fit $\dd{R_\text{fit}(t)}/\dd{t}$ with the approximate numerical derivative from finite differences of the same numerical samples.
Besides excellent agreement between the two, we observe a clear cusp at $t=0$ in the derivative, which signals a discontinuity in the second derivative.
We further note that $c_{1/2}=0$ implies non-trivial cancellations between different contributions in \cref{eq:remainder}.
Other partial helicity remainders, as well as the two-loop hard function (defined in \cref{sec:numeval}), exhibit similar behavior locally around the threshold.
We therefore confirm that the two-loop finite remainders exhibit non-analytic behavior at $\sif{3} = 0$.
This condition is satisfied by non-degenerate momenta configurations, \emph{i.e.}\ momenta which are linearly independent and do not correspond to the physical region boundary.
Equivalently, Gram determinants of the external momenta are not required to vanish at this threshold.
To the best of our knowledge, an explicit example of such an ``anomalous'' threshold in \emph{massless} scattering has not been previously reported.

We continue with an interesting observation regarding the differences between function spaces of the two-loop finite remainders with different helicities.
We find that both $R^{(2)}_1(1^+,2^+,3^-,4^-,5)$ and $R^{(2)}_1(1^+,2^-,3^+,4^-,5)$ do not contain any pentagon function monomials that are odd under $\sqrtsi{i}$ sign flips (although they do contain functions with letters depending on $\sif{i}$).
Equivalently, no monomials like in \cref{eq:sig5pent} appear in these helicity remainders.
On the other hand, $R^{(2)}_1(1^+,2^+,3^+,4^-,5)$ contains odd monomials \eqref{eq:sig5pent} with one of $\sqrtsi{i}$, while $R^{(2)}_1(1^+,2^+,3^+,4^+,5)$ contains odd monomials \eqref{eq:sig5pent} with two different $\sqrtsi{i}$.

Finally, we discuss the structure of divergences of the rational coefficients in \cref{eq:remainder}.
We observe that the set of denominator factors encountered in rational coefficients is identical, up to
permutations, to the one appearing in the LCA of the $Vjj$ amplitudes \cite{Abreu:2021asb,DeLaurentis:2025dxw}.
This is surprising, since the latter contain only contributions from planar diagrams, while the amplitudes considered here receive contributions from non-planar integral topologies.
In particular, we find that non-planar letters involving new quadratic and cubic polynomials in Mandelstam invariants appear in the contributing pentagon functions, but are absent in the denominators of rational coefficients.
This also implies that neither $\Delta_5$ nor $\Sigma_5^{(i)}$ appear as denominators in \cref{eq:ratFunc}.

\subsection{Numerical evaluation}\label{sec:numeval}

The main perturbative input for observable predictions at NNLO is the two-loop hard function,
defined for a given process from helicity- and color-summed squared finite remainders as
\begin{equation}\label{eq:hard-function}
  \mathcal{H} = \frac{1}{\mathcal{G}} \sum_{\text{color}, \text{helicity}} \mathcal{R}^\star \mathcal{R} \,, \qquad \mathcal{G} = \sum_{\text{color}, \text{helicity}} \abs{\mathcal{A}^{(0)}}^2 \,,
\end{equation}
In the sums we use the fact that all helicity configurations can be mapped to the chosen basis configurations listed in~\eqref{eq:listHelConfs}.
We expand it in powers of $\frac{\alpha_s}{2 \pi}$ through two loops as
\begin{subequations}
  \label{eq:hard-function-partial}
  \begin{align}
    \mathcal{H}^{(0)} &= 1  \,,  \\
    \mathcal{H}^{(1)} &= H^{(1)[0]} ~+~ \NF H^{(1)[1]} \,, \\
    \mathcal{H}^{(2)} &= H^{(2)[0]} ~+~ \NF H^{(2)[1]}~+~ \NF^2 H^{(2)[2]} \,.
  \end{align}
\end{subequations}

\begin{figure}[ht]
  \centering
  \begin{subfigure}[c]{0.9\textwidth}
        \centering
        \includegraphics[width=1\linewidth]{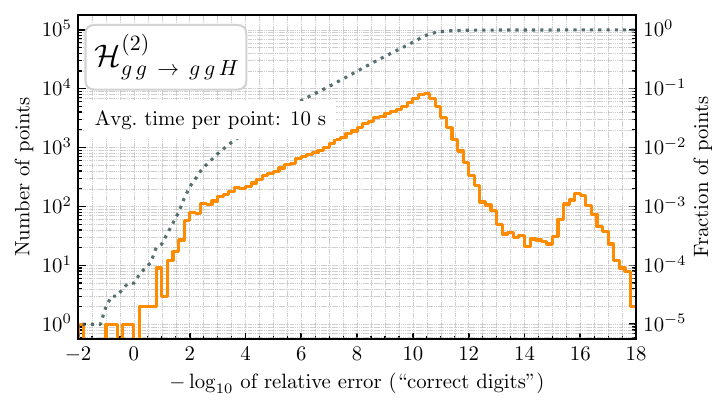}
  \end{subfigure}
  \begin{subfigure}[c]{0.9\textwidth}
        \centering
        \includegraphics[width=1\linewidth]{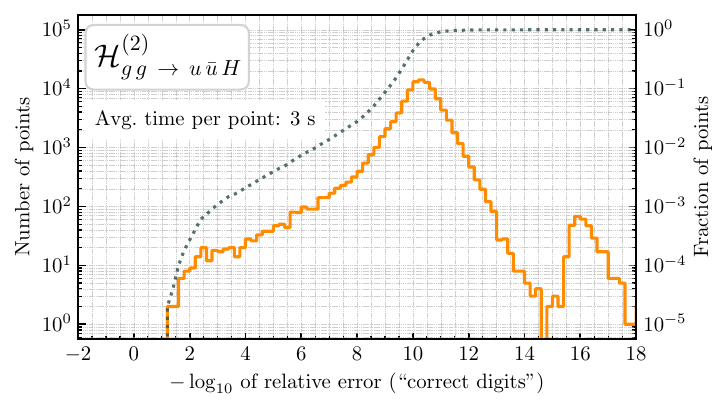}
  \end{subfigure}
  \caption{
    Distributions of correct digits for the two-loop hard functions of four-gluon and two-quark two-gluon representative partonic channels.
    The dashed curve shows the corresponding cumulative distribution.
  }
  \label{fig:stability}
\end{figure}

We have implemented the numerical evaluation of our analytic finite
remainders \eqref{eq:finRemainderAnalytic}, as well as the hard
functions in a \texttt{C++} library \cite{FivePointAmplitudes}.
For convenience, the reference evaluations of the hard functions in all independent partonic channels are given in \cref{sec:referenceEvaluations}.

To detect numerical instabilities of the evaluations, we employ the strategy described in \cite{DeLaurentis:2025dxw}. 
For each phase-space point, a second evaluation with a random boost applied to the momenta and a changed renormalization scale $\mu^\prime$ is performed.
The first evaluation is evolved using the relations in~\cref{eq:scaling-remainders} to the $\mu^\prime$, and the difference between the two is taken as an error estimate.
If the estimated error is above a given threshold, an evaluation in quadruple precision is performed.

\begin{figure}[ht]
  \centering
  \begin{subfigure}[c]{0.9\textwidth}
        \centering
        \includegraphics[width=1\linewidth]{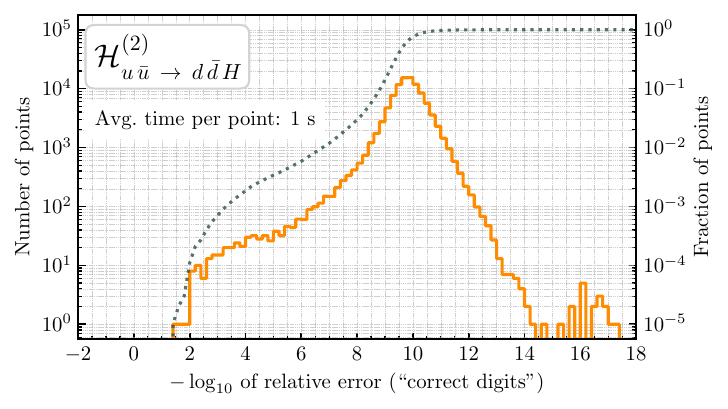}
  \end{subfigure}
  \begin{subfigure}[c]{0.9\textwidth}
        \centering
        \includegraphics[width=1\linewidth]{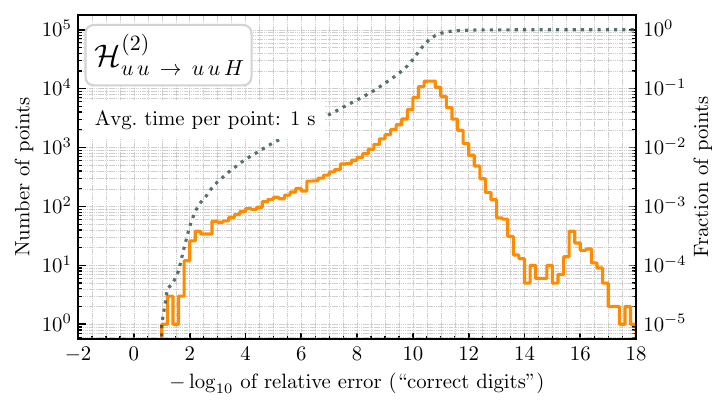}
  \end{subfigure}
  \caption{
    Distributions of correct digits for the two-loop hard functions of four-quark representative partonic channels.
    The dashed curve shows the corresponding cumulative distribution.
  }
  \label{fig:stability2}
\end{figure}

To assess the numerical performance of our implementation, we sample the hard functions over $100\,$k phase-space points generated with \texttt{NNLOjet} \cite{NNLOJET:2025rno}.
The points are generated for proton collisions with $\sqrt{s} = 13\,$TeV, with restrictions on jet transverse momenta $p_{T,j} > 30\,$GeV, and rapidities $\abs{y_{\text{jet}}} < 4.4$.
Here we set $\NC=3$ and $\NF=5$, and the renormalization scale is set dynamically to $\frac{1}{2}\qty(\sqrt{\mH^2 + p_{T,{\rm H}}^2} + \sum_{j} p_{T,j})$.

In \cref{fig:stability,fig:stability2} we show distributions of correct digits (base $10$ logarithm of relative error) and average evaluation times, for the two-loop hard functions of four representative partonic channels.
The average timings of single-threaded evaluation of two-loop hard functions are measured while running 16 threads on \texttt{AMD Ryzen 7 5700X} CPU with $3400\,$MHz frequency.
The timings include a second evaluation to estimate the accuracy and an additional quad precision evaluation whenever required.
The memory requirement (which includes quad-precision initialization) are around $1.4\,$GB for the four-gluon, $0.8\,$GB for two-gluon two-quark, and $0.5\,$GB for the identical four-quark channels.
The relative error is determined by comparing the result of running the library in ``production'' mode as described above with a separate quadruple-precision evaluation on a perturbed phase-space point.
Overall we observe good numerical performance, with peaks at around 10 digits and steeply falling distributions.
In the most difficult channel, $99\%$ of points could still be evaluated in double precision to reach the target of 2 digits.
This demonstrates that our results are suitable for application to NNLO QCD  computations.

The four-gluon channel is clearly numerically more challenging compared to the other partonic channels: both the average evaluation time and numerical stability are noticeably worse.
Let us comment on a few evaluations of $\mathcal{H}^{(2)}_{gg\to ggH}$ that have no correct digits.
Our error estimation strategy successfully determined that these evaluations are unstable, but the reevaluation in quadruple precision was not sufficient to obtain the desired precision.
We observe that such points have small $p_{T,{\rm H}}$, while not corresponding to any soft or collinear divergences.
This implies almost linearly dependent momenta, and therefore small $\Delta_5$.
Since this affects only a negligible fraction of evaluations, and such momentum configurations are phase-space suppressed, we expect that this issue would not hinder the feasibility of phenomenological applications through NNLO.

Finally, we note that while we computed the finite remainders in Catani's scheme for IR singularities, 
the library also provides evaluations of the hard functions in the MS scheme of refs.~\cite{Becher:2009cu,Becher:2009qa}.
The scheme change is calculated as described in ref.~\cite{DeLaurentis:2023izi}.

\subsection{Validation}

We performed several non-trivial checks to validate our results.

First, we confirmed internal consistency of the extensions implemented
in the \Caravel{} program as discussed in \cref{sec:numComputation}.
To test the implementation of the dimensional reduction of the gluon-Higgs vertices, 
we compared to an alternative computation which applies dimensional reconstruction and found agreement.
The newly obtained surface terms were confirmed to be IBP relations
using \FIRE5 \cite{Smirnov:2014hma} at a fixed numerical phase-space point.
Finally, we explicitly observed expected analytic properties
of the integral coefficients of the amplitude decomposed in the pure integral basis.
The denominators of the coefficients were found
to factorize into a product of a kinematics-independent polynomial in the dimensional regulator $\eps$
and the factors of the alphabet of the integral from eq.~\eqref{eq:denomfactors}.

Next, we performed checks on the finite remainders:
the cancellation of UV and IR poles in the assembly of the finite remainders was explicitly observed at every
finite-field sample that was used in the analytic reconstruction.

To confirm the correctness of the analytic reconstruction in \cref{sec:functionalReconstruction},
we compared the analytic coefficients to a reevaluation with \Caravel{} in a finite-field that was
not used for the reconstruction.
We also checked that the analytic form of the finite remainder
satisfy the expected scale-variation relations in eq.~\eqref{eq:scaling-remainders}.

Furthermore, we investigated if the finite remainders have the expected pole structure.
Due to the ordering of colored states in the LCA, some of these poles of the coefficient functions $r_i$ are spurious and cancel in the finite remainders.
We numerically checked this cancellation in the limits where the following denominator polynomials vanish:
$\{s_{13},s_{24}\}$ for the $ggggH$ channel, $\{s_{14},s_{23}\}$ for the $u\bar{u}ggH$ channel and $\{s_{14},s_{23}\}$ for $u\bar{u}d\bar{d}H$.
In order to analyze the scaling behavior of the remainders in collinear limits,
we evaluated the remainders on a sequence of points that approach collinear limits $p_i||p_j$ for massless $i,j=1,\dots,4$.
We observed the correct scaling behavior
\begin{equation}
	R^{(2)} \sim \frac{1}{\sqrt{s_{ij}}} \left(c_0 + c_1\log(s_{ij})+c_2\log^2(s_{ij})\right)   \qquad \text{for} \quad s_{ij} \to 0
\end{equation}
for adjacent $i,j$ with constants $c_k$ w.r.t.~$s_{ij}$.

To ensure correct assembly and evaluation of the hard functions,
which involves the maps of helicity amplitudes to our chosen basis in
\cref{eq:listHelConfs}, we checked their symmetries:
we confirmed invariance under Lorentz transformations, Bose symmetry for the exchange of gluon states and
charge conjugation symmetry.
The latter correspond to an invariance under the exchange of momenta $1\leftrightarrow2$ for the $\bar{u}u\to ggH$ channel and
$1\leftrightarrow2$ combined with $3\leftrightarrow 4$ for $\bar{u}u\to d\bar{d}H$.

Finally, we validated our results against independent computations: 
we compared the one-loop hard functions for all three processes
with the analytic leading-color expressions available in
\texttt{NNLOjet}~\cite{NNLOJET:2025rno}, finding full agreement.
The two-loop hard functions for the two-quark and four-quark channels
also agree with the results of ref.~\cite{Hartanto:2026xjz}
at individual phase-space points.
This comparison required a conversion to the IR subtraction scheme used in that reference.

\section{Conclusions}
\label{sec:conclusions}

We have computed the five-point two-loop amplitudes required for NNLO
QCD predictions for Higgs$+$2-jet production at hadron colliders. The
results are obtained in the leading-color approximation in the
heavy-top-quark limit. We provide both analytic expressions and an
efficient numerical implementation in terms of one-mass pentagon
functions \cite{Abreu:2023rco}, with evaluation times ranging
from 1 to 10 seconds per phase-space point across all channels.

The new amplitudes are among the most complicated analytic results
currently known for two-loop five-point one-mass kinematics. Even in
the leading-color approximation, they receive non-planar
contributions, and the effective Higgs interaction raises the
loop-momentum power counting beyond that of pure QCD. Both features
significantly increase the difficulty of the computation at the level
of integral reduction, numerical evaluation, and analytic
reconstruction. To overcome these challenges, we built on the existing
numerical unitarity framework implemented in \textsc{Caravel}. In
particular, we extended the underlying integrand decomposition by
master integrands and surface terms to cover the relevant
non-planar topologies and the enhanced momentum dependence of the
effective $ggH$ interaction. We also applied the dimensional-reduction
approach
\cite{Anger:2018ove,Abreu:2018jgq,Abreu:2019odu,Sotnikov:2019onv} to
the effective Higgs theory in order to determine the regulator
dependence more efficiently.

A central part of the present work is the analytic reconstruction of
the rational coefficient functions from finite-field evaluations of
the amplitudes. To this end, we proceeded in two stages. First, we
identified improved bases of coefficients with simpler denominator
structure through a basis-change algorithm.  Second, we constructed
compact partial-fraction ans\"atze in spinor-helicity variables. We
improved the basis-change algorithm in several ways, including
determining the basis-change matrix in a single finite field,
excluding certain pole residues from the search, and introducing a
more efficient search strategy for the most complex functions. A
further central ingredient is the construction of multivariate partial
fraction decompositions directly from numerical data in a finite
field, which we determine from a bivariate slice using functional
reconstruction in two variables.
These
improvements lead to a substantial reduction in the number of
numerical samples required for reconstruction, limiting ansatz sizes
to at most 12k unknowns.

The amplitudes computed in this work are expected to be closely related to the 
four-point form factor for the stress-tensor multiplet in $\mathcal{N}=4$ 
super Yang--Mills (SYM) \cite{Kotikov:2001sc}, which was recently determined at 
two-loop order \cite{Guo:2024bsd,Dixon:2024yvq}. For processes with fewer external states, 
this connection is well-established: the maximal transcendental part of the QCD 
amplitudes directly corresponds to the $\mathcal{N}=4$ form factor 
\cite{Kotikov:2001sc}, providing a powerful constraint on QCD results. While the 
relationship is less clear for higher-point cases, the amplitudes presented here 
provide the essential ingredients for a rigorous study into the  
correspondence between five-point QCD amplitudes and four-point $\mathcal{N}=4$ 
SYM form factors.

The analytic results reveal several interesting structural features of
the Higgs-plus-four-parton amplitudes. First, although non-planar
integrals enter already at leading color, the denominator structure of
the rational coefficients remains unexpectedly close to that of the
planar five-point $Vjj$ amplitudes
\cite{Abreu:2021asb,DeLaurentis:2025dxw}. In particular, the mostly
spurious poles appearing in the finite remainders match, up to
permutations, those observed in the planar case. Moreover, the
multivariate partial fraction decomposition succeeds in separating a
large number of denominator factors, confirming patterns observed in
earlier calculations. Second, we find a new type of non-analytic
behavior associated with the vanishing of $\Sigma_5^{(i)}$. Although
the finite remainders and hard functions remain continuous there, they
exhibit a cusp in the first derivative, equivalent to a discontinuity
in the second derivative. This is an
explicit example of such threshold behavior in a
massless scattering process. It would be interesting to understand
whether this reflects a more general feature of perturbative hard
functions in massless scattering, and whether it survives in more
physical quantities beyond fixed-order perturbation theory.

Looking ahead, together with integral computation and integration-by-parts reduction,
functional reconstruction remains one of the central bottlenecks in
analytic amplitude computations. The methods presented here are to a
large extent process independent, and should therefore apply equally
well to a broader class of two-loop scattering amplitudes. At the same
time, we have taken several steps towards automating the
reconstruction procedure. In particular, the determination of
multivariate partial fraction decompositions from a bivariate slice
replaces a substantial part of what would otherwise be educated
guesses by a more systematic algorithmic procedure.  We expect that
such progress will further improve the reconstruction strategy and
extend its scope towards subleading-color contributions and processes
with even more intricate kinematic dependence.

\acknowledgments
We thank Maximilian Klinkert for collaboration at early stages of this project. 
G.D.L.\ thanks the CERN-TH group for access to computing resources.
M.R.\ thanks Bernhard Mistlberger for useful discussion and sharing expressions used in checks of the collinear limits. 
We acknowledge access to Alps at the Swiss National Supercomputing Centre, Switzerland under the University of Zurich share with the project ID uzh41.
G.D.L.'s work is supported in part by the U.K.\ Royal Society through Grant URF\textbackslash R1\textbackslash 20109. 
M.R. is supported by the United States Department of Energy, Contract DE-AC02-76SF00515.
V.S.\ is supported by the European Research Council (ERC) under the European Union's Horizon 2020 research and innovation programme grant agreement 101019620 (ERC Advanced Grant TOPUP).

\appendix
\section{Reference evaluations}
\label{sec:referenceEvaluations}

We present in~\tab{tab:Hjj-benchmark} reference evaluations of the hard functions at the kinematic point
\begin{align} \label{eq:referencePointBis}
    p_1 &= \{-88.000000000000000,0,0,88.000000000000000\},\\\nn
    p_2 &= \{-88.000000000000000,0,0,-88.000000000000000\},\\\nn
    p_3 &= \{12.392045454545455,4.3855252841337518,8.2930798971534504,-8.0965909090909091\},\\\nn
    p_4 &= \{54.130681818181818,37.047749948312009,-8.2930798971534504,38.585227272727273\},\\\nn
    p_5 &= \{109.47727272727273, -41.43327523244576, 0, -30.488636363636363 \} \,, \nn
\end{align}
with the renormalization scale set to $\mu=63$, and $\NC=3$.
We remind the reader that whenever we are using the $\to$ notation, the first two
particles are to be understood as crossed to be incoming.

\begin{table}[H]
  \renewcommand{\arraystretch}{1.2}
  \centering
  \begin{adjustbox}{width=1\textwidth}
  \begin{tabular}{c*{5}{C{15ex}}}
    \toprule
 Channel  &  $H^{(1)[0]}$  &  $H^{(1)[1]}$  &  $H^{(2)[0]}$  &  $H^{(2)[1]}$  &  $H^{(2)[2]}$  \\
 \midrule
$g\,g\rightarrow g\,g\,H\,$ & $21.60693008$ & $-0.2042948925$ & $758.1516119$ & $-59.89542405$ & $0.1057461955$ \\
  \midrule
$g\,g\rightarrow u\,{\bar u}\,H\,$ & $23.79850997$ & $-0.8520056545$ & $871.0125340$ & $-73.34374616$ & $-0.02463963835$ \\
$g\,u\rightarrow g\,u\,H\,$ & $35.75432424$ & $-1.667934932$ & $1346.890607$ & $-146.8303568$ & $2.080699995$ \\
${\bar u}\,u\rightarrow g\,g\,H\,$ & $-0.7439021475$ & $-0.4143437544$ & $323.4456429$ & $-16.17511090$ & $-1.000952210$ \\
\midrule
${\bar u}\,u\rightarrow d\,{\bar d}\,H\,$ & $13.52201772$ & $-1.387398374$ & $500.2142706$ & $-45.24344286$ & $-0.5366784223$ \\
$d\,u\rightarrow d\,u\,H\,$ & $65.11323369$ & $-3.575719056$ & $3180.810875$ & $-369.2573793$ & $8.083011923$ \\
${\bar d}\,u\rightarrow {\bar d}\,u\,H\,$ & $38.89392891$ & $-3.575719056$ & $1532.625650$ & $-254.1536590$ & $8.083011923$ \\
${\bar u}\,u\rightarrow u\,{\bar u}\,H\,$ & $31.68055000$ & $-1.355979469$ & $1049.884750$ & $-103.8069713$ & $0.7504319454$ \\
$u\,u\rightarrow u\,u\,H\,$ & $64.67929567$ & $-3.497230387$ & $3141.865903$ & $-362.0898486$ & $7.842910126$ \\
\bottomrule
  \end{tabular}
  \end{adjustbox}
  \caption{
   Reference evaluations of hard functions for the different $pp \to H\,jj$ partonic channels.
  }
  \label{tab:Hjj-benchmark}
\end{table}

\section{Ancillaries}\label{app:ancillaries}

The results are subdivided by folders into the three
channels,\footnote{For \texttt{Python}, these are in the
\texttt{antares\_results/Hjj/HTL/} folder.}
\begin{center}
  1. \texttt{ggggH/}  $\kern10mm$ 2. \texttt{uubggH/} $\kern10mm$ 3. \texttt{uubddbH/}
\end{center}
each one of them containing a single file
\begin{flushleft}
  $\quad$ a. $\;$ \texttt{basis\_transcendental.txt}
\end{flushleft}
where the vector of pentagon functions, $G_j$, is given with one
monomial per line.

The results are then further organized by representative helicities,
\noindent
\begin{center}
\begin{minipage}[t]{0.22\textwidth}
\begin{center}
  \texttt{ggggH/ppmm/} \\
  \phantom{ggggH}\texttt{/pppm/} \\
  \phantom{ggggH}\texttt{/pppp/}
\end{center}
\end{minipage}\hspace{0.02\textwidth}
\begin{minipage}[t]{0.22\textwidth}
\begin{center}
  \texttt{uubggH/pmpp/} \\
  \phantom{uubggH}\texttt{/pmpm/}
\end{center}
\end{minipage}\hspace{0.02\textwidth}
\begin{minipage}[t]{0.22\textwidth}
\begin{center}
  \texttt{uubddbH/pmpm/}
\end{center}
\end{minipage}
\end{center}
Within each of these folders, for each helicity that is mapped onto
the helicity of reference by permutations, and for each leading-color
partial amplitude, uniquely labelled by $\ell$-loop, $n_c$ power of
$\NC$ and $n_f$ powers of $\NF$, we have a file
\begin{flushleft}
  $\quad $b. $\;$ \texttt{matrix\_\{$helicity$\}\_\{$\ell$\}L\_Nc\{$n_c$\}\_Nf\{$n_f$\}}
\end{flushleft}
where the curly brackets are to be replaced with appropriate
values. These sparse matrices $M_{ij}$ are provided in coordinate
format with three columns: row index, column index, value.

Lastly, the basis $\tilde r_i$ of spinor-helicity rational functions
is subdivided by mass dimension, provided in three directories
\begin{center}
  \texttt{md\_0/} \hspace{10mm} \texttt{md\_2/} \hspace{10mm} \texttt{md\_m4/}
\end{center}
with each subspace organized as
\begin{flushleft}
  $\quad$ c. $\;$ \texttt{basis.txt} master file, referring to the
  corresponding \texttt{coeff\_\{$\#$\}.tex} files.
\end{flushleft}
The \texttt{\_\_init\_\_.py} files show how to load the results in
\texttt{Python} using the \textsc{antares} library, and make the
expressions available for import. For example, from the parent folder,
the vector space obtained by combining the three disjoint subspaces
can be imported as
\begin{flushleft}
  \texttt{from antares\_results.Hjj.HTL.\{$channel$\}.\{$helicity$\} import lTerms}
\end{flushleft}

The file \texttt{Hjj/remainders} provides a function to evaluate the
remainders numerically or semi-numerically, the latter leaving the
pentagon functions symbolic. It can be imported and inspected as
\begin{flushleft}
  \texttt{from antares\_results.Hjj.remainders import remainder}\\
  \texttt{help(remainder)}
\end{flushleft}
As part of the GitHub Actions continuous-integration workflow,
\texttt{Hjj/test\_remainders} is run to verify the expressions against
cached target values.

For simplicity, in the \texttt{Mathematica} version, we provide the
rational functions as a single file for each representative helicity,
\begin{flushleft}
  $\quad$ c. $\;$ \texttt{basis\_rational\_\{$helicity$\}.m}
\end{flushleft}
where permutations have also been removed. Files \texttt{assembly.m}
and \texttt{functions.m} are provided for evaluating the results.

\section{Tree amplitudes}\label{app:trees}

We provide here the tree amplitudes of
\cref{eq:4g-color-dec,eq:2q2g-color-dec,eq:4q-color-dec}
in spinor-helicity variables.
Respectively, the gluon amplitudes are,
\begin{equation}
	A_1^{(0)}(1^+,2^+,3^+,4^+,5) = \frac{(p_h^2)^2}{⟨12⟩⟨23⟩⟨34⟩⟨41⟩}\, ,
\end{equation}
\begin{gather}
	A_1^{(0)}(1^+,2^+,3^+,4^-,5) = \frac{[23]⟨4|2+3|1]^2}{⟨23⟩⟨34⟩[34]s_{234}} +
                                      \frac{[12]⟨4|1+2|3]^2}{⟨12⟩⟨14⟩[14]s_{124}} +
                                      \frac{[13]^2⟨4|1+3|2]^2}{⟨14⟩[14]⟨34⟩[34]s_{134}} + \nonumber \\
                                      \kern20mm\frac{[13]⟨24⟩(⟨12⟩[12]^2⟨14⟩+⟨23⟩[23]^2⟨34⟩)}{⟨12⟩⟨14⟩[14]⟨23⟩⟨34⟩[34]} +
                                      \frac{[13]^2(s_{13}-2s_{123})}{⟨12⟩[14]⟨23⟩[34]} \, ,
\end{gather}
\begin{equation}
	A_1^{(0)}(1^+,2^+,3^-,4^-,5) = \frac{⟨34⟩^3}{⟨12⟩⟨23⟩⟨41⟩}+\frac{[12]^3}{[23][34][41]} \,,
\end{equation}
\begin{equation}
	A_1^{(0)}(1^+,2^-,3^+,4^-,5) = \frac{⟨24⟩^4}{⟨12⟩⟨23⟩⟨34⟩⟨41⟩}+\frac{[13]^4}{[12][23][34][41]} \,.
\end{equation}
The 2-quark amplitudes are,
\begin{equation}
\begin{split}
	A_3^{(0)}(1^+,2^-,3^+,4^+,5) = \frac{[14]⟨2|1+4|3]^2}{⟨12⟩[12]⟨24⟩s_{124}} -
                                       \frac{[13]⟨1|2+3|4]⟨2|1+3|4]}{⟨12⟩[12]⟨13⟩s_{123}} + \\
                                       \frac{[14]⟨2|1+4|3]}{[12]⟨13⟩⟨24⟩} +
                                       \frac{⟨1|3+4|1]⟨2|3+4|1]}{[12]⟨13⟩⟨24⟩⟨43⟩} \,,
\end{split}
\end{equation}
\begin{equation}
	A^{(0)}_3(1^+,2^-,3^+,4^-,5) = \frac{⟨14⟩⟨24⟩^2}{⟨12⟩⟨13⟩⟨43⟩}-\frac{[13]^2[23]}{[12][24][43]} \,,
\end{equation}
\begin{equation}
	A^{(0)}_3(1^+,2^-,3^-,4^+,5) = \frac{⟨23⟩^3}{⟨12⟩⟨24⟩⟨43⟩}-\frac{[14]^3}{[12][13][43]} \,,
\end{equation}
Finally, the 4-quark tree amplitudes are,
\begin{equation}
	A^{(0)}_5(1^+,2^-,3^+,4^-,5) = -\frac{⟨24⟩^2}{⟨12⟩⟨34⟩}-\frac{[13]^2}{[12][34]}  \,,
\end{equation}
\begin{equation}
	A^{(0)}_5(1^+,2^-,3^-,4^+,5) = \frac{⟨23⟩^2}{⟨12⟩⟨34⟩}+\frac{[14]^2}{[12][34]} \,.
\end{equation}
We verified these expressions against those available in the literature
\cite{Budge:2020oyl, Kauffman:1996ix, Dawson:1991au,
  Badger:2009hw}. For the amplitude $\mathcal{A}_3$ we apply charge
conjugation to compare a $\bar u u ggH$ to a $u\bar u ggH$ amplitude.
Due to the color structure $\bigl(T^{a_3}
T^{a_4}\bigr)^{\,\,\,\bar{i}_2}_{i_1}$ this operation results in a
$3\leftrightarrow 4$ swap in the color-stripped amplitudes.

\section{Infrared operators}
\label{sec:CataniOperators}

Here we list the expressions of the IR pole operators in Catani's scheme with the 
conventions of the $\beta$-function coefficients and amplitude normalization given in~\cref{sec:renormalization}.
They act on the partial  amplitudes for the default LC color structures 
$\tracep{T^{a_1} T^{a_2} T^{a_3} T^{a_4}}$, $\left(T^{a_3}T^{a_4}\right)^{\,\,\,\bar{i}_2}_{i_1}$ and 
$\delta^{\bar{i}_4}_{i_1}\delta^{\bar{i}_2}_{i_3}$ of the 
$ggggH$, $u\bar uggH$ and $u\bar u d \bar dH$ channels, respectively.

In the LC limit, the operator $\mathbf{I}^{(1)}$ reads for the three channels
\begin{align}
    \mathbf{I}^{(1)}_{4gH} =&\ N_\epsilon\,\left(\frac{\NC}{2\epsilon^2} + \frac{11\NC - 2 \NF}{12\epsilon}\right) \left(\left(-\tfrac{\mu^2}{s_{12}}\right)^\epsilon + \left(-\tfrac{\mu^2}{s_{23}}\right)^\epsilon + \left(-\tfrac{\mu^2}{s_{34}}\right)^\epsilon + \left(-\tfrac{\mu^2}{s_{14}}\right)^\epsilon \right)\,,\\
\begin{split}
    \mathbf{I}^{(1)}_{2q2gH} =&\ \frac{N_\epsilon
    }{2\epsilon^2}\,\NC\,\left( \left(-\tfrac{\mu^2}{s_{13}}\right)^\epsilon + \left(-\tfrac{\mu^2}{s_{24}}\right)^\epsilon + \left(-\tfrac{\mu^2}{s_{34}}\right)^\epsilon \right) \\
    &\ + \frac{ N_\epsilon}{12\epsilon}\biggl[ \NC \left( 10\left(-\tfrac{\mu^2}{s_{13}}\right)^\epsilon + 10\left(-\tfrac{\mu^2}{s_{24}}\right)^\epsilon + 11\left(-\tfrac{\mu^2}{s_{34}}\right)^\epsilon \right) \\
    &\ \qquad\qquad - \NF \left( \left(-\tfrac{\mu^2}{s_{13}}\right)^\epsilon + \left(-\tfrac{\mu^2}{s_{24}}\right)^\epsilon + 2\left(-\tfrac{\mu^2}{s_{34}}\right)^\epsilon \right)\biggr]\,,
\end{split}\\
   \mathbf{I}^{(1)}_{4qH} =&\ N_\epsilon\, \frac{\NC}{2} \left(\frac{1}{\epsilon^2} + \frac{3}{2\epsilon}\right)\, \left(\left(-\tfrac{\mu^2}{s_{14}}\right)^\epsilon +\left(-\tfrac{\mu^2}{s_{23}}\right)^\epsilon\right) \,,
\end{align}
with $N_\epsilon = \frac{e^{\epsilon\gamma_{\mathrm{E}}}}{\Gamma(1-\epsilon)}$ and $s_{ij} = 2p_i \cdot p_j + i0$, where the momenta $p_{i,j}$ are understood to be in the all-outgoing convention.

The operator $\mathbf{I}^{(2)}$ is of the general form\footnote{Note that the expressions given here assume that $\alphas/2\pi$ is used as expansion parameter for the amplitude, beta function and IR operator, whereas $\alphas/4\pi$ is used in ref.~\cite{Becher:2009qa}.}~\cite{Catani:1998bh, Becher:2009cu, Becher:2009qa}
\begin{equation}
\label{eq: Catani Operator I2}
\begin{split}
	\mathbf{I}^{(2)}(\mu, \epsilon) = & \frac{1}{2} \mathbf{I}^{(1)}(\mu, \epsilon) \left(\mathbf{I}^{(1)}(\mu, \epsilon) - \frac{2\beta_0}{\epsilon}\right)\\
		& + \frac{N_\epsilon}{N_{2\epsilon}} \left(\frac{\gamma_1^{\mathrm{cusp}}}{8} + 
		\frac{\beta_0}{\epsilon}\right) \mathbf{I}^{(1)}(\mu, 2\epsilon) - \frac{1}{\epsilon} \mathbf{H}^{(2)}_{\mathrm{R.S.}}
\end{split}
\end{equation}
where the cusp anomalous dimension $\gamma_1^{\mathrm{cusp}}$ is given by
\begin{equation}
	\gamma_1^{\mathrm{cusp}} = \left(\frac{268}{9}-\frac{4\pi^2}{3}\right)\NC -\frac{40}{9}\NF.
\end{equation}
The operator $\mathbf{H}^{(2)}_{\mathrm{R.S.}}$ comes with a simple pole in $\epsilon$, made explicit in eq.~\eqref{eq: Catani Operator I2}.
For the three channels $ggggH$, $u\bar{u}ggH$ and $u\bar{u}d\bar{d}H$, its leading color expression read
\begin{align}
	\mathbf{H}^{(2)}_{\mathrm{R.S.}, 4gH} &= \frac{1}{3\epsilon}\left(\frac{60 + 11\pi^2 + 72 \zeta_3}{48}\NC^2  - \frac{178 + 3\pi^2}{72} \NC\NF + \tfrac{5}{9}\NF^2\right)\,,\\
	\mathbf{H}^{(2)}_{\mathrm{R.S.}, 2q2gH} &= \frac{1}{54\epsilon} \left(\frac{769 - 3(11\pi^2 - 648\zeta_3)}{32} \NC^2-\frac{406-3\pi^2}{16} \NC\NF + 5 \NF^2 \right)\,,\\
	\mathbf{H}^{(2)}_{\mathrm{R.S.}, 4qH} &= \frac{1}{864\epsilon} \left((409 - 9 (11\pi^2 - 168\zeta_3)) \NC^2 - 2(50 - 9 \pi^2) \NC\NF\right)\,,
\end{align}
respectively.

\section{Parametric slices of phase space} \label{sec:slices}
In this appendix, we review the parametric univariate and bivariate
slices used in the reconstruction. 

\paragraph{Univariate slice}
The univariate slice is obtained in analogy with a BCFW shift
\cite{Britto:2005fq}, now involving all variables
\cite{PageSAGEXLectures,Abreu:2021asb,Abreu:2023bdp,Elvang:2008vz},
\begin{equation}
  \lambda_i \rightarrow \lambda_i + t \, x_i \eta \, , \qquad
  \tilde\lambda_i \rightarrow \tilde\lambda_i + t \, y_i \tilde\eta \, ,
\end{equation}
where $x_i$ and $y_i$ are chosen at random, subject to momentum
conservation,
\begin{equation}
  \sum_i y_i \lambda_i \tilde\eta + x_i \eta\tilde\lambda_i = \sum_i
  x_i y_i \eta\tilde\eta = 0 \, .
\end{equation}
This construction can also be generalized to arbitrary quotient rings
\cite{Campbell:2024tqg}.

\paragraph{Bivariate slice}
Similarly to the univariate slice, we construct a bivariate slice by
shifting all spinors,
\begin{align}\label{eq:bivariate-spinor-slice}
  \lambda_i &\rightarrow \lambda_i + u \, x_{1,i} \eta + v \, x_{2,i} \theta \, , \\
  \tilde\lambda_i &\rightarrow \tilde\lambda_i + u \, y_{1,i} \tilde\eta + v \, y_{2,i} \tilde\theta \, ,
\end{align}
where $\eta$, $\tilde\eta$, $\theta$, and $\tilde\theta$ are fixed
random reference spinors, and $x_{1,i}$, $x_{2,i}$, $y_{1,i}$, and
$y_{2,i}$ are chosen at random such that momentum conservation is
satisfied for any value of $u$ and $v$. At six-point massless
kinematics, which we use as a representation of five-point one-mass
kinematics, this results in 20 equations defining a codimension-10
variety in a 24-dimensional space.\footnote{A random point on this
variety is selected using \texttt{syngular.Ideal.point\_on\_variety}
\cite{giuseppe_de_laurentis_2026_18881385}.  The full construction
from \cref{eq:bivariate-spinor-slice} is implemented in
\texttt{lips.Particles.bivariate\_slice} \cite{
  giuseppe_de_laurentis_2026_18890565}}

\bibliographystyle{JHEP}
\bibliography{main.bib}

\end{document}